\documentclass[aps,prd,superscriptaddress,10]{revtex4-2}
\pdfoutput=1
\def\bng{\bngx}

\font\bngx=bang10

\def\*#1*#2{o\null{#2}{#1}}

\def\sh#1{\setbox0=\hbox{#1}%
     \kern-.02em\copy0\kern-\wd0
     \kern.04em\copy0\kern-\wd0
     \kern-.02em\raise.0433em\box0 }   
\usepackage{graphicx,color}
\usepackage{amssymb,amsmath,amsfonts}              
\usepackage{orcidlink}             
\usepackage{bm}
\usepackage{booktabs}
\usepackage{slashed}
\usepackage{hyperref}
\usepackage[caption=false]{subfig}
\usepackage{xspace}
\usepackage{physics}
\usepackage{mathrsfs}

\def \beq  {\begin{equation}}
\def \eeq  {\end{equation}}
\def \beqar {\begin{eqnarray}}
\def \eeqar {\end{eqnarray}}
\def\be {\begin{equation}}
\def\ee {\end{equation}}
\def\bea {\begin{eqnarray}}
\def\eea {\end{eqnarray}}
\def\bc {\begin{center}}
	\def\ec {\end{center}}
\def\nn {\nonumber}
\def\eps {\epsilon}

\def\mn {\mu\nu}

\def\({\left(}
\def\){\right)}
\def\[{\left[}
\def\]{\right]}

\def \bfL{\mathbf{L}\xspace}
\def \suN{$SU(N_c)$\xspace}
\def \zN{{$Z(N_c)$}\xspace}
\def \Td{$T_d$\xspace}
\def \begb{\text{\bng b}}
\def \begd{\text{\bng d}}
\def \begt{\text{\bng 3}}
\def\signf{{\mathcal S}}
\def \hatl{\hat{l}}

\def\sqr#1#2{{\vcenter{\vb*ox{\hrule height.#2pt
\hbox{\vrule width.#2pt height#1pt \kern#1pt
\vrule width.#2pt}\hrule height.#2pt}}}}

\definecolor{refcol}{RGB}{178,34,34}
\usepackage{microtype}

\begin{document}
\preprint{YITP-24-150, RIKEN-iTHEMS-Report-25}

\title{Shear and bulk viscosity for a pure glue theory using an effective matrix model}
\author{Manas Debnath\,\orcidlink{0009-0003-8488-7146}}
\email{manas.debnath@niser.ac.in}
	\affiliation{School of Physical Sciences, National Institute of Science Education and Research, An OCC of Homi Bhabha National Institute,  Jatni, Khurda 752050, India}
\author{Ritesh Ghosh\,\orcidlink{0000-0002-6740-7038}}
\email{Ritesh.Ghosh@asu.edu}
\affiliation{College of Integrative Sciences and Arts, Arizona State University, Mesa, Arizona 85212, USA}
\author{Najmul Haque\,\orcidlink{0000-0001-6448-089X}}
\email{nhaque@niser.ac.in}
	\affiliation{School of Physical Sciences, National Institute of Science Education and Research, An OCC of Homi Bhabha National Institute,  Jatni, Khurda 752050, India}
\author{Yoshimasa Hidaka\,\orcidlink{0000-0001-7534-6418}}
\email{yoshimasa.hidaka@yukawa.kyoto-u.ac.jp}
\affiliation{Yukawa Institute for Theoretical Physics, Kyoto University, Kyoto 606-8502, Japan}
\affiliation{RIKEN Center for Interdisciplinary Theoretical and Mathematical Sciences (iTHEMS), RIKEN, Wako 351-0198, Japan}
\author{Robert D. Pisarski
\,\orcidlink{0000-0002-7862-4759}
}
\email{pisarski@bnl.gov}
\affiliation{Department of Physics, Brookhaven National Laboratory, Upton, NY 11973}

\begin{abstract}
At nonzero temperatures, the deconfining phase transition
can be analyzed using an effective matrix model to characterize the change in holonomy. The model includes gluons and two-dimensional ghost fields in the adjoint representation, or ``teens''.  As ghosts, the teen fields are responsible for the decrease of the pressure as $T \rightarrow T_d$, with \Td\  the transition temperature for deconfinement. Using the solution of this matrix model for a large number of colors, the parameters of the teen fields are adjusted so that the expectation value of the Polyakov loop is close to the values from the lattice. The shear, $\eta$, and bulk, $\zeta$, viscosities 
{\color{black} are computed at nonzero holonomy
to leading logarithmic order in weak coupling}.
In the pure glue theory, the value of the Polyakov loop is relatively large in the deconfined phase, $\approx 1/2$ at \Td.  Consequently, if $s$ is the entropy density, while $\eta/s$ decreases as $T\rightarrow T_d$, it is still well above the conformal bound. In contrast, $\zeta/s$ is largest at \Td, comparable to $\eta/s$, then falls off rapidly with increasing temperature and is negligible by $\sim 2 T_d$.
\end{abstract}

\maketitle

\section{Introduction} 
The behavior of gauge theories at nonzero temperature is a question of fundamental importance, affecting both the behavior of the collisions of heavy ions at high energies and the evolution of the early universe. While quantities in thermal equilibrium can be computed from numerical simulations on the lattice, quantities near equilibrium are not easy to measure. Notably, this includes the
transport coefficients, for which, at present, the lattice can only provide rough estimates.

In a \suN gauge theory without dynamical quarks, the
simplest way to measure the deconfining phase
transition is through the Polyakov loop, $\ell$,
which is the trace of the Wilson line in the direction of imaginary time,
$\ell = \tr\bfL/N_c$.
When $N_c>2$, the trace of higher powers of the Wilson line,
$\ell_j = \tr\bfL^j/N_c$, $j=1 \ldots (N_c-1)$
are also independent, gauge invariant quantities, and so should be included for a 
complete description of deconfinement. Alternatively, one can consider the $N_c-1$ 
eigenvalues of the thermal Wilson line. These are gauge-invariant and measure the
holonomy of the Wilson line in imaginary time. They can naturally be used to construct a
matrix model for deconfinement.

If the deconfining phase transition occurs at a temperature \Td, the Polyakov loops, or 
equivalently the holonomy, is non-trivial from \Td to a couple of 
times \Td. This region 
can be described as a semi-quark gluon plasma (semi-QGP), where the 
effects of partial deconfinement play a dominant role.

In this paper we use an effective matrix model,
which describes a partially
deconfined phase with nonzero holonomy~\cite{Gross:1980br,Weiss:1980rj,Enqvist:1990mm,Bhattacharya:1990hk,Bhattacharya:1992qb,Gocksch:1993iy,KorthalsAltes:1999cp,Pisarski:2000eq,Dumitru:2002cf,Dumitru:2003hp,Dumitru:2004gd,Dumitru:2005ng,Oswald:2005vr,Pisarski:2006hz}. 
Since we compute scattering processes,
we need the generalization of 
hard thermal loops (HTL's)~\cite{Pisarski:1988vd,Braaten:1989mz,Braaten:1989kk,Braaten:1990it,Braaten:1991gm,Kelly:1994ig,Kelly:1994dh,Blaizot:2001nr,Bellac:2011kqa,Haque:2014rua,Haque:2024gva} to nonzero holonomy~\cite{Hidaka:2008dr,Hidaka:2009ma,Hidaka:2009hs,Hidaka:2009xh,Dumitru:2010mj,Dumitru:2012fw,Kashiwa:2012wa,Pisarski:2012bj,Bicudo:2013yza,Kashiwa:2013rm,Lin:2013efa,Lin:2013qu,Smith:2013msa,Bicudo:2014cra,Gale:2014dfa,Hidaka:2015ima,Pisarski:2016ixt,KorthalsAltes:2020ryu,Hidaka:2020vna}.
The computation of the shear~\cite{Arnold:2000dr,Arnold:2003zc}
and bulk \cite{Arnold:2006fz} viscosities
follows the computation to leading logarithmic order
in ordinary perturbation theory, at
zero holonomy.  

Besides using a model with nonzero holonomy, our
principal assumption is that the gauge coupling is
of moderate strength, even down to $T \sim T_d$.  For
example, in an effective three-dimensional theory,
the coupling constant runs not as $\alpha_s(T_d)$,
but from two loop calculations, closer to $\alpha_s(2 \pi T_d)$ 
\cite{Braaten:1995jr,Laine:2005ai}.  We then assume
that both gluons and teen fields act
like well defined quasiparticles, even down to temperatures
$\sim T_d$.

At nonzero holonomy, besides the usual gluons, it is also necessary to add new, fictional degrees of freedom, which we
call ``teens''.  We denote this by the Bengali character $\begt$~\cite{Ghosh_pdfLaTeX_Bengali_2022}, which is pronounced teen, and represents the number three.
This is because our theory has gluons, Faddeev-Popov ghosts, and teens.  For future reference, we also use
the Bengali characters $\begb$, pronounced ``baw'', and
$\begd$, as ``daw''.

At large $N_c$, in the deconfined phase, the pressure is $\sim N_c^2$, while it is $\sim N_c^0$ in the confined phase.  Thus, to model the theory as $T \rightarrow T_d$, it is necessary to add something so
that the leading term in the pressure vanishes.  We take the teen fields to lie in the adjoint representation, so their pressure is $\sim N_c^2$.  For the term in the pressure
$\sim N_c^2$ to vanish, these teen fields must be
ghosts, with negative pressure.
Because the leading nonperturbative term in the pressure is $\sim T^2$ \cite{Pisarski:2006hz}, we take
the teens as two-dimensional fields, isotropically embedded in four dimensions.  

The outline of the paper is as follows. In Sec.~\ref{sec:matrix}, we discuss how to characterize non-trivial holonomy. In Sec.~\ref{sec:matrix_model} we review the solution of the matrix model at large $N_c$~\cite{Pisarski:2012bj,Lin:2013qu,Nishimura:2017crr,KorthalsAltes:2019yih,KorthalsAltes:2020ryu}. We describe how the previous solution can
be modified to give close agreement to lattice values for
the Polyakov loop. In Sec.~\ref{sec:pert_thy_nonzero_holonomy}, we describe how to compute in perturbation theory at nonzero holonomy. In Secs.~\ref{sec:shear_viscosity} and~\ref {sec:bulk_viscosity}, we calculate the shear viscosity and bulk viscosities respectively, considering $2\rightarrow2$ scattering involving gluon and teen,
{\color{black}
computing to leading logarithmic order in weak coupling.
}
In Sec.~\ref{sec:summary}, we summarize the results.
\section{The thermal Wilson line and non-trivial holonomy}
\label{sec:matrix}
At a nonzero temperature $T$, the thermal Wilson line is
\begin{equation}
    \bfL(\bm{x})
    =  \mathcal{P} \exp \left( i g \int^{1/T}_0  
    A_0(\bm{x},\tau) d\tau \right) , 
    \label{thermal_Wilson_line}
\end{equation}
where $\cal P$ refers to path, which is here time ordering, and the
Euclidean time $\tau : 0 \rightarrow 1/T$.  Under a gauge transformation
$\Omega(\bm{x},\tau)$ this transforms as 
\begin{equation}
\bfL(\bm{x}) \rightarrow \Omega^\dagger(\bm{x},1/T)\,
\bfL(\bm{x}) \, \Omega(\bm{x},0) \; .
\end{equation}
%
%
We can define local gauge transformations as those
which fall off at spatial infinity, $\Omega(\bm{x}, \tau) \rightarrow
\vb*{1}$ as $\bm{x} \rightarrow \infty$.  There are also global gauge rotations, which are constant in space.  Normally these are innocuous,
as they just rotate the gauge fields.  
However, in a pure gauge theory, we can take a global gauge rotation in the center of the gauge group.  For \suN these are
elements of \zN, where $\Omega_k = \mathrm{e}^{2 \pi i k/N_c} \vb*{1}$,
with $k= 1\ldots N_c$.  Since the $\Omega_k$ commute with all
group elements, the gluons remain unchanged.  Thus we can take the gauge
transformation at $\tau = 1/T$ equal to that at $\tau =0$, times
an element of \zN:
and thus, the traced Polyakov loop, $\tr\,\bfL(\bm{x})$, is a gauge invariant operator.
The gauge transformations can be generalized into aperiodic transformations
up to a constant element in the center of the gauge group. 
\begin{equation}
\Omega(\bm{x},1/T) = \mathrm{e}^{2 \pi i k/N_c}\, \Omega(\bm{x},0) \; ; \;
k = 1, \ldots, (N_c-1) \; .
\end{equation}
While locally, the gluons are insensitive to the $\Omega_k$,
the thermal Wilson line is not:
\begin{equation}
    \bfL(\bm{x}) \rightarrow \mathrm{e}^{2 \pi i  k/N_c} 
    \Omega^\dagger(\bm{x},0) \, \bfL(\bm{x}) \, \Omega(\bm{x},0) \; .
\end{equation}
From this form, we can take traces of the powers of the thermal Wilson line,
\begin{equation}
\ell_j(\bm{x}) = \frac{1}{N_c}\,  \tr \, \bfL(\bm{x})^j \; .
\end{equation}
These are all invariant under gauge transformations, but
transform under global \zN rotations as 
\begin{equation}
\ell_j(\bm{x}) \rightarrow \mathrm{e}^{2 \pi i j k/N_c} \, \ell_j(\bm{x}) \;.
\end{equation}
%
This global \zN symmetry is related to a one-form
\zN symmetry~\cite{Gaiotto:2014kfa}, with the thermal Wilson line the one-form operator.
For the propagation of an infinitely massive test quark, the thermal Wilson line describes the evolution of the color charge through the thermal medium. The Polyakov loops $\ell_j$ represent the trace of this propagator, as it winds around in imaginary time $j$ times.

Test quarks do not propagate in the confined phase of a gauge theory, and so the expectation value of \zN charged
Polyakov loops, where $j = 1, \ldots, (N_c-1)$, vanish. 
Of course, $\ell_{N_c}$ is a singlet under \zN, and has
a nonzero expectation value at any $T \neq 0$.
Conversely, the expectation values of \zN charged
loops are nonzero in the deconfined phase, above the temperature for the deconfining phase transition, at \Td.

Numerical simulations on the lattice typically measure the expectation
value of the loop with unit charge, $\ell_1$~\cite{Gupta:2007ax}.  
To measure loops with higher charge, it is necessary
to measure loops in an irreducible representation of
the gauge group.  After the fundamental representation,
given by $\ell_1$, the next simplest is the adjoint, which is
\zN neutral, and representations with two indices
\cite{Dumitru:2003hp}, which carry charge two under \zN.  
See, {\it e.g.}, Eq.~(21) of Ref.~\cite{Dumitru:2003hp}.
Polyakov loops with higher charge play an essential
role in the effective theory we construct.  
Models employing just the first Polyakov loop 
were first proposed by Fukushima~\cite{Fukushima:2003fw},
and are extremely useful. Since only the first loop is
included, however, they cannot be completed.

An alternate approach is to notice that since the thermal Wilson line
transforms homogeneously under (periodic) gauge transformations,
then, we can always rotate it to a diagonal form:
\begin{equation}
    \bfL(\bm{x}) = \exp(2 \pi i \, \vb*{q}/N_c) \,,
    \label{diagonal_ell}
\end{equation}
where $\vb*{q}$ is a diagonal matrix.  As an element of \suN, $\vb*{q}$
is traceless, with $N_c-1$ independent elements.
The form of Eq.~\eqref{diagonal_ell} can be described by expanding about a constant value of the timelike component of the gauge field,
\begin{equation}
    A_0^{a b}(\bm{x},\tau) = \frac{2 \pi i T}{g} \; q_a \; \delta^{a b}
    \; ,
    \label{A0ansatz}
\end{equation}
where $a,b=1,\ldots, N_c$ refer to color indices in the fundamental representation.  We note that at the leading order, the $q_a$ are automatically gauge invariant; beyond the leading loop order, the gauge-invariant quantities are related to the
``bare'' $q_a$ through finite terms~\cite{Bhattacharya:1992qb}.
We only compute the potential for the $\vb*{q}$ at one-loop order,
so this will not arise in our analysis.  We simply mention it
to stress that the eigenvalues of the thermal Wilson line form
gauge invariant quantities.  
At the tree level, a constant value for the gauge potential has zero
action. A potential for $\vb*{q}$ is generated at one-loop order~\cite{Gross:1980br,Weiss:1980rj},
\begin{equation}
\mathcal{V}_4 = \frac{2 \pi^2 T^4}{3} 
\sum_{a,b = 1}^{N_c} \; \mathcal{P}^{a b , b a} \;
B_4(q_a - q_b) \; .
\label{eq:V4}
\end{equation}
%
We introduce the color projection operator
\begin{equation}
{\mathcal P}^{ab,cd} = \delta^{a d} \delta^{b c} - \frac{1}{N_c}\,
\delta^{a b} \delta^{c d} \; .
\label{eq:def_P}
\end{equation}
This operator is transverse in the color indices, 
$\sum_{a}{\mathcal P}^{a a,cd} = 0$.  It is 
convenient to introduce in Eq.~\eqref{eq:V4}, so that we sum
only over the $N_c^2 -1 $ degrees of freedom for
$SU(N_c)$, and not those of $U(N_c)$.  It also enters frequently later.
Also, 
\begin{equation}
B_4(q) = - \frac{1}{30} + q^2 (1 -q)^2 \; ; \; q \equiv |q|_{\mathrm{mod} 1}
\end{equation}
is the fourth Bernoulli polynomial.
The potential $\mathcal{V}_4$ is elementary to compute, using a constant background field
for $A_0$.  Because the $q_a$'s enter through the adjoint covariant
derivative, the quantities $q_a - q_b$ enter automatically. 
Since the
$q_a$'s only enter into $\bfL$ through an exponential, the potential is periodic, so the $q$'s
enter as $|q|$, modulo unity.

This potential is minimized at $q_a = 0$, which is the usual
perturbative vacuum, or at \zN transforms thereof~\cite{Bhattacharya:1990hk,Bhattacharya:1992qb}.  
This potential can then be used to compute the \zN interface tension, which is
equivalent to a `t Hooft loop around a spatial
boundary~\cite{Korthals-Altes:1999cqo}.
Thus, while this one-loop potential is useful at high temperatures,
it manifestly cannot describe the transition to a confining phase.
This behavior is manifestly non-perturbative. To describe this, we add the following term:
\begin{equation}
\mathcal{V}_2 = c \, T^2 T_d^2 \; \sum_{a,b=1}^{N_c} 
\; \mathcal{P}^{a b , b a} \; B_2(q_a- q_b) \; ,
\label{teen_potential}
\end{equation}
where $B_2$ is the second Bernoulli polynomial,
\begin{equation}
B_2(q) = \frac{1}{6}- q(1 -q) \; ; \; q \equiv |q|_{\mathrm{mod}\,1} \; ,
\end{equation}
%
and the constant $c$ is adjusted so that the transition
occurs at \Td.
This is non-perturbative, as it is proportional to \Td, the temperature
for deconfinement.  

This term is certainly not unique. It is motivated on two grounds.
First, from $\approx 1.2 \, T_d$ to $\approx 4 \, T_d$, the leading
correction to the pressure in a pure gauge theory is $\sim T^2$.  
This is not exact, however, only approximate.  
Secondly, it is necessary to ensure that the non-perturbative part
of the potential is such that there is no first-order transition from
a truly perturbative phase, where all $q_a = 0$, to the semi-QGP,
where $q_a \neq 0$.  The second Bernoulli polynomial is particularly
useful to ensure this, since $B_2 \sim q$ at small $q$.  

\section{Matrix solution at large \texorpdfstring{$N_c$}{Nc}}
\label{sec:matrix_model}
In Refs.~\cite{Dumitru:2010mj,Dumitru:2012fw}, matrix models were
analyzed by fitting to the pressure.  In computing the shear
and bulk viscosity, however, we found it convenient to simply 
by taking the limit of large $N_c$.  This is simply because, at
finite $N_c$, one has to take into account that the generators of
$SU(N_c)$ are traceless, and this rather complicates the analysis.
Since the model, as we discuss, has obvious limitations, we ease our computational burden by considering the limit of 
large $N_c$. Thus, in this section, we review the solution at large $N_c$~\cite{Pisarski:2012bj}; see, also, Refs.~\cite{Lin:2013qu,Nishimura:2017crr,KorthalsAltes:2019yih,KorthalsAltes:2020ryu}.  

At large $N_c$, the sum over the color indices can be replaced by an integral,
\begin{equation}
    \sum_{a=1}^{N_c} \rightarrow N_c \int^{+q_0}_{- q_0} d q \; \rho(q) \; .
\end{equation}
Here we introduce $y=a/N_c$, with $q_a \rightarrow q(y)$.  Instead of
integrating over $y$, however, we can introduce the spectral density,
$\rho(q) = dy/dq$.  The solution at large $N_c$ involves a spectral
density that has a finite range, $-q_0 \rightarrow q_0$.

At large $N_c$, we then need to minimize the potential
\begin{equation}
    \mathcal{V}_\mathrm{eff}(q) = d_2(T) V_2(q) - d_1(T) V_1(q) \; ,
\end{equation}
where
\begin{equation}
    V_n(q) = \int^{+q_0}_{-q_0} dq \int^{+ q_0}_{- q_0} dq'
    \rho(q) \, \rho(q') \, |q-q'|^n (1 - |q-q'|)^n \; .
\end{equation}
%
To one-loop order, $d_2 = (2 \pi^2/3) \,  T^4$.
Previously 
\cite{Pisarski:2012bj,Lin:2013qu,Nishimura:2017crr,KorthalsAltes:2019yih,KorthalsAltes:2020ryu},
it was assumed that $d_1 \sim T^2 T_d^2$.
In this work, in order to fit the Polyakov loop,
we adopt a more general ansatz.
By taking derivatives of the equation of motion, the solution is
found to be the following.  The temperature dependence only enters
through the ratio of the coefficients of the Bernoulli polynomials,
\begin{equation}
    d(T)^2 = \frac{ 12 d_2(T)}{d_1(T)} \; .
    \label{definition_d(T)}
\end{equation}
The eigenvalue density has an elementary form,
\begin{equation}
    \rho(q) = 1 + b(T) \cos(d(T) \, q) \; ; \;
    q: -q_0(T) \rightarrow q_0(T) \; .
    \label{eq:solution_eig_density}
\end{equation}
This eigenvalue density must also satisfy the normalization condition,
\begin{equation}
    \int^{+ q_0}_{-q_0} dq \; \rho(q) = 1 \; .
\end{equation}
The solution is
\begin{equation}
    \cot\left(d(T) q_0(T) \right) 
    = \frac{d(T)}{3} \left( \frac{1}{2} - q_0(T) \right)
    - \frac{2}{d(T) (1 - 2 q_0(T))} \; ,
\end{equation}
and
\begin{equation}
    b(T)^2 = \frac{d(T)^4}{9} \left(
    \frac{1}{2} - q_0(T) \right)^4 + \frac{d(T)^2}{3}
    \left( \frac{1}{2} - q_0(T) \right)^2 + 1 \; .
\end{equation}
The properties of this solution are roughly in accord with the numerical simulations from the lattice. The loop with charge $k$ is given by
\begin{equation}
    \ell_k = \int^{+ q_0}_{- q_0} dq \; \rho(q) \cos(2 \pi k q) \; .
\end{equation}
Notably, at the deconfining phase transition, $d(T_d) = 2 \pi$, and the value of the first loop is 
\begin{equation}
\ell_1(T_d) = \frac{1}{2} \; .
\end{equation}
Unexpectedly, the values of all higher loops vanish at this point,
\begin{equation}
    \ell_k(T_d) = 0 \; ; \; k \geq 2 \; .
    \label{eq:higher_loops_Td}
\end{equation}
This follows directly from the form of the solution at $T_d$, Eq.\eqref{eq:solution_eig_density}, which equals
\begin{equation}
    \rho(q) = 1 + \cos(2 \pi q) \; ; \; T = T_d\;.   
\end{equation}
%
\begin{figure}[ht]
	\begin{center}
	\includegraphics[scale=1.2]{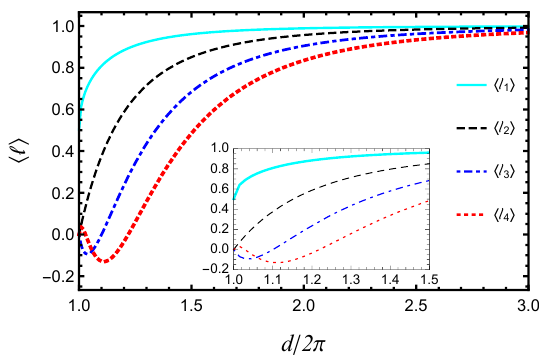}
	\caption{ Plots of the first four Polyakov loops as a function of $d/(2 \pi)$ in the matrix model.}
	\label{fig:several_loops_model}
	\end{center}
\end{figure}
We do not see a general reason why all loops with charge two and higher should vanish at $T_d$, Eq.~\eqref{eq:higher_loops_Td}, but it obviously follows from the simple form of the eigenvalue density at the transition.  

Away from the transition, plots of the charged loops are shown in Fig.~\ref{fig:several_loops_model}, plotted versus
$d/(2 \pi)$.  Loops with charge two or greater vanish at the transition, where $d/(2 \pi) = 1$, but they also begin to oscillate in sign.  Again, this is presumably a detailed feature of the model.

\begin{figure}[ht]
		\begin{center}
		\includegraphics[scale=1.2]{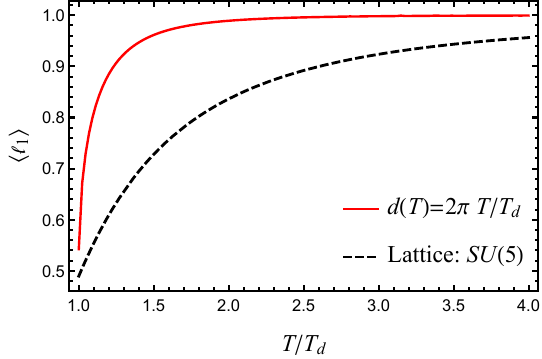}
			\caption{Plot of the Polyakov loop using the solution $d(T) = 2 \pi T/T_d$, versus that for $N_c=5$.}
			\label{fig:lattice_loop_simple_model}
		\end{center}
	\end{figure}

In Fig.~\ref{fig:lattice_loop_simple_model}, we show a plot of
the value of the first Polyakov loop, $\ell_1(T)$, versus results
from the lattice. If we require that the transition take place
at $T_d$, then we must fix $d(T) = 2 \pi T/T_d$, as
the transition occurs when $d(T) = 2 \pi$.
Results for the renormalized Polyakov loop are available for
$N_c=3$~\cite{Gupta:2007ax} and higher $N_c$~\cite{Mykkanen:2012ri}.  The results are similar, if
not identical.  For $N_c=3$, the value of the renormalized loop
is smaller at $T_d$, $\ell_1(T_d) \approx 0.4$. For
$N_c=5$, $\ell_1(T_d) \approx 0.5$.  
	\begin{figure}[ht]
		\begin{center}
		\includegraphics[scale=1.6]{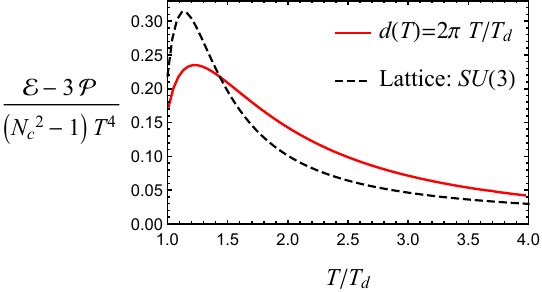}
			\caption{Plot of the interaction measure,
    divided by the number of gluons, $N_c^2-1$, using the solution $d(T) = 2 \pi T/T_d$, versus that for $N_c=3$.}
			\label{fig:interaction_measure_simple_model}
		\end{center}
	\end{figure}
The potential is proportional to the pressure, so one can also immediately compute the entropy density, $s(T) = \partial \mathscr{P}(T)/\partial T$, and from that, the energy density, $\mathscr{E}(T) = T s(T) - \mathscr{P}(T)$.
The deviation from conformality is most easily measured through the ``interaction measure'', which is 
\begin{equation}
\frac{\mathscr{E}-3\mathscr{P}}{T^4} = T \frac{\partial}{\partial T} 
\left(\frac{\mathscr{P}}{T^4}\right) \; .
\end{equation}
This is plotted in Fig.~\ref{fig:interaction_measure_simple_model}.  
The pressure from the matrix model is compared to the lattice results for
$N_c=3$~\cite{Caselle:2018kap}, as that is the most accurate.
From Ref.~\cite{Caselle:2018kap}, for three colors the pressure
can be fit by the function
\begin{eqnarray}
	\mathscr{P}_\mathrm{lat}(T)=	8 \, T^4\; 
 \frac{p_1 + p_2\ln(t) + p_3\ln^2(t)}{8 (1 + p_4 \ln(t) + p_5 \ln^2(t))}
 \; , \; t = \frac{T}{T_d} \;,
 \label{eq:pressure_N=3}
	\end{eqnarray}
%
where $p_1=0.0045$, $p_2 = 1.76$, $p_3 = 10.6$, $p_4 = 2.07$, 
and $p_5 = 5.8$. The overall factor of $8$ on the right-hand side is for the number of gluons.
We comment that there are results for
the pressure from the lattice for higher $N_c$, but the results 
do not differ significantly from $N_c=3$~\cite{Datta:2010sq},
at least to the accuracy at which we work.
We note that very recent work~\cite{Giusti:2025fxu} computes thermodynamic
quantities across the $SU(3)$ deconfinement
transition and presents an improved equation of state, with a similar fit function for the
pressure in Eq.~\eqref{eq:pressure_N=3}. 
We mention that in the analysis for three colors, the pressure was fit to rather good accuracy but at the expense
of adding further terms to the pressure~\cite{Dumitru:2010mj,Dumitru:2012fw}.  
In any case, while the fit to the pressure is approximate,
as is obvious from Fig.~\ref{fig:lattice_loop_simple_model},
there is a sharp discrepancy between the Polyakov loop
in the matrix model versus the renormalized Polyakov loop
from the lattice for $N_c=5$.
In particular, the Polyakov loop on the lattice approaches unity 
{\it much} slower than in the matrix model.
This is not an artifact of large $N_c$, and is present in the solution of the matrix model
for three colors~\cite{Dumitru:2010mj,Dumitru:2012fw}.
This is a serious deficiency of the matrix model, for which we 
do not have an explanation.
\begin{figure}[ht]
\begin{center}
\includegraphics[scale=1.2]{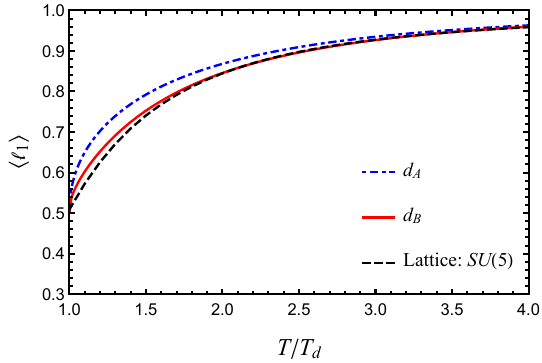}
\caption{Results for the Polyakov loop with the ansatz of Eq.~\eqref{improved_loop} versus the lattice for $N_c=5$.}			\label{fig:improved_loop}
\end{center}
\end{figure}
Since we can compute using the matrix model, however, we adopt
the following approach.  This is admittedly a heuristic approach
until said time that a better matrix model can be developed.

We take the solution of the matrix model and adjust
$d(T)$ so that the Polyakov loop agrees with the values from the
lattice.  Accordingly, the terms for the pressure also need to be
adapted.
We take two ansatzes:
	\begin{eqnarray}
		d_A(T)&=&1.08\, t+5.2032\nonumber\;,\\
		d_B(T)&=& \frac{0.26}{t^3}+1.105\,  t +4.9182 \; , \; t = \frac{T}{T_d} \; .
  \label{improved_loop}
	\end{eqnarray}
We take $d_2(T) \sim T^4$, as for an ideal gas, and define
$d_1(T)$ from the definition of $d(T)$ in Eq.~\eqref{definition_d(T)}.

\begin{figure}[ht]
\begin{center}
\includegraphics[scale=1.5]{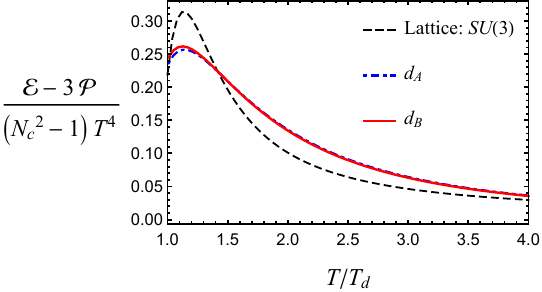}
\caption{Results for the interaction measure, divided by the number of degrees of freedom, $N_c^2-1$, with the ansatz of Eq.~\eqref{improved_loop} versus that, from the lattice, for $N_c=3$, Eq.~\eqref{eq:pressure_N=3}. }
\label{fig:improved_pressure}
\end{center}
\end{figure}

The results for the first Polyakov loop are plotted in Fig.~\ref{fig:improved_loop}, where they are compared to the lattice Polyakov loop for $N_c=5$. The first form, $d_A(T)$, is
relatively close to the lattice loop.  
Given the definition of $d(T)$, $d_1(T) \sim d_2(T)/d_A(T)^2$, and so while at large $T$, $d_1(T) \sim T^2$, as expected from
the matrix model, near the transition, the form of $d_1(T)$ cannot be motivated except by the fit to the loop. This is even more true for the second form, $d_B(T)$, which contains a somewhat arbitrary coefficient $\sim 1/t^3$, which is chosen so that there is close agreement
with the lattice value of $\ell_1$.
The results for the interaction measure are shown in Fig.~\ref{fig:improved_pressure}.  The agreement is satisfactory,
although the matrix model does not describe the 
a sharp peak in the interaction
measure at $\sim 1.1 T_d$.  

%
\section{Perturbation theory at nonzero holonomy}
\label{sec:pert_thy_nonzero_holonomy}
It is straightforward to compute at nonzero holonomy by expanding about the background field of Eq.~\eqref{A0ansatz}. For a detailed discussion, see Ref.~\cite{Hidaka:2009hs}.  The computation for the gluons is standard so that here we simply establish notation. 

We compute at nonzero temperature in Euclidean spacetime, where
\begin{equation}
\int \frac{d^4 K}{(2 \pi)^4} =
T \sum_{n= - \infty}^{+ \infty} \frac{d^3 k}{(2 \pi)^3} \; ; \;
K_\mu = (k_0,\bm{k}) \; ; \; k_0 = 2 \pi n T \; .
\label{def_nonzeroT_int}
\end{equation}
%
In the background field of Eq.~\eqref{A0ansatz}, 
for a gluon the time-like component of the momentum is
\begin{equation}
K_\mu^{a b} = (k_0^{a b}, \bm{k}) \; , \;
k_0^{ a b} =k_0 + Q^{ab} =k_0 + Q^a - Q^b = 2 \pi T (n + q_a - q_b) \; .
\label{k0_definition}
\end{equation}
The difference $q_a - q_b$ enters because the adjoint covariant derivative involves a commutator of the background field, $A_0^\mathrm{cl}$. This only happens for the quantum fluctuations with color components
that are off-diagonal. For fluctuations that are color diagonal, the background field is not entered.  This is the principal reason why computations at nonzero holonomy are more complicated at finite $N_c$ than at infinite $N_c$.

In the Landau choice of background field gauge,
the gluon propagator is 
\begin{equation}
    D^{\mu \nu}(K^{a b}) =
    \left( \delta^{\mu \nu} - \frac{K_\mu^{ab} K_\nu^{ab}}{(K^{a b})^2} \right)
     \frac{1}{(K^{a b})^2} \; {\mathcal P}^{a b, c d} \; .
\label{eq:gluon_propagator}
\end{equation}
This involves the color projection operator, $\mathcal{P}^{ a b, c d}$, Eq.~\eqref{eq:def_P}.
The three and four-gluon vertices,
and the coupling to Faddeev-Popov ghosts, follow directly by replacing the momentum with color flow momenta.

\subsection{Teen fields}
We concentrate on the effective theory for the two-dimensional ghost fields~\cite{Hidaka:2020vna}, which we denote as \begt.  
This is the Bengali character for three, pronounced teen. We chose this because it is simple to write, and it is the third field after gluons and Faddeev-Popov ghosts.
We introduce a unit spatial vector $\hatl$, 
and to respect rotational invariance, integrate over all directions of $\hatl$. The longitudinal and transverse coordinates with respect to $\hatl$ are
\begin{equation}
x^i = ({x}_\parallel, \bm{x}_\perp) \;\;\; , \;\;\;
x_\parallel = \bm{x} \cdot \hatl \;\;\; , \;\;\; 
\bm{x}_\perp \cdot \hatl = 0 \; .
\end{equation}
We then introduce gauge covariant derivatives for the fermionic field, $\phi$:
%
\begin{equation}
S_\begt = \; \int^{1/T}_0 d\tau \int \frac{d\Omega_{\hatl} }{4 \pi}
\int^\infty_{-\infty} d x_\parallel \int_{|x_\perp^2| > 1/T_d^2} d^2x_\perp\;\mathcal{L}_\begt\;,
\label{teen_lag}
\end{equation}
where the Lagrangian associated with the $2$D ghost field, i.e.,``teen'' field is given by
%
\bea
\mathcal{L}_\begt=\; 2 \; \tr \left( 
(D_0 \bar{\phi})( D_0 \phi) + 
(D_\parallel \bar{\phi})(D_\parallel\phi) + (\bm{D}_\perp \bar{\phi})\cdot(\bm{D}_\perp \phi) \right) \; ,
\eea
%
and $\phi$ and $\bar{\phi}$ are taken to lie
in the adjoint representation.  
The overall factor of $2$ is because we use generators normalized such that $\tr t^{ab} t^{cd} = \mathcal{P}^{ a b, c d}/2$.
The integral over transverse momenta is explicitly the cutoff
at $T_d$, and so is manifestly non-perturbative.

%
In practice, we do not evaluate the effective Lagrangian of 
Eq.~(\ref{teen_lag}) exactly, but directly in momentum space,
putting a cutoff on the transverse directions $=T_d$.  This violates gauge invariance.  
We show that at one-loop order, this violation does not appear in the
teen contribution to the gluon self-energy, but does in the teen self-energy.  However, the teen self-energy does not contribute to 
either the shear or bulk viscosities 
{\color{black} 
to leading logarithmic order in weak coupling}, 
and so for now, we 
ignore how to implement the teen fields in a more consistent manner.

Now, the teen propagator is 
\bea
D^{ab,cd}_\begt(K)=\frac{1}{(K^{ab})^2} \; \mathcal{P}^{ab,cd} \; .
\eea
As for the gluons,
the interactions follow directly using color flow momenta.  The three-point vertex connecting a teen, an anti-teen, and one gluon is
\bea
-\frac{\delta S_{\begt}}{\delta \phi^{fe}(R)\, \delta  B_\mu^{dc}(Q)\, \delta \bar{\phi}^{ba}(P) }=i g f^{ab,cd,ef}\,
\left(P_\mu^{ab}- R^{ef}_\mu \right) \; .
\eea
Here, $f^{ab,cd,ef}=i(\delta^{ad} \delta^{cf} \delta^{eb} -\delta^{af} \delta^{cb} \delta^{ed})/\sqrt{2}$ is the structure constant.
Now, the four-point vertex connecting two gluons and two teen fields is
\bea
-\frac{\delta S_\begt}{\delta B_\mu^{hg}(S)\, \delta B_\mu^{fe}(R)\, \delta \bar{\phi}^{dc}(Q)  \delta \phi^{ba}(P)\,}=- g^2\sum_{ij}( f^{ab,ef,ij}f^{cd,gh,ji}+ f^{ab,gh,ij}f^{cd,ef,ji})\,\delta_{\mn}\;.
\eea
%
\subsection{Teen contribution to gluon self-energy}
\label{teen_cont_gluon_self_energy}
\begin{figure}[ht]
	\begin{center}
		\includegraphics[scale=0.7]{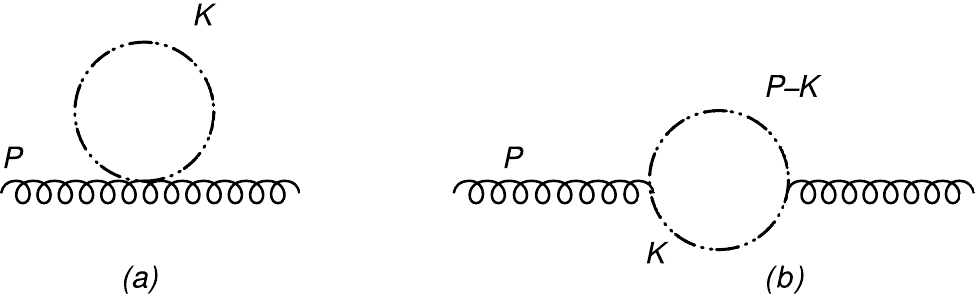}
		\caption{Teen ({\bng 3}) contribution to the gluon self-energy}
		\label{fig:se_teen}
	\end{center}
\end{figure}

The two contributions to the gluon self-energy from teen fields
are illustrated in Fig.~\ref{fig:se_teen}.  As usual, there is
a tadpole term, and a term involving two teen fields.
From Fig.~\ref{fig:se_teen}a, the tadpole diagram contributes
\bea
J_1^{\mn}&=& f^{ab,ef,gh} f^{cd,fe,hg} X_1^{\mu \nu} \;,
\nonumber \\
X_1^{\mu \nu} &=& -
2g^2 \int_\begt \frac{d^4 k}{(2\pi)^4} \frac{\delta^{\mn}}{(K^{fe})^2} \; .
\eea
%
The projection operators $\mathcal{P}^{ab,cd}$ are multiplied with traceless generators $t^{ab}$, 
so the second term $\sim -\delta^{ab}\delta^{cd}/N_c$ 
drops out, and we can take $\mathcal{P}^{ab,cd} \rightarrow \delta^{a c}\delta^{b d}/N_c$.

The momentum integral for the teen field is
%
\begin{equation}
\int_\begt \frac{d^4 K}{(2\pi)^4}
= T \sum_{n = -\infty}^{+\infty} \int_\begt \frac{d^3 k}{(2 \pi)^3} \; , \; \int_\begt \frac{d^3 k}{(2 \pi)^3} = \int_{-\infty}^{+\infty}
\frac{d k_\parallel}{2 \pi} \int_{|k_\perp| < T_d} \frac{d^2k_\perp}{(2 \pi)^2} \; .
\label{eq:def_teen_integral}
\end{equation}
Here $k_\parallel = \bm{k} \cdot \hatl$, and 
$\bm{k}_\perp = \bm{k} - \hatl (\bm{k} \cdot \hatl)$
for a given $\hatl$, with $\hatl^{ \, 2} = 1$.  The integration over all directions of
$\hatl$ then restores rotational invariance. 
To simplify the integral over the teen momenta, we will
usually assume that $T \gg T_d$.  This allows us to separate
the sum over $n$, with momenta $\sim 2 \pi n T$, and $k_\parallel$,
from the two-dimensional integral over $k_\perp$.

The contribution to the self-energy from the self-energy diagram can be written from Fig.~\ref{fig:se_teen}b as
\bea
J_2^{\mn}&=&f^{ab,ef,gh} f^{cd,fe,hg} \, X_2^{\mu \nu} \; , \nonumber \\ 
X_2^{\mu \nu} &=& g^2 \int_\begt \frac{d^4 K}{(2\pi)^4} 
\frac{(2K^{fe}-P^{ab})^\mu(2K^{fe}-P^{ab})^\nu}{(K^{fe})^2(P^{ab}-K^{fe})^2}\; .
\eea
%
The combined contribution to the gluon self-energy from the two diagrams becomes
%
\bea
X^{\mn}&=& X_1^{\mn}+X_2^{\mn}\nonumber\\
&=&-g^2  \int_\begt \frac{d^4 K}{(2\pi)^4} \bigg\{\frac{2\delta^{\mn}}{(K^{fe})^2} -\frac{4(K^{fe})^\mu(K^{fe})^\nu}{(K^{fe})^2(P^{ab}-K^{fe})^2}\bigg\}\;.
\eea
Here, we neglected the external momentum in the numerator, as the calculation is performed using the HTL approximation in the following.
%
%
%
We start with the spatial component of $X_2^{\mn}$
\bea
X_2^{ij}(P^{12},Q_1,Q_2)=\frac{1}{2}\; g^2  \int_\begt  \frac{d^3 k}{(2\pi)^3}\frac{4k^i k^j}{4E_k E_{p-k}}\left[ (\mathcal{I}_1+\mathcal{I}_2+\mathcal{I}_3+\mathcal{I}_4)+(Q_1\leftrightarrow Q_2)\right] \; .
\label{Xmunu_def}
\eea
Here, $E_k=|\bm{k}|$, and $E_{p-k}=|\bm{p}-\bm{k}|$.
We have symmetrized with respect to $Q_1$ and $Q_2$, which accounts
for the overall factor of $1/2$.  In detail,
 \begin{align}
\mathcal{I}_1&=\frac{-1}{ip^{12}_0-E_k-E_{p-k}}\Bigl(1+n(E_k-iQ_1)+n(E_{p-k}-iQ_2)\Bigr)\; ,\\
\mathcal{I}_2&=\frac{1}{ip^{12}_0-E_k+E_{p-k}}\Bigl(n(E_k-iQ_1)-n(E_{p-k}+iQ_2)\Bigr)\; ,\label{landau2}\\
\mathcal{I}_3&=\frac{-1}{ip^{12}_0+E_k-E_{p-k}}\Bigl(n(E_k+iQ_1)-n(E_{p-k}-iQ_2)\Bigr)\; ,\label{landau3}\\
\mathcal{I}_4&=\frac{1}{ip^{12}_0+E_k+E_{p-k}}\Bigl(1+n(E_k+iQ_1)+n(E_{p-k}+iQ_2)\Bigr) \; .
\end{align}
To simplify things we write $Q_1=Q^f - Q^e$ and $Q_2 = Q^a - Q^b+Q^f - Q^e$,
with $p_0^{12} = p_0 - Q_1 + Q_2$.
 
The terms that do not involve Landau-damping,
and so are due to quasi-particle (QP), can be written as
\bea
&&X_{2,\mathrm{QP}}^{ij}(P^{12},Q_1,Q_2)=\frac{1}{2}g^2  \int_\begt  \frac{d^3 k}{(2\pi)^3}\frac{{4}k^i k^j}{4E_k E_{p-k}} \left[(\mathcal{I}_1+\mathcal{I}_4)+(Q_1\leftrightarrow Q_2)\right] \nn\\
&=&g^2 \int_\begt  \frac{d^3 k}{(2\pi)^3}\frac{2\delta^{ij}}{4E_k }\frac{\bm{v}_k^{\, 2}}{3}\bigg[n(E_k-i Q_1)+n(E_k+i Q_1)+n(E_k-i Q_2)+n(E_k+i Q_2)\bigg]\;,
\label{non_LD}
\eea
where $\bm{v}_k=\bm{k}/E_k$. Now, the terms involving Landau damping (LD) are
\bea
&&X_{2,\mathrm{LD}}^{ij}(P^{12},Q_1,Q_2) = {\frac{1}{2}}g^2  \int_\begt  \frac{d^3 k}{(2\pi)^3}\frac{{4}k^i k^j}{4E_k E_{p-k}}\left[(\mathcal{I}_2+\mathcal{I}_3)+(Q_1\leftrightarrow Q_2)\right]\nn\\
&=&\frac{1}{2}g^2  \int_\begt  \frac{d^3 k}{(2\pi)^3}v^i_k v^j_k\bigg[\frac{{2}}{ip_0^{12}+\bm{p}\cdot \hatl}\big\{n(E_k-i Q_1)-n(E_k+i Q_1)+n(E_k-i Q_2)-n(E_k+i Q_2)\big\}\nn\\
&-&\frac{\bm{p}\cdot \hatl}{ip_0^{12}+\bm{p}\cdot \hatl}\,\frac{\partial}{\partial E_k}\big\{n(E_k-i Q_1)+n(E_k+i Q_1)+n(E_k-i Q_2)+n(E_k+i Q_2)\big\}\bigg]\;.
\label{eq:X2ld}
\eea
The second line of Eq.~\eqref{eq:X2ld} can be simplified as
\bea
&-&\frac{g^2}{2} \int_\begt  \frac{d^3 k}{(2\pi)^3}\bigg[v^i_k v^j_k
\frac{-ip_0^{12}}{ip_0^{12}+\bm{p}\cdot \hatl}\,\big\{n'(E_k-i Q_1)+n'(E_k+i Q_1)+n'(E_k-i Q_2)+n'(E_k+i Q_2)\big\}\nn\\
&&\hspace{2cm}+\ v^i_k v^j_k\,\big\{n'(E_k-i Q_1)+n'(E_k+i Q_1)+n'(E_k-i Q_2)+n'(E_k+i Q_2)\big\}\bigg]\nn\\
&=&-\frac{g^2}{2} \int_\begt  \frac{d^3 k}{(2\pi)^3}\bigg[v^i_k v^j_k\frac{-ip_0^{12}}{ip_0^{12}+\bm{p}\cdot \hatl}\,\big\{n'(E_k-i Q_1)+n'(E_k+i Q_1)+n'(E_k-i Q_2)+n'(E_k+i Q_2)\big\}\nn\\
&&\hspace{1cm}-\delta^{ij} \bigg(\frac{1}{E_k}-\frac{v_k^2}{3E_k}\bigg)\big\{n(E_k-i Q_1)+n(E_k+i Q_1)+n(E_k-i Q_2)+n(E_k+i Q_2)\big\}\bigg]
\; ,
\label{LD}
\eea
where $n'(E- i Q_1) = \partial n(E-i Q_1)/\partial E$, {\it etc.}
%

Now, the spatial components of the tadpole term are
\begin{align}
X_1^{ij}&=-g^2\int \frac{d^4 K}{(2\pi)^4} \delta^{ij}\bigg[\frac{1}{K_1^2}+\frac{1}{K_2^2}\bigg]\nn\\
&\hspace{-0.4cm}= -g^2\int \frac{d^3k}{(2\pi)^3} \delta^{ij}\frac{1}{2E_k}\bigg[n(E_k-i Q_1)+n(E_k+i Q_1)+n(E_k-i Q_2)+n(E_k+i Q_2)\bigg] .
\label{tadpole}
\end{align}
If we add Eqs.~\eqref{non_LD}, \eqref{LD}, and~\eqref{tadpole}, the terms $\sim \delta^{ij}$ vanish.

Collecting the non-vanishing terms, the spatial components of $X^{\mu\nu}$ become
\begin{align}
 X^{ij}&=g^2\int_0^{T_d}\frac{k_\perp \,dk_\perp}{2\pi}\int_0^\infty\frac{2dk_\parallel}{2\pi}\int \frac{d\Omega_{\hatl}}{4\pi}\,\frac{\hat k^i \hat k^j}{ip_0^{12}+\bm{p}\cdot \hatl}\nn\\
&\times\bigg[\big\{n(E_k-i Q_1)-n(E_k+i Q_1)+n(E_k-i Q_2)-n(E_k+i Q_2)\big\}\nn\\
&\hspace{0.5cm}+ip^{12}_0 \frac{1}{2} \bigg\{n'(E_k-i Q_1)+n'(E_k+i Q_1)+n'(E_k-i Q_2)+n'(E_k+i Q_2)\bigg\}\bigg] \;.
\end{align}
%
Now, the temporal component of $X^{\mu\nu}$ is
\bea
 X^{00}&=& -g^2 \int \frac{d^4 K}{(2\pi)^4} \bigg[\frac{1}{K_1^2}+\frac{1}{K_2^2}\bigg]
+ \frac{1}{2}g^2 \int \frac{d^4 K}{(2\pi)^4} \bigg[\frac{4 (k_1^0)^2}{K_1^2(P^{12}-K_1)^2}+\frac{4 (k_2^0)^2}{K_2^2(P^{12}-K_2)^2}\bigg]\nn\\
&=&g^2 \int \frac{d^4 K}{(2\pi)^4} \bigg[\frac{1}{K_1^2}+\frac{1}{K_2^2}\bigg]
- \frac{1}{2}g^2 \int \frac{d^4 K}{(2\pi)^4} \bigg[\frac{4 ({\bm{ k}_1})^2}{K_1^2(P^{12}-K_1)^2}+\frac{4 ({\bm {k}_2})^2}{K_2^2(P^{12}-K_2)^2}\bigg] \;.
\eea
%

Collecting the temporal and spatial components, the total $X^{\mu\nu}$ defined in Eq.~\eqref{Xmunu_def} becomes
\bea
 X^{\mn}&=& \frac{g^2}{2}\int_0^{T_d}\frac{k_\perp \,dk_\perp}{2\pi^2}\int_0^\infty2dk_\parallel\int \frac{d\Omega_{\hatl}}{4\pi}\,\nn\\
&\times&\bigg[\frac{\bar K^\mu \bar K^\nu}{ip_0^{12}+\bm{p}\cdot \hatl}\big\{n(E_{k_\parallel}-i Q_1)-n(E_{k_\parallel}+i Q_1)+n(E_{k_\parallel}-i Q_2)-n(E_{k_\parallel}+i Q_2)\big\}\nn\\
&+&\frac{1}{2}\bigg(u^\mu u^\nu+\frac{ip_0^{12}\bar K^\mu \bar K^\nu}{ip_0^{12}+\bm{p}\cdot \hatl}\bigg)\big\{n'(E_{k_\parallel}-i Q_1)+n'(E_{k_\parallel}+i Q_1)+n'(E_{k_\parallel}-i Q_2)+n'(E_{k_\parallel}+i Q_2)\big\}\bigg]\nn\\
&=&\frac{g^2}{2\pi^2}\frac{T_d^2}{2}\bigg[-2\pi i\big(B_1(q_1)+B_1(q_2)\big)T\int \frac{d\Omega_{\hatl}}{4\pi}\frac{\bar K^\mu \bar K^\nu}{\bar{K} \cdot P^{12}}\bigg]
+\frac{g^2}{2\pi^2}\frac{T_d^2}{2}\bigg[u^\mu u^\nu+ip_0^{12}\int \frac{d\Omega_{\hatl}}{4\pi}\frac{\bar K^\mu \bar K^\nu}{\bar{K} \cdot P^{12}}\bigg]\nn\\
&=&\frac{g^2}{2\pi^2}\frac{T\,T_d^2}{2}\bigg[-2\pi i\big(B_1(q_1)+B_1(q_2)\big)\bigg]\delta \Gamma^{\mn}(P^{12})
-\frac{g^2}{2\pi^2}\frac{T_d^2}{2}\delta\Pi^{\mn}(P^{12}) \; .
\label{eq:X_munu}
\eea
%
%
This involves the functions $\delta \Gamma$ and $\delta \Pi$.
$\delta \Pi$ is the usual hard thermal loop function,
\begin{equation}
\delta \Pi^{\mu \nu}(P)
= - u^{\mu} u^{\nu} - i p_0 \int \frac{d \Omega_{\hatl}}{4 \pi}
\frac{\bar K^\mu \bar K^\nu}{P \cdot \bar K} \; ,
\label{eq:def_HTL_Pi}
\end{equation}
where $u^\mu = \delta^{\mu 0}$ is a vector in the rest frame
of the medium.  The other function is
\begin{equation}
\delta \Gamma^{\mu \nu}(P) = 
-\frac{1}{i p_0} \left( \delta \Pi^{\mu \nu}(P) + u^{\mu} u^{\nu}
\right) \; .
\label{eq:def_delta_gamma}
\end{equation}
The null vector $\bar K^\mu\equiv (i,\hatl)$, and $B_1$ is the first Bernoulli polynomial,
\begin{equation}
    B_1(q) = |q|_{\mathrm{mod} \, 1} - \frac{1}{2} \; .
\end{equation}
%
So, the teen contribution to the gluon self-energy becomes
\bea
\Pi_\begt^{\mn;ab, cd}(P^{ab})
&=&J_1^{\mn}(P^{ab})+J_2^{\mn}(P^{ab}) \nn\\
&\stackrel{\text{HTL}}\approx&  f^{ab,ef,gh}f^{cd,fe,hg} X^{\mn}(P^{ab},Q^{fe},Q^{hg})\nn\\
&=& \delta^{ad,bc}X^{\mn}(P^{ab},Q^{ae},Q^{eb})-\delta^{ab,cd}X^{\mn}(P^{ab},Q^{ca},Q^{ac})\;.
\label{eq:teen_cont_gluon_pi}
\eea
For the gluon self-energy, the largest terms 
are $\sim g^2 T^3$ times $\delta \Gamma$, and
$\sim g^2 T^2$ times $\delta \Pi$.
For soft momenta $P_\mu \sim g T$, 
the term $\sim g^2 T^2$ is as large as the
term at the tree level.  Remember that after analytic continuation,
this holds as well for the time-like component,
$p_0^{12} = - i \omega + \epsilon$, with $|\omega| \sim g T$.

For the teen loop in the gluon self-energy, 
we keep the largest terms, which are $\sim T T_d^2$
times $\delta \Gamma$, and $\sim T_d^2$ times $\delta \Pi$.
We show shortly when {\it both} the gluon and teen
contributions are included, that the terms $\sim \delta \Gamma$
cancel~\cite{Hidaka:2009hs,Hidaka:2020vna}, as a consequence
of the equations of motion.  The term $\sim \delta \Pi$ contributes to the gluon
self-energy.  In keeping the largest terms, we write
$\stackrel{\text{HTL}} \approx$ in Eq.~\eqref{eq:teen_cont_gluon_pi} to represent that only the largest
terms for soft momenta have been kept.

The contribution to the gluon self-energy from a gluon
loop at non-zero holonomy is given in Ref.~\cite{Hidaka:2009hs}. The total is
\bea
\Pi^{\mn;ab, cd}_{\mathrm{total}}(P^{ab})&=&\Pi_{\mathrm{gl}}^{\mn;ab, cd}(P^{ab})+
\Pi_\begt^{\mn;ab, cd}(P^{ab}) \nn\\
&=& - \mathcal{K}^{ab,cd} \delta \Gamma^{\mn}(P^{ab})-(m^2)^{ab,cd} \delta \Pi^{\mn}(P^{ab})
\; .
\label{eq:gluon_self_energy}
\eea
%

Now, $\mathcal{K}^{ab,cd}$ in Eq.~\eqref{eq:gluon_self_energy} is 
%
\bea
\mathcal{K}^{ab,cd} &=&\frac{2\pi i g^2}{3} \, \delta^{ad}\delta^{bc}
\bigg[\sum_{e=1}^{N_c}\bigg(\mathcal{A}_0(Q^{ae})+\mathcal{A}_0(Q^{eb})\bigg)T^3\nn\\
&&+\, 3 \sum_{e=1}^{N_c}\frac{1}{8\pi^3}\left\{
\big(B_1(q_{a}-q_e)+B_1(q_e-q_b)\big)\right\}T_d^2 T\bigg] \; ,
\label{eq:tadpole_term}
\eea
where
\bea
\mathcal{A}_0(Q)=\frac{3}{2\pi i T^3}\int \frac{d^3k}{(2\pi)^3}\left(n(E_k-iQ)-n(E_k+iQ)\right)  \; .
\eea
It is direct to show that Eq.~\eqref{eq:tadpole_term} vanishes by the equations of motion~\cite {Hidaka:2020vna}. Physically, the sum of the currents induced
by the gluon and teen field cancel~\cite{Hidaka:2020vna}. This is an essential check
on the consistency of the model.

The second term in Eq.~\eqref{eq:gluon_self_energy} involves the usual hard thermal loop in the gluon 
self-energy, $\delta \Pi(P)$, times the Debye mass squared~\cite{Hidaka:2009hs,Hidaka:2020vna}. The latter includes contributions from the gluons and teen fields,
\bea
(m^2)^{ab,cd}=(m^2_{\mathrm{gl}})^{ab,cd}+\frac{g^2 T_d^2}{4 \pi^2}\mathcal{P}^{ab,cd}\;,
\eea
where
\bea
(m^2_{\mathrm{gl}})^{ab,cd}=\frac{g^2T^2}{6}\left(\delta^{ad}\delta^{bc}\sum_{f=1}^{N_c}\left\{\mathcal{A}(Q^{af})+\mathcal{A}(Q^{fb})\right\}-2\delta^{ab}\delta^{cd}\mathcal{A}(Q^{ac})\right)\;,
\eea
with
\bea
\mathcal{A}(Q)=\frac{12}{T^2}\int \frac{d^3k}{(2\pi)^3}\frac{1}{2E_k}\left(n(E_k-iQ)+n(E_k+iQ)\right)\;.
\eea
\subsection{Teen self-energy}
\label{sec:teen_self_energy}
\begin{figure}[ht]
	\begin{center}
		\includegraphics[scale=0.75]{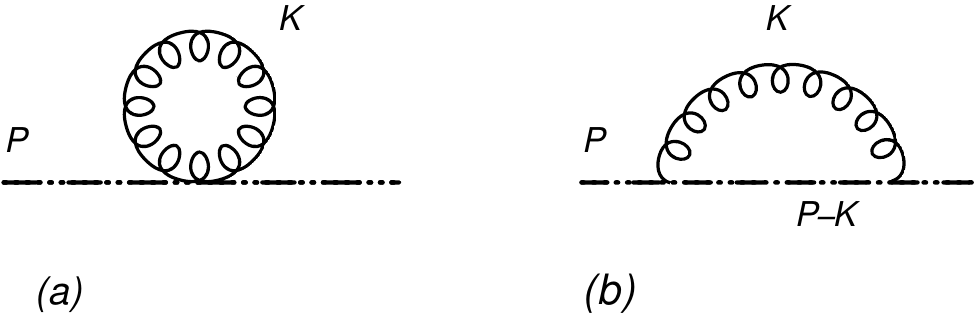}
		\caption{Teen self-energy diagrams}
		\label{fig:teen_SE}
	\end{center}
\end{figure}

The diagrams which contribute to the teen self-energy are
those of Fig.~\ref{fig:teen_SE}.  There is a tadpole diagram,
plus a diagram with a virtual gluon teen pair.
As for the usual scalar self-energy, the teen self-energy is gauge-dependent. Thus we content ourselves with power counting.
The tadpole diagram, Fig.~\ref{fig:teen_SE}a, only involves
a virtual gluon, and so is proportional to $\sim g^2 T^2$.  
The diagram with a gluon-teen pair involves a virtual teen field,
Fig.~\ref{fig:teen_SE}b involves an integral over a teen field,
and so is automatically $\sim g^2 T_d^2$. We shall see, however, that the teen self-energy does not enter into the computation of the shear and bulk viscosities at leading logarithmic order.  
As we demonstrate in the following, this is not obvious, and due to which the diagrams in detail contribute at leading logarithmic order.

\section{Shear viscosity}
\label{sec:shear_viscosity}
\subsection{Scattering amplitude}
\label{sec:shear_scat_amp}
We follow the perturbative computation of the shear viscosity
following Arnold, Moore, and Yaffe 
\cite{Arnold:2000dr,Arnold:2003zc,Jackson:2017hfz,Ghiglieri:2018dib,Moore:2020pfu,Danhoni:2022xmt,Danhoni:2024ewq,MacKay:2024jus},
modified to the case of nonzero holonomy~\cite{Hidaka:2008dr,Hidaka:2009xh}.
\begin{figure}
	\begin{center}
		\includegraphics[scale=0.2]{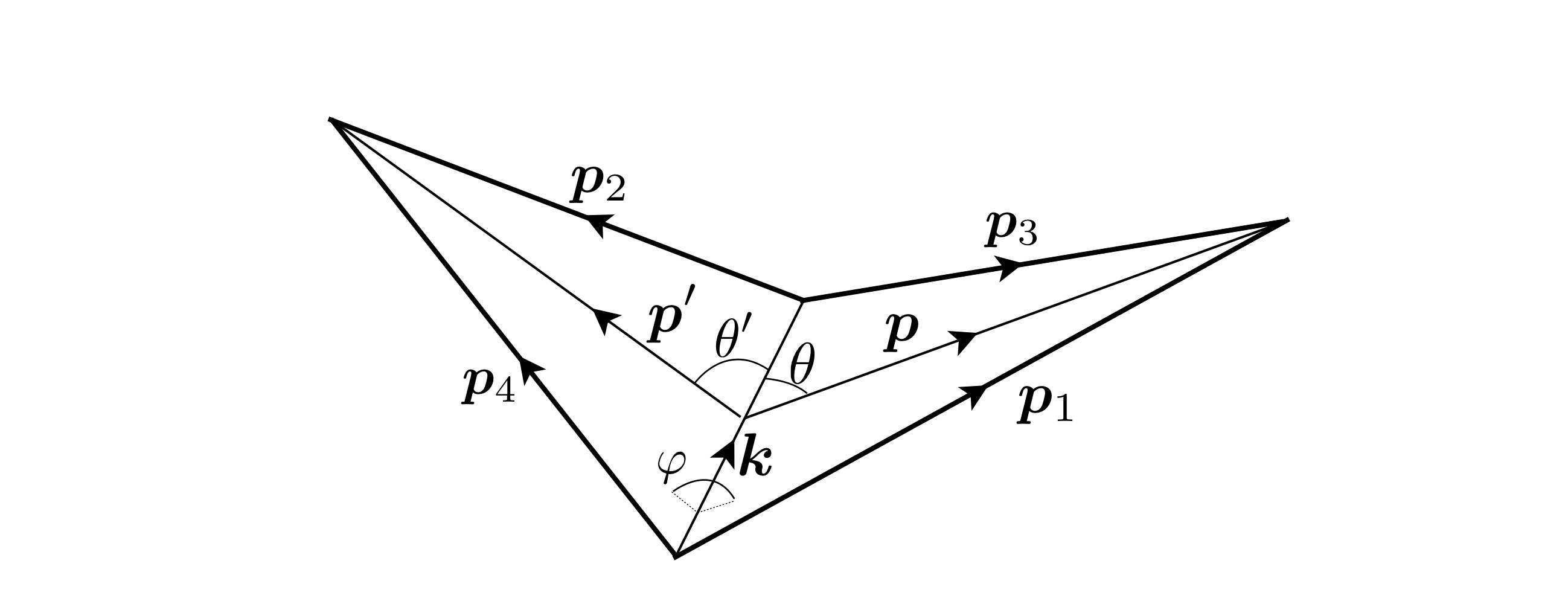}
		\caption{Kinematics of 2-2 scattering, from Ref.~\cite{Hidaka:2009ma}}
		\label{fig:kinematic}
	\end{center}
\end{figure}
{\color{black}
At leading logarithmic order,
}
what contributes is the scattering of 
two to two scattering: two incident particles,
with momenta $P_1$ and $P_2$, go into fields with
momenta $P_3$ and $P_4$.  The particles can be either gluon or teen fields.
For $2 \rightarrow 2$ scattering, the Mandelstam variables are 
\bea
s=(P_1+P_2)^2 \; , \; 
t=(P_1-P_3)^2 \; , \;
u= (P_1-P_4)^2 \;  .
\eea
%
While we begin with Euclidean momenta, when evaluating
the matrix elements, and quantities such as the 
Mandelstam variables, we blithely analytically continue
to Minkowski momenta.

It is convenient to parametrize these momenta as 
\bea\label{mom_variable}
P_1= P+\frac{K}{2} \; , \; P_2= P'-\frac{K}{2} \; ,
P_3= P-\frac{K}{2} \; ,  P_4= P'+\frac{K}{2} \; .
\eea
The dominant contribution to the shear and bulk viscosities is forward scattering. The momenta $P_1$ and $P_2$, and $P_3$ and $P_4$, are hard, with the components of the momenta on the order of the temperature, $T$. The exchanged momentum in the $t$-channel, $K$, is soft,
on the order of $g T$. We introduce the angles
$\theta $ and $\theta'$
\bea
\cos \theta= \hat{\bm k}\cdot \hat{\bm p}\;, \hspace{2 cm} \cos \theta'=\hat{\bm k}\cdot \hat{\bm p}'\;,
\eea
as illustrated in Fig.~\ref{fig:kinematic}.
We define the unit vector $\hat{\bm k}=\bm{k}/k$, where $k = |\bm{k}|$, and $\phi$ as the angle between two planes spanned by vectors $(\bm{p}_1, \bm{p}_3)$ and $(\bm{p}_2, \bm{p}_4)$. 

For forward scattering, 
 \begin{equation}
     \cos \theta \simeq x \equiv \frac{k_0}{k} \; ,
     \label{eq:define_x}
 \end{equation}
and $\theta \approx  \theta'$, with
 \bea
 \cos \phi\simeq\frac{\hat{\bm p} \cdot {\hat{\bm p}'}-x^2}{1-x^2}\;.
 \eea
 The Mandelstam variables become
 \bea
 s&\simeq&2 p p' (1-x^2)(1-\cos \phi)\;,\nn\\
 u&\simeq&-2 p p' (1-x^2)(1-\cos \phi)\;,\nn\\
 t&\simeq& -k^2 (1-x^2)\;.
 \eea
%
We only consider the diagrams that contribute
to leading logarithmic order for the shear and bulk viscosity.
These are the $t$- and $u$-channels, where a soft field
is exchanged in the $t$-channel; the $u$-channel is related
to the $t$-channel by exchanging $P_3 \leftrightarrow P_4$.
All diagrams in the $t$- and $u$-channels were taken into
account in computing the full result.
Thus, we neglect diagrams where soft fields are exchanged in
other channels, and diagrams with four-point interactions,
which include those between four gluons, or two gluons and two teens.

The diagrams that contribute to leading logarithmic order are of two types.
There are those involving the exchange of a soft gluon,
illustrated in Figs.~\ref{fig:ABCdiagram}a-c, and
those involving the exchange of a soft teen field,
illustrated in Figs.~\ref{fig:DEdiagram}a,b.   

\subsubsection{Exchange of a soft, virtual gluon}
\begin{figure}
	\begin{center}
		\includegraphics[scale=.9]{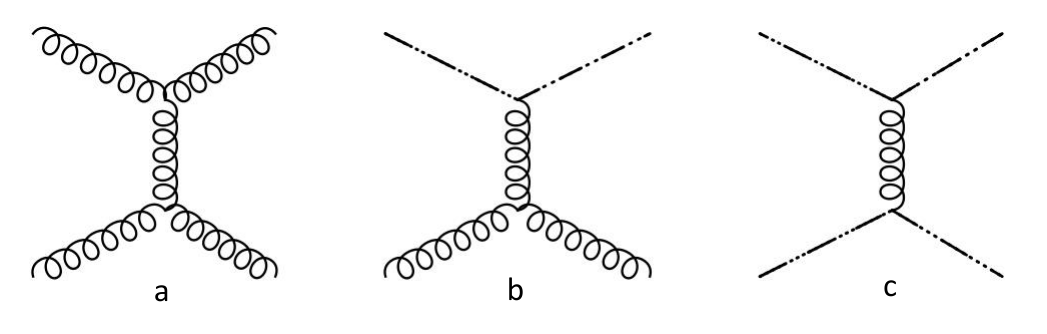}
		\caption{The diagrams involving the exchange of a virtual soft
  gluon: gluon-gluon scattering (a), gluon-teen scattering (b),
  and teen-teen scattering (c).}
		\label{fig:ABCdiagram}
	\end{center}
\end{figure}
The amplitude for two gluons scattering into two gluons, 
Fig.~\ref{fig:ABCdiagram}a, is
%
\begin{eqnarray}   
\mathcal{M}_{{\mathrm{gg}}\rightarrow {\mathrm{gg}}}=\epsilon_1^{\mu_1}(P_1^{a_1b_1}) \, \epsilon_3^{*\mu_3}(P_3^{a_3b_3}) &\;& 
J_{{\mathrm{g}}, \mu_1 \mu_3}^{\mu, a b}(P_1^{a_1 b_1},P_3^{ a_3 b_3}) 
\,D_{\mn}^{ba,dc}(K^{ba})\,  \nonumber \\
&&\times\ J_{\mathrm{g}, \mu_2 \mu_4}^{\nu, c d}(P_2^{a_2 b_2},
P_4^{a_4 b_4}) \; \epsilon_2^{\mu_2}(P_2^{a_2b_2}) \, \epsilon_4^{*\mu_4}(P_4^{a_4b_4})\; .
\label{eq:gl-gl_scat}
\end{eqnarray}
Here $P_i^{\mu, a_i,b_i}$ are the momenta 
for particles $i= 1, \ldots, 4$.  The polarization
tensor are $\epsilon_{1,2}^{\mu}$ for incoming particles,
and $\epsilon_{3,4}^{*\mu}$ for outgoing fields.
They satisfy
\begin{equation}
    \epsilon_i^{\mu}(P_{i}^{a_i b_i}) \, P_{\mu,i}^{a_i b_i} =
    \epsilon_i^{* \mu}(P_i^{a_i b_i}) \, P_{\mu,i}^{a_i b_i} =0\;.
    \label{eq:polarization_tensor}
\end{equation}
In these expressions, we keep the color indices because 
especially Eq.~\eqref{eq:gl-gl_scat} makes it clear how the color flow works.
They really are not
necessary in Eq.~\eqref{eq:polarization_tensor}, though,
since the on-shell condition only applies after
analytic continuation to Minkowski momenta,
$p_0^{a b} \rightarrow - i \omega$.

In Eq.~\eqref{eq:gl-gl_scat}, 
$D_{\mn}^{ba,dc}(K^{ba})$ is the gluon propagator of Eq.~\eqref{eq:gluon_propagator}.
Now, the color current for a gluon is 
%
\bea
 J_{{\mathrm{g}},\mu_1 \mu_3}^{\mu, a b}(P_1^{ a_1 b_1},P_3^{a_3 b_3}) 
=2i g P^{\mu,ab}\; 
\delta^{\mu_1 \mu_3} \; t^{ab}_{a_1 b_1, a_3 b_3} \; .
\label{eq:gluon_current}
\eea
We suppress the two spin indices for a spin-one gluon since that
is just an overall degeneracy.
The generator in the adjoint representation is
%
\bea
t^{ab}_{cd, ef} =i\,f^{(ab,cd,ef)}  = 
\frac{i}{\sqrt{2}} \left( \delta^{a d} \delta^{c f} \delta^{b e}
- \delta^{a f} \delta^{b c} \delta^{d e} \right) \; ,
\eea
as defined in Eq.~(24) of Ref.~\cite{Hidaka:2009hs}.
In the matrix element, we only consider 
forward scattering, where $P^{a_1b_1}_1\simeq P^{a_3b_3}_3 \simeq P^{ab}$.  
All of the external gluon momenta are hard, $P^{a_ib_i} \sim T$.
 
The scattering for a teen plus gluon scattering into the same is shown in Fig.~\ref{fig:ABCdiagram}b.  The teen field has
spin-zero, and so no spin degeneracy.
The scattering amplitude in the $t$-channel is
%
\bea
\mathcal{M}_{\begt {\mathrm{g}} \rightarrow \begt{\mathrm{g}}}=
J_\begt^{\mu, ab}(P_1^{ a_1 b_1},P_3^{ a_3 b_3}) \,D_{\mn}^{ba,dc}(K^{ba})\, 
J_{{\mathrm{g}}, \mu_2 \mu_4}^{\nu, cd}(P_2^{a_2 b_2},P_4^{a_4 b_4}) \, 
\epsilon^{\mu_2}(P^{a_2b_2}_2) \, \epsilon^{* \mu_4}(P^{a_4b_4}_4) \; ,
\eea
where $J_ \begt^{\mu; ab}$ is the color current associated with teen ({\bng 3})-gluon vertex, 
\bea
J_ \begt^{\mu, ab}(P_1^{ a_1 b_1},P_3^{ a_3 b_3}) =2i g P^{\mu,ab}\, 
t^{ab}_{a_1 b_1, a_3 b_3}\; .
\label{eq:ghost_current}
\eea
The color structure is identical to that of the gluon, 
Eq.~\eqref{eq:gluon_current}, as both lie in the adjoint representation.  Since the teen particle
is a scalar, though, there is no factor of
$\delta^{\mu_1 \mu_3}$, as in the gluon current of
Eq.~\eqref{eq:gluon_current}.  

There is another implicit difference: the perpendicular component of the teen momenta $P^{a_1b_1}_1$ and $P^{a_2b_2}_2$ are of the order of \Td.
Thus, when a soft gluon is exchanged with
$K\sim gT$, $k_\perp$ is also restricted to be $\sim T_d$. 

The scattering amplitude for teen plus teen scattering
into the same, Fig.~\ref{fig:ABCdiagram}c, is
\bea
\mathcal{M}_{\begt\begt\rightarrow \begt\begt}
= J_\begt^{\mu; ab}(P_1^{ a_1 b_1},P_3^{ a_3 b_3}) \,D_{\mn}^{ba,dc}(K^{ba})\, 
J_\begt^{\nu;cd}(P_2^{a_2 b_2},P_4^{a_4 b_4}) \; .
\eea
\subsubsection{Exchange of a soft teen field}
\begin{figure}
	\begin{center}
		\includegraphics[scale=.7]{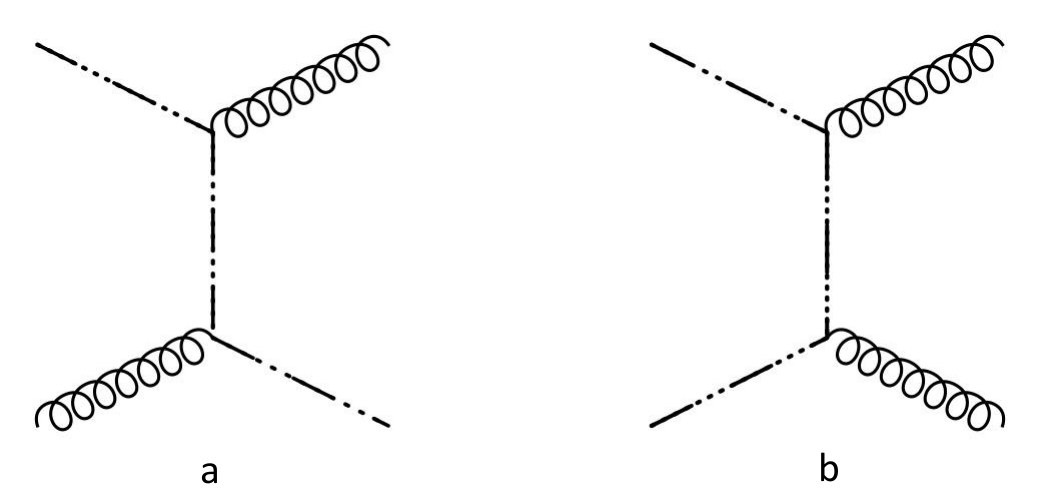}
		\caption{The diagrams involving the exchange of a soft, virtual teen
  particle: Compton scattering (a) and teen-teen annihilation (b).}
		\label{fig:DEdiagram}
	\end{center}
\end{figure}
There are two diagrams involving the exchange of a teen particle.
One contribution is where a teen plus a gluon scatters
into a gluon plus a teen,
Fig.~\ref{fig:DEdiagram}a.  This is analogous
to Compton scattering.  The amplitude is 
\bea
\mathcal{M}_{\begt {\mathrm{g}} \rightarrow {\mathrm{g}} \begt}
= \epsilon^{\mu}(P^{a_1b_1}_1) \, J_{\begt}^{\mu, ab}(P^{a_1b_1}_1,P^{a_3b_3}_3) \,D_{\begt}^{ba,dc}(K^{ba})\, 
J_{\begt}^{\nu, cd}(P^{a_2b_2}_2,P^{a_4b_4}_4) \, \epsilon^{*\nu}(P^{a_4b_4}_4)\;.
\eea
Lastly, there is the scattering of two
teens into two gluons, illustrated in
Fig.~\ref{fig:DEdiagram}b.
The amplitude is 
\bea
\mathcal{M}_{\begt \begt \rightarrow {\mathrm{g g}}}
= J_\begt^{\mu, ab}(P^{a_1b_1}_1,P^{a_3b_3}_3) \,D_{\begt}^{ba,dc}(K^{ba})\, 
J_\begt^{\nu, cd}(P^{a_2b_2}_2,P^{a_4b_4}_4)
\epsilon^{*\mu}(P^{a_3b_3}_3) \, \epsilon^{*\nu}(P^{a_4b_4}_4) \; .
\eea
These are identical to the diagrams involving gluon exchange, with the replacement of the gluon by the teen particle.

The matrix elements for the teen particle, however,
simplify significantly.  Consider the factors of
the polarization tensor for the scattering of
two gluons into two gluons, 
Eq.~(\ref{eq:gl-gl_scat}).  This involves
\begin{equation}
    \mathcal{M}_{{\mathrm{gg}}\rightarrow {\mathrm{gg}}} \sim
    g^2 \; \left(
    \epsilon^{\mu_1}(P^{a_1b_1}_1) \epsilon^{*}_{\mu_1}(P^{a_3b_3}_3)
    \right) \;
\left( \epsilon^{\mu_2}(P^{a_2b_2}_2) 
\epsilon^{*}_{ \mu_2}(P^{a_4b_4}_4) \right) \; \ldots \; ,
\end{equation}
%
where for simplicity, we drop the other factors.  For
all of the other diagrams involving gluon exchange,
the polarization tensors are contracted with one another,
and so give contributions to leading order in the
coupling constant, $\sim g^2$.

In contrast, consider the scattering of a teen plus
gluon into a gluon plus teen, Fig.~\ref{fig:DEdiagram}a.   Including just the
factors of the polarization tensor, which is
\begin{equation}
\mathcal{M}_{\begt {\mathrm{g}} \rightarrow {\mathrm{g}} \begt} \; \sim \; g^2 \; 
\left(\epsilon^\mu(P^{a_1b_1}_1) P^{ab}_\mu \right) \;
\left(\epsilon^{*\nu}(P^{a_4b_4}_4) P^{ab}_\nu \right)\; \sim g^4 \; .
\end{equation}
Each of the polarization vectors dotted into the momentum of that field vanishes, such as $\epsilon^\mu_1 P^{a_1b_1}_{1 \mu} = 0$, {\it etc.}, Eq.~\eqref{eq:polarization_tensor}.  
Then the momentum that appears in the current,
$\sim P^{ab}_\mu$, is approximately equal to the
external momenta, such as $P^{a_1b_1}_{1 \mu}$,
up to soft corrections, $\sim K^{ba}_\mu$.  These
terms are, by assumption, suppressed by $\sim g$.
Thus in all, the matrix elements involving the exchange of a teen particle,
Fig.~\ref{fig:DEdiagram}, are only $\sim g^4$,
relative to those involving gluon exchange,
Fig.~\ref{fig:ABCdiagram},
which are $\sim g^2$.

This cancellation for the scattering of the scalar field is well known.
As discussed in Sec.~\ref{sec:teen_self_energy}, this is most
fortunate, and these subtleties associated with constructing a complete,
and gauge invariant description of the teen fields can thus be
overlooked to leading logarithmic order.

\subsection{Collision integrals}

To evaluate the transport coefficients, we use kinetic theory.
The Boltzmann equation described the evolution of the 
particle distribution function $ f_{ab}(\vb*{r}, \vb*{p}, t)$ in phase space.
The Boltzmann equation accounts for advection due to motion, external forces, and collisions among particles. 

The Boltzmann equation is evaluated using the Chapman-Enskog method.
The distribution function $ f_{a b}(\vb*{r}, \vb*{p}, t)$ is expanded in a series in powers of the Knudsen number $ \epsilon$, which quantifies the ratio of the mean free path to the characteristic length scale of the system. This expansion assumes that the system is close to local equilibrium, allowing for the separation of equilibrium and non-equilibrium contributions to the distribution function.

At zeroth order $\epsilon^0$, the collisions among particles are neglected, and the equilibrium distribution function $f_{ab}^0$ is obtained. 
Deviations from equilibrium enter at $\epsilon^1$. By substituting the zeroth-order solution into the Boltzmann equation and retaining terms linear in $ \epsilon$, a linearized Boltzmann equation is obtained. Solving this equation iteratively yields the first-order correction to the distribution function $ f^1_{a b}$, capturing the effects of collisions on the distribution of particles.

The shear viscosity $\eta$ is then extracted from the correction
at first order, $f^1_{a b} $. This involves calculating the momentum flux tensor and comparing it to the gradient of the velocity field, quantifying the resistance to shear flow in the plasma.

Finally, the obtained expression for shear viscosity undergoes various consistency checks to ensure its physical validity. These checks may involve verifying that $\eta$ is positive,
and exhibits the correct behavior in the limit of a small Knudsen number.


The energy-momentum tensor for the gluon is given by
%
\bea
T^{\mn}_{\mathrm{g}}(\bm{r}, t)= 2
\int_{\mathrm{g}} d\Gamma_{a b} \; P^{\mu, a b} 
P^{\nu, a b}\; f_{{\mathrm{g}},ab}(\vb*{r}, \bm{p}, t) \;,
\eea
where $f_{\mathrm{g},ab}(\bm{r}, \bm{p}, t)$ is the statistical distribution function for the gluon.
In thermal equilibrium, this is the Bose-Einstein distribution function in a background field $Q^a-Q^b$,
\bea
f^0_{{\mathrm{g}},ab}=f^0_{ab}(E_p) \equiv \frac{1}{\mathrm{e}^{(E_p - i (Q^a-Q^b))/T} - 1} \; ,
\eea
with $E_p = \sqrt{\bm{p}^{\, 2}}$ the energy.
The phase space integral is
\begin{equation}
\int_{\mathrm{g}} d\Gamma_{a b} = \sum_{s}\sum_{a, b = 1}^{N_c} \; 
\mathcal{P}^{a b, b a} \; 
\int \frac{d^3 p}{(2\pi)^3 2E_p} \; .
\end{equation}
This includes a sum over the colors $a,b$, and the helicity $s$.
The factor $\mathcal{P}^{a b, b a}$
ensures that the overall number of degrees of freedom
is $N_c^2-1$, as appropriate for $SU(N_c)$.
%

The teen field is analogous to a Faddeev-Popov ghost field
and is an anti-commuting field in the adjoint representation.  An anti-commuting field is normally a fermion.  Fermions, however, are anti-periodic in the imaginary time, $\tau$,
under the shift $\tau \rightarrow \tau + 1/T$.  Instead, in thermal 
equilibrium, the teen field should be periodic in
$\tau$, like a gluons. Consequently, for the teen field, we start with
the statistical distribution function for a 
fermion, after shifting $(Q^a- Q^b)$ by $\pi T$:
\begin{equation}
    \frac{1}{\mathrm{e}^{(E_p - i (Q^a-Q^b + \pi T))/T} + 1}
    = {\color{black} (-)} f^0_{ab}(E) \; .
    \label{eq:sign_teen_function}
\end{equation}
Thus absorption of a teen field from a thermal bath produces
a relative minus sign, versus that of a gluon.  This shows
that the teen field is a ghost.  
To keep track of the signs, we introduce a sign function,
which $=+1$ for gluons, and $=-1$ for teens, 
\begin{equation}
\signf_{\mathrm{g}} = +1 \;\; ; \;\;
\signf_{\begt} = - 1 \; .
\end{equation}
In equilibrium, we find it useful to take the statistical distribution function for the teen equal to
that for the gluon, 
\begin{equation}
    f^0_{\begt, a b}(E_p) = {\color{black} (+)}
    f_{{\mathrm{g}}, a b}^0(E_p) \; .
    \label{eq:equality_gluon_teen_equi}
\end{equation}
If we take the distribution functions to be the same, 
then because there is a $(-)$ sign on the right hand side
of Eq.~(\ref{eq:sign_teen_function}), whenever
a single factor of a statistical distribution function
enters for absorption from the medium, $= f^0_{ab}(E)$,
we have to multiply that by the sign function:
$\signf_\begt= - 1$ for a teen field,
and $\signf_{\mathrm g} = +1$ for a gluon.

This is {\it not} true for emission into the medium, however.
For a fermion with a statistical distribution function
$f^0_{\mathrm{ferm},ab}(E)$, emission 
is associated with the factor $1 - f^0_{\mathrm{ferm},ab}(E)$.
In this case, the minus sign reflects the Pauli exclusion
principle.  For a teen field, then, by 
Eq.~(\ref{eq:sign_teen_function})
this becomes $1 + f^0_{\mathrm{g},ab}(E)$.  Thus for a teen field, there is {\it no} sign function, which arises when it is emitted into the medium.  
This is why we prefer to keep the signs of the teen and
gluon distribution functions the same, Eq.~(\ref{eq:equality_gluon_teen_equi}), and instead introduce
the sign functions $\signf_\begb$, {\it etc.}, for absorption.

It is useful to compare these signs to the case where gluons
scatter off of physical fermions, which we do in Appendix~\ref{sec:ghost_sign}.
Because they are physical, fermions
do not have a $(-)$ sign when they are absorbed from the medium,
while teen particles, which are ghosts, do.

The teen contribution to the stress-energy tensor is
\bea
T^{\mn}_{\begt}=  {2}
\int_\begt d\Gamma_{a b} \; P^{\mu, ab} 
P^{\nu, ab} \; 
\, { \signf_\begt} \; 
f_{\begt, ab} \; . 
\eea
%
This contributes to the pressure with a negative sign,
which is what is expected for a ghost field.  As explained previously, we need a ghost to decrease the pressure
as the temperature is lowered, and the confined phase is
approached.

While the gluon and teen distribution
functions are equal in equilibrium, they differ away from equilibrium. Thus, there is a Boltzmann
equation describing the evolution of each distribution
function.  

In the absence of collisions, for a system in the background gauge
fields $A_\mu$ and with total color charge ${\cal Q}$,
the gluon statistical distribution function satisfies the Vlasov equation \cite{Kelly:1994ig,Kelly:1994dh,Blaizot:2001nr},
\begin{equation}
    P^\mu \left(
    \frac{\partial}{\partial x_\mu} 
    - g \,\tr \left({\cal Q} G_{\mu \nu} \right)\frac{\partial}{\partial p_\nu} - i g \left[ {\cal Q} ,\frac{\partial}{\partial {\cal Q}}\right] \right) f(x,p,{\cal Q}) = 0 \; .
    \label{eq:vlasov}
    \end{equation}
In the semi-QGP, however, since $A_0^{\rm cl}$ is constant,
the field strength vanishes, $G_{\mu \nu} = 0$.
Further, as shown by Eq.~\eqref{eq:tadpole_term}, the
total color charge vanishes ${\cal Q} = 0$.
Thus, only the first term on the left hand side of
Eq.~\eqref{eq:vlasov} contributes, and we must consider the effect of collisions in the Boltzmann equation.

For collisions, we include the scattering
of $2 \rightarrow 2$ particles to lowest order,
\begin{equation}
    1_\begb \; 2_\begd \rightarrow 3_{{\begb'}} \,\; 4_{{\begd'}} \; .
\end{equation}
The associated momenta are $P^{\mu, a_1 b_1}_{1}$, {\it etc.}
We introduce the symbols $\begb$, $\begd$, $\begb'$, and $\begd'$ to denote that these particles can be either gluon or teen fields.
The Boltzmann equation is 
\bea
\signf_{\begb} \; 2P^{\mu, a_1 b_1}
\partial_\mu f_{\begb, a_1 b_1} (\bm{x},\bm{p}, t)=-
{\mathcal C}_{\begb,a_1 b_1}[f_\begb] \;,
\label{eq:coll_term}
\eea
where the collisional kernel is
\begin{align}
{\mathcal C}_{\begb, a_1 b_1}[f_\begb]&=
{\frac{1}{2}}\, 
{\sum_{\begd, \begb', \begd'}}
\signf_\begb \, \signf_\begd
\int_{\begd} d\Gamma_{a_2 b_2} \,\int_{\begb'} d\Gamma_{a_3 b_3} \,
\int_{\begd'} d\Gamma_{a_4 b_4} \,(2\pi)^4 
\nn\\
&\times \delta^{(4)}(P^{a_1b_1}_1+P^{a_2b_2}_2-P^{a_3b_3}_3-P^{a_4b_4}_4)\; |\mathcal{M}_{\begb \begd \rightarrow \begb' \begd'}|^2\nn\\
&\hspace{0cm}\times\left[f_{\begb,a_1 b_1} f_{\begd,a_2 b_2} 
\left(1+f_{\begb', a_3 b_3}\right)\left(1+f_{\begd', a_4 b_4}\right)\right.\nn\\
&\left.\hspace{1cm}-
f_{\begb',a_3 b_3} f_{\begd',a_4 b_4} \left(1+f_{\begb,a_1 b_1}\right)\left(1+f_{\begd,a_2 b_2}\right)\right] .
\label{eq:collision_term}
\end{align}
The only subtlety is the sign factors.  
The sign factor $\signf_{\begb}$ enters on the left hand side
of Eq.~(\ref{eq:coll_term}), as that describes the distribution
function for the initial particle, $1_{\begb}$.  

On the right-hand side, in the collision term of Eq.~\eqref{eq:collision_term}, there are only symmetry factors for
particles absorbed from the medium. This is $\signf_\begb \, \signf_\begd$ for the 
first term, and $\signf_{\begb'} \signf_{\begd'}$ for the second. Since teen number is conserved, however, the sign factors satisfy
\begin{equation}
    \signf_\begb \; \signf_\begd = \signf_{\begb'} \;\signf_{\begd'} \; .
\end{equation}
It is easy to convince oneself by working out all of the examples, such as $\begt {\mathrm g} \rightarrow \begt {\mathrm g}$, {\it etc.}
Because of this, in Eq.~\eqref{eq:collision_term} we can write the sign factors of both
terms as a common prefactor, $= \signf_\begb \, \signf_\begd$.

With the Chapman-Enskog method, the time derivative is expanded in the power of $\epsilon$
\bea
\partial_t=\sum_{n=1}^\infty \eps^n \partial_t^{(n)}.
\eea
We expand the distribution function around the 
distribution function in thermal equilibrium, $f^{0}_{ab}$,
\bea
f_{\begb, ab}= f^{0}_{ ab}+
\eps f^{1}_{\begb, ab}+\eps^2 f^{2}_{\begb, ab}+\cdots \;.
\eea
%
Notice that the index $\begb$ does not enter into the
distribution function as lowest order, because in equilibrium
the gluon and teen distribution functions are the same as given in
Eq.~\eqref{eq:equality_gluon_teen_equi}.
Thus we generally replace $f_{\begb, ab}^0(E)$ by $f_{ab}^0(E)$.
We parametrize $f^{n}_{\begb, ab}$ as  
\bea
f^{n}_{\begb, ab}=\ f^{0}_{ab}
(1+f^{0}_{ab})\phi_{\begb,ab}^{(n)} \;.
\label{eq:sign_f}
\eea
Because we introduced the sign functions above,
there is no
need to introduce a sign function in the definition of
$\phi_{\begb,ab}^{(n)}$. 

To $\eps^n$ order, the Boltzmann equation is 
%
\bea
\signf_\begb\, 
2P^{\mu,{a_1b_1}} \partial_\mu^{(n)}f_{\begb, a_1 b_1}=-(\mathcal{L}
\phi^{(n)})_{\begb, a_1 b_1}+K^{(n)}_{\begb, a_1 b_1} \; , \;
K^{(n)}_{\begb, a_1 b_1}=-
{\mathcal{C}}^{(n)}_{\begb, a_1 b_1}+
(\mathcal{L}\phi^{(n)})_{\begb, a_1 b_1} \; .
\eea
%
The linearized operator $\mathcal{L}$ is 
%
\begin{align}
\label{shear_L}
 (\mathcal{L}
\phi^{(n)})_{\begb, a_1 b_1}&= 
\frac{1}{2}
\sum_{\begd, \begb', \begd'}
\signf_\begb \signf_\begd
\int_\begd d\Gamma_{a_2 b_2}
\int_{\begb'} d\Gamma_{a_3 b_3} \int_{\begd'}d\Gamma_{a_4 b_4} \nn\\
&\times
\,(2\pi)^4 \delta^{(4)}(P^{a_1b_1}_1+P^{a_2b_2}_2-P^{a_3b_3}_3-P^{a_4b_4}_4)  |\mathcal{M}_{\begb \begd \rightarrow \begb' \begd'}|^2
 \nn\\
&\hspace{.15cm}
\times \left[f_{a_1 b_1}^{0} f_{a_2 b_2}^{0} 
(1+f_{a_3 b_3}^{0})(1+f_{a_4 b_4}^{0})\right]
\left(\phi_{\begb, a_1b_1}^{(n)}+\phi_{\begd,a_2b_2}^{(n)}-
\phi_{\begb',a_3b_3}^{(n)}-\phi_{\begd',a_4b_4}^{(n)}\right).
\end{align}
%
The right-hand side includes a sum over 
both gluons and teens. 
To order $\eps^n$,
%
\bea
P^{\mu, a_1 b_1} \partial_\mu^{(n)}f_{\begb,a_1b_1}=p^{0,a_1 b_1}\sum_{k=1}^n \partial_{t}^{(k)} 
f_{{\begb,t},a_1b_1}^{n-k}+p^i\partial_i f_{{\begb,t},a_1b_1}^{n-1} \;.
\eea
%
{\color{black} To leading logarithmic order in weak coupling
}, the viscosities are computed from 
\bea
\signf_\begb \; 
2P^{\mu,a_1,b_1} \partial_\mu^{(1)}f_{\begb,a_1 b_1}=
-(\mathcal{L}
\phi^{(1)})_{\begb,a_1 b_1} \; .
\label{leading}
\eea 
%
We set $K^1 = 0$, so that the right hand side of Eq.~\eqref{leading} is just the collision kernel, 
$(\mathcal{L} \phi^{(1)})_{\begb,a_1 b_1}
= {\mathcal{C}}^{(1)}_{\begb, a_1 b_1}$. 
%

We assume that near equilibrium, that the four
velocity of the medium is $\beta u^\mu$, where
$\beta = 1/T$ is the temperature, and 
$u^\mu = (1, \bm{u})/\sqrt{1-\bm{u}^2}$ the velocity.  It suffices
to work to linear order in $|\bm{u}|$.  Then, except for an overall factor of $\signf_\begb$, the left-hand side of the Boltzmann equation is
\begin{align}\label{BE_LHS2}
 2P^{\mu, a_1 b_1} \partial_\mu^{(1)}f^{0}_{\begb, a_1 b_1}&=-2P^{\mu, a_1 b_1}f^0_{a_1 b_1}(1+f^0_{a_1 b_1})
 (P^{\nu, a_1b_1}\partial^{(1)}_\mu(\beta u_\nu) 
 -i\partial^{(1)}_\mu\nu^{a_1b_1}
 )\nn\\
 &=-f^0_{a_1 b_1}(1+f^0_{a_1 b_1})
 \Bigl[
2(P^{0, a_1 b_1})^2\partial^{(1)}_t\beta
+2p^{i}P^{0, a_1 b_1}(\partial_i\beta +\beta\partial^{(1)}_t u_i)\nn\\
&\quad +p^jp^i\beta\sigma_{ij}-\frac{2}{3}p^2\beta\partial_k u^k -2P^{\mu, a_1 b_1}i\partial^{(1)}_\mu\nu^{a_1b_1}
 \Bigr]\;,
\end{align}
where
\begin{equation}
\sigma_{ i j} = \partial_i u_j + \partial_j u_i - \frac{2}{3} 
\, \delta^{i j} \, \partial_k u_k \; ,
\label{eq:sigma_{ij}}
\end{equation}
and $\nu^{ab}=\beta(Q^a-Q^b)=2\pi(q_a-q_b)$.

Using energy-momentum conservation or hydrodynamic equations to leading order,
\begin{subequations}\label{BE_LHS3}
\bea
\partial^{(1)}_t\beta &=& \beta v_s^2\partial_i u^i\;,\\
\beta\partial^{(1)}_t u_i &=&-\partial_{i}\beta\;,
\eea
\end{subequations}
where the speed of sound, squared, is
\begin{equation}
v_s^2= \frac{\partial \mathscr{P}}{\partial \mathscr{E}}\;,
\label{eq:speedofsound}
\end{equation}
we obtain
\begin{equation}
 2P^{\mu, a_1 b_1} \partial_\mu^{(1)}f_{\begb, a_1 b_1}
 =f^0_{a_1 b_1}(1+f^0_{a_1 b_1})
 \Bigl[2 p^2 \qty(\frac{1}{3}
- v_s^2)\beta\partial_iu^i
  -p^ip^j\beta\sigma_{ij}
 \Bigr]\;.
\label{eq:BEfinal}
\end{equation}
To obtain this expression, we have analytically continued
$P^{0, a_1 b_1}\rightarrow E_p = p$.  

We have also dropped a term
$\sim\partial^{(1)}_\mu\nu^{ab}=2\pi\partial^{(1)}_\mu(q_a-q_b)$.
By Eq.~\eqref{BE_LHS3}
this term is $\sim \partial_i u^i$ times
$i \partial_T q_{a_1 b_1} f^0_{a_1 b_1} (1+ f^0_{a_1 b_1})$.
As it is proportional to $\sim \partial_i u_i$, it only
contributes to the bulk viscosity. Since
$f^0_{a_1 b_1} (1+ f^0_{a_1 b_1})$, and all other
terms which contribute, are even under
$a_1 \leftrightarrow b_1$, 
since $q_{a_1 b_1} = - q_{b_1 a_1}$ is odd,
this cancels after summing over $a_1$ and $b_1$.

We also neglected the contribution of the thermal mass to the bulk viscosity. This contribution becomes non-negligible at high temperatures where  $v_s^2\sim 1/3$; however, near $T_d$, 
as discussed in Sec.~\ref{sec:bulk_viscosity}, the deviation of $v_s^2$ from $1/3$ provides the dominant contribution in our model.

For the shear viscosity, we can drop the term 
$\sim \partial_i u^i$, and we are left with
\bea
2P^{\mu, a_1 b_1} \partial_\mu^{(1)}f_{\begb,a_1 b_1}&=&-\beta 
f^{0}_{a_1 b_1}(1+f^{0}_{a_1 b_1}) p^i p^j \sigma_{ij} \;.
\eea
The term $\sim \partial_i u^i$ does contribute to the
bulk viscosity, which we compute in the next section.

The Boltzmann equation becomes
\bea \label{lhs=rhs}
\beta f^{0}_{a_1b_1}(1+f^{0}_{a_1b_1}) p^i p^j \sigma_{ij}= \signf_{\begb}
(\mathcal{L} \phi^{(1)})_{\begb, a_1 b_1} \; .
\eea
%
Now, $\phi^{(1)}_{\begb,ab}$  can be parameterized as 
\bea
\phi^{(1)}_{\begb,ab}
=\sigma_{ij}\bigg(p^i p^j -\frac{1}{3}p^2\delta^{ij}\bigg) 
\chi_{\begb,ab}(p)=\sigma_{ij}
\chi_{\begb,ab}^{ij}(p) \; ,
\eea
and so we can write
\bea
\beta\bigg(p^i p^j -\frac{1}{3}p^2\delta^{ij}\bigg) f^{0}_{a_1b_1}(1+f^{0}_{a_1b_1}) =
\signf_{\begb} 
(\mathcal{L} \chi^{ij})_{\begb,a_1b_1} \;.
\label{Eq:BE_inter}
\eea
%
The dissipative part of the energy-momentum tensor becomes
\bea
\delta T^{ij}&=& 
\sum_{\begb}2\signf_{\begb} \int_{\begb} 
d\Gamma_{ab}\; p^i p^j f^{(1)}_{\begb, ab}=\sum_{\begb} 2\signf_{\begb}  \int_{\begb} d\Gamma_{ab}
\; p^i p^j f^{0}_{ab}\Big(1+f^{0}_{ab}\Big)\phi^{(1)}_{\begb,ab}\nn\\
&=& \sigma^{ij}\sum_{\begb} \signf_{\begb}\; \frac{2}{5}\;
\int_{\begb} d\Gamma_{ab} \; \chi^{kl}_{\begb,ab}\bigg(p^k p^l-\frac{1}{3}p^2\delta^{kl}\bigg) f^{0}_{a b}(1+f^{0}_{a b}) \;.
\eea 
Using Eq.~\eqref{Eq:BE_inter} in the last equation, we can write
\bea
\delta T^{ij}&=& \sigma^{ij} \frac{2T}{5} \sum_{\begb}
\int_{\begb} d\Gamma_{ab} 
\chi_{\begb,ab}^{kl}
(\mathcal{L} \chi^{kl})_{\begb,ab} \; .
\eea
The expression for the shear viscosity then reduces from $\delta T^{ij}=\eta \sigma^{ij}$ to
\bea
\eta=\frac{2T}{5} \sum_{\begb} \int_{\begb} d\Gamma_{ab}  \,
\chi_{\begb,ab}^{ij} (\mathcal{L}
\chi^{ij})_{\begb,ab} \; .
\eea
Thus we need to determine the functions $\chi^{ij}_{\begb,ab}$
for both gluons and teens. The case of gluons is familiar~\cite{Arnold:1994ps,Arnold:1995bh}.
Expanding $\chi_{\mathrm{g},ab}$ in orthogonal polynomials,
\begin{equation}\label{orthogonal_expansion}
    \chi_{\mathrm{g},ab}
    =\sum_{n=0}^\infty c^{\mathrm{g}}_n  \; \chi^{\mathrm{g}}_n(p) \; .
\end{equation}
We define the $d^{\mathrm{g}}_n$'s as
\bea
d^{\mathrm{g}}_n \; \delta_{mn} = \frac{1}{T^6}
\sum_{ab,s}\mathcal{P}^{a b, b a} \int \frac{d^3 p}{(2\pi)^3 2E_p}f^0_{ab}(1+f^0_{ab}) p^4 \; \chi^{\mathrm{g}}_m(p) \; \chi^{\mathrm{g}}_n(p) \;.
\label{normalization_d}
\eea
Using Gram-Schmidt orthogonalization, we can then solve
for the $\chi^{\mathrm{g}}_n$'s. We choose 
\begin{equation}
    \chi^{\mathrm{g}}_n=\sum_{m=0}^n b^{\mathrm{g}}_{nm}
    \left( \frac{p}{T} \right)^m
    \; ,
\end{equation}
with leading coefficients $b^{\mathrm{g}}_{nn}=1$. The first few terms are
%
\begin{subequations}
\bea
\chi_0^\mathrm{g}&=&1 \;,\\
\chi_1^\mathrm{g}&=& \frac{p}{T}+b^{\mathrm{g}}_{10} \;,\\
\chi_2^\mathrm{g}&=& \left( \frac{p}{T}\right)^2
+\left( \frac{p}{T}\right)\;b^{\mathrm{g}}_{21} +b^{\mathrm{g}}_{20} \; .
\eea
\end{subequations}
This involves the dimensionless constants, 
\begin{subequations}
\bea
b^{\mathrm{g}}_{10}&=&-\frac{h_1}{h_0} \;,\\
b^{\mathrm{g}}_{20}&=&\frac{h_1 h_3-h_2^2}{h_0 h_2-h_1^2} \;, \\
b^{\mathrm{g}}_{21}&=& \frac{h_1 h_2-h_0 h_3}{h_0 h_2-h_1^2} \;.
\label{defined_b}
\eea
\end{subequations}
The $h_n$'s are then 
\bea
h_n&=&\frac{1}{ T^{6+n} } \sum_{ab,s}\mathcal{P}^{a b, b a} \int \frac{d^3 p}{(2\pi)^3 2E_p}f^0_{ab}(1+f^0_{ab}) p^{4+n}\nn\\
&=&\frac{1}{T^{6+n}} \sum_{ab,s} \mathcal{P}^{a b, b a} \int \frac{d^3 p}{(2\pi)^3 2E_p} \frac{\text{e}^{-(E_p- i Q^{ab})/T}}{(1-\text{e}^{(E_p - i Q^{ab})/T})^2} p^{4+n} \nn \\
&=& \frac{\Gamma(6+n)}{2\pi^2}\sum_{k=1}^\infty \frac{1}{k^{5+n}}
\left(|\Tr  \bfL^k|^2-1 \right) \; .
\eea
%
$\bfL$ is the thermal Wilson line, Eq.~\eqref{thermal_Wilson_line} of Sec.~\ref{sec:matrix}.
The lowest $d^{\mathrm{g}}_n$ can be written in terms of $h_n$ as
\begin{subequations}
\bea
d^{\mathrm{g}}_0&=&h_0 \;,\\
d^{\mathrm{g}}_1&=&h_2+b_{10} h_1 \;,\\
d^{\mathrm{g}}_2&=& h_4+ h_3 b^{\mathrm{g}}_{21}+h_2 b^{\mathrm{g}}_{20} \;.
\label{dh_relation}
\eea
\end{subequations}
%
For the teen field, we proceed similarly.
We expand $\chi_{\begt,ab}$ in orthogonal polynomials,
\begin{equation}
\chi_{\begt,ab}
=\sum_{n=0}^\infty c^\begt_n  \,\chi^\begt_n(p) \; ,
\end{equation}
and define
\bea
d^{\begt}_n \; \delta_{mn} = \frac{1}{T^4 T_d^2}
\sum_{a,b}\mathcal{P}^{a b, b a} 
\int_\begt \frac{d^3 p}{(2\pi)^3 2E_p}f^0_{ab}(1+f^0_{ab}) \, p^4 \; \chi^\begt_m(p) \; \chi^\begt_n(p) \;.
\eea
The momentum integral is that for a teen field,
Eq.~\eqref{eq:def_teen_integral}.  
Again by using Gram-Schmidt orthogonalization,
\begin{subequations}
\bea
\chi_0^\begt&=&1 \;,\\
\chi_1^\begt&=& \frac{p}{T}+b^{\begt}_{10} \;,\\
\chi_2^\begt&=& \left( \frac{p}{T}\right)^2+\left( \frac{p}{T}\right)\;b^{\begt}_{21} +b^{\begt}_{20} \;,
\eea
\end{subequations}
and
\bea
b^{\begt}_{10}&=&-\frac{t_1}{t_0} \;,\nn\\
b^{\begt}_{20}&=&\frac{t_1 t_3-t_2^2}{t_0 t_2-t_1^2} \;, \nn\\
b^{\begt}_{21}&=& \frac{t_1 t_2-t_0 t_3}{t_0 t_2-t_1^2} \;,
\label{defined_d}
\eea
where 
\bea
t_n&=&\frac{1}{T_d^2 T^{4+n} }
\sum_{a,b} \mathcal{P}^{a b, b a} \int_\begb \frac{d^3 p}{(2\pi)^3 2E_p}f_{ab}(1+f_{ab}) p^{4+n}\nn\\
&=&
\frac{1}{T_d^2 T^{4+n} } 
\sum_{a,b} \mathcal{P}^{a b, b a} \int \frac{ p_\perp dp_\perp dp_\parallel}{(2\pi)^2 2E_{p_\parallel}}f_{ab}(1+f_{ab}) p_\parallel^{4+n}\nn\\
&\approx& \frac{\Gamma(4+n)}{{8}\pi^2}\sum_{k=1}^\infty \frac{1}{k^{3+n}}\left(|\Tr  \bfL^k|^2-1 \right) \;.
\eea
The normalization constants $d^\begt_n$ are obtained similarly as Eq.~\eqref{dh_relation}.
We have used the approximation $p_\parallel(\sim T)\gg p_\perp(\sim T_d)$ for the teen field.

Here one should note that for the integration over the teen field momentum we have used,
\begin{equation}
\label{teen_mom_int}
\int_0^{\infty} dp\, p^2 = \int_0^{T_d}p_{\perp}\, dp_{\perp} \int_0^{\infty} dp_{\parallel}=\frac{T_d^2}{2} \int_0^{\infty} dp_{\parallel} \,.
\end{equation}
Using the Boltzmann equation,  
\bea \label{S_equal_Lc}
S_n^{\begb}=(\mathcal{L} \,c)^\begb_n \;,
\label{eq:Boltmzann_Sn}
\eea
where 
\bea
S_n^{\begb}=\signf_{\begb} \; 
\beta \int_\begb d\Gamma_{ab} \; \chi_{\begb ,n}^{ij} f^{0}_{ab} \big(1+f^{0}_{ab} \big)\bigg(p^i p^j-\frac{1}{3}p^2\delta^{ij}\bigg) \; ,
\eea
and
\bea
(\mathcal{L} \,c)^\begb_n
&=&  \int_\begb d\Gamma_{a_1b_1} \; \chi_{\begb ,n}^{ij} 
 (\mathcal{L} \chi^{ij})_{\begb,a_1b_1} \nn\\
 &=&
{\frac{1}{2}}\sum_{m=0}^\infty
{\sum_{\begd, \begb', \begd'}}
\signf_\begb \; \signf_\begd \;
\int_\begb d\Gamma_{{a_1b_1}}
\int_\begd d\Gamma_{a_2 b_2}
\,\int_{\begb'} d\Gamma_{a_3 b_3} \,\int_{\begd'}d\Gamma_{a_4 b_4} (2\pi)^4
\nn\\
&\times&  \delta^{(4)}(P^{a_1b_1}_1+P^{a_2b_2}_2-P^{a_3b_3}_3-P^{a_4b_4}_4) \; |\mathcal{M}_{\begb \begd \rightarrow \begb' \begd'}|^2 \left[f_{a_1 b_1}^{0} f_{a_2 b_2}^{0} 
(1+f_{a_3 b_3}^{0})(1+f_{a_4 b_4}^{0})\right]
\nn \\
\; &\times & \chi_{{\begb} ,n}^{ij} 
\bigg(\chi_{\begb,m}^{ij}c^{\begb}_m+\chi_{\begd,m}^{ij}c^{\begd}_m-
\chi_{\begb',m}^{ij}c^{\begb'}_m-\chi_{\begd',m}^{ij}c^{\begd'}_m\bigg)
\,.
\eea
The shear viscosity is then
\bea
\eta=
\frac{2T}{5} \sum_{\begb} 
\sum_{n=0}^\infty c_n^{\begb}
(\mathcal{L}c)_n^{\begb}
=\frac{2T}{5} \sum_{\begb} 
\sum_{ n = 0}^\infty S_n^{\begb}
 (\mathcal{L}^{-1}S)_n^{\begb}\;.
 \label{eq:eta_S}
\eea
To simplify the computation, we only take the lowest elements,
\begin{subequations}
\bea
S_n^\mathrm{g}&=& \frac{2T^5}{3} \,d^\mathrm{g}_0\, \delta_{n,0}\;,\\
S_n^\text{{\bng 3}}&=& -
\frac{2T_d^2 T^3}{3}\, d^\mathrm{\begt}_0\, \delta_{n,0} \;.
\eea
\end{subequations}
This approximation is valid to a few percent for the 
theory with pure glue~\cite{Arnold:2000dr,Arnold:2003zc}
and with non-zero holonomy~\cite{Hidaka:2009hs}.
Given the coarseness of our approximations, we assume the
same here.
\subsection{Kinematics}
We consider the scattering of two to two particles,
$\begb\begd \rightarrow \begb\begd$ processes in Fig.~\ref{fig:ABCdiagram}. 
In Eq.~\eqref{eq:eta_S} we can symmetrize between
the $c_n^{\begb}$'s, and rewrite
%
\begin{equation}\label{eq:sum_cLc}
\sum_{n,\begb}c_n^{\begb}(\mathcal{L}c)_n^{\begb}=\sum_{n,m,\begb,\begd}\Lambda_{nm}^{\begb,\begd} \; ,
\end{equation}
where
\bea \label{shear_matrix_1}
\Lambda_{nm}^{\begb,\begd}
&=& \left( 2 \, \frac{1}{4} \right)\frac{1}{2}\;
\signf_{\begb}\;  \signf_{\begd} \;
\int_{\begb} d\Gamma_{{a_1b_1}}
\int_{\begd} d\Gamma_{a_2 b_2}
\,\int_{\begb} d\Gamma_{a_3 b_3} \,
\int_{\begd} d\Gamma_{a_4 b_4}  (2\pi)^4 \nn \\
&&\times\delta^{(4)} (P^{a_1b_1}_1+P^{a_2b_2}_2-P^{a_3b_3}_3-P^{a_4b_4}_4)
|\mathcal{M}_{\begb \begd \rightarrow \begb \begd}|^2 
f_{a_1 b_1}^{0}f_{a_2 b_2}^{0}
\big(1+f_{a_3 b_3}^{0}\big)\big(1+f_{a_4 b_4}^{0}\big)\nn\\
&&\times \left[\big(\chi_{\begb,n}^{ij}(p_1)-\chi_{\begb,n}^{ij}(p_3)\big)c_n^\begb
+\bigl(\chi_{\begd,n}^{ij}(p_2)-\chi_{\begd,n}^{ij}(p_4)\bigr)c_n^\begd\right]\nn\\
&&\times 
\left[\big(\chi_{\begb,m}^{ij}(p_1)-\chi_{\begb,m}^{ij}(p_3)\big)c_m^\begb
+\bigl(\chi_{\begd,m}^{ij}(p_2)-\chi_{\begd,m}^{ij}(p_4)\bigr)c_m^\begd\right]\;. 
\eea
%
%
%
In these expressions, the first factor of $2$ is because
$\begb\begd \rightarrow \begd \begb$ gives the same contribution as $\begb\begd \rightarrow \begb\begd$ to leading logarithmic order~\cite{Arnold:2000dr}.  The factor of $1/4$ is
because we have symmetrized with respect to the $c_n^{\begb}$'s.

Assuming that the dominant scattering is in the forward direction, with the momenta in the $t$-channel small,
%
\bea
\Lambda^{\begb,\begd}_{nm}
&\simeq& \frac{1}{4} 
\signf_\begb \; \signf_\begd \;
\int_\begb d\Gamma_{{a_1b_1}}\int_\begd d\Gamma_{a_2 b_2}\int_{\begb} d\Gamma_{a_3 b_3} \,\int_{\begd}d\Gamma_{a_4 b_4}
\nn \\
&& \times (2\pi)^4 \delta^{(4)}(P^{a_1b_1}_1+P^{a_2b_2}_2-P^{a_3b_3}_3-P^{a_4b_4}_4)
 |\mathcal{M}_{\begb \begd \rightarrow \begb \begd}|^2 \nn\\
&& \times f_{a_1 b_1}^{0}f_{a_2 b_2}^{0}\big(1+f_{a_3 b_3}^{0}\big)\big(1+f_{a_4 b_4}^{0}\big)k^2 (F_{nm}^{\begb}(p,x)c^{\begb}_nc^{\begb}_m+F_{nm}^{\begd}(p',x)c^{\begd}_{n}c_m^{\begd})\;,
\eea
with
\bea
F_{nm}^{\begb}(p,x)=2(1-x^2)p^2\chi_n^\begb(p) \chi_m^\begb(p)+\frac{2x^2}{3}(p^2 \chi_n^\begb(p))'(p^2\chi_m^\begb(p))' \;,
\eea
where $x=k_0/k$ as defined in Eq.~\eqref{eq:define_x}.
In the limit of $k_0 \rightarrow 0$, we can write
\begin{align}
f_{a_1 b_1}^{0}&f_{a_2 b_2}^{0}\big(1+f_{a_3 b_3}^{0}\big)\big(1+f_{a_4 b_4}^{0}\big)=\sum_{n_4=0}^\infty\sum_{n_3=0}^\infty\sum_{n_2=1}^\infty\sum_{n_1=1}^\infty \exp\bigg[-\beta\sum_{j=1}^4n_j(E_j-iQ^{a_j b_j}) \bigg]\nn\\
&\simeq\sum_{n_3,n_4=0}^\infty\sum_{n_1,n_2=1}^\infty \mathcal{G}(Q,n_1,\ldots, n_4)\exp\big[-\beta\{ (n_1+n_3) p+(n_2+n_4) p'\}\big] \;,
\label{eq:ff_expansion}
\end{align}
where
\begin{equation}
\mathcal{G}(Q,n_1,\ldots, n_4) = \exp\left[ i \beta \sum_{j=1}^4 n_j Q^{a_j b_j}
\right] \; ,
\end{equation}
and we employed the expansion of the distribution function,
\begin{equation}
f^0_{ab}(E)
=\sum_{n=1}^\infty \mathrm{e}^{-\beta n (E-iQ^{ab})}
\; .
\end{equation}
%
%
The form in Eq.~\eqref{eq:ff_expansion} is useful, as we have factored the dependence upon the momenta from that upon the color charge.

\subsubsection{Gluon-gluon scattering}
At nonzero holonomy, the matrix elements for
$\mathrm{g}\mathrm{g} \rightarrow \mathrm{g}\mathrm{g}$ are 
%
\bea
\Lambda^{\mathrm{g}\mathrm{g}}_{nm}&=& 2\frac{g^4}{(2\pi)^5}\prod_{i=1}^4 \sum_{a_i b_i} \sum_{n_1,n_2=1}^\infty \sum_{n_3, n_4=0}^\infty \mathcal{G}(Q,n_1,\ldots, n_4) \,t_{13}^{ab} \,t_{24}^{ba} \,t_{13}^{cd} \,t_{24}^{dc}\,
\int_0^{\infty}  dp  \,p^2e^{-\beta p (n_1+n_3)}\nn\\
&\times& \int_0^\infty dp' p'^2 e^{-\beta p' (n_2+n_4)}\int_{-1}^1 dx\, (F_{nm}^{\mathrm{g}}(p,x)+F_{nm}^{\mathrm{g}}(p',x))c^{\mathrm{g}}_nc^{\mathrm{g}}_m \int_{-\pi}^{\pi}\frac{d\phi}{2\pi}  \int_0^\infty dk\, k^3\nn\\
&\times& \bigg\{D_L^{ab}(K)+(1-x^2)\cos \phi D_T^{ab}(K)\bigg\}\bigg\{D_L^{cd}(K)+(1-x^2)\cos \phi D_T^{cd}(K)\bigg\}^* .
\label{eq:Lambdagg}
\eea
The detailed definitions of the longitudinal and transverse
propagators, $D_L^{ab}(K)$ and $D_T^{ab}(K)$, are not required, since to leading logarithmic order~\cite{Bellac:2011kqa}, 
	\bea \label{prop_integral}
	\int dk\, k^3 \, |D_L^{ab}(K)+(1-x^2)\cos \phi D_T^{ab}(K)|^2 \simeq \frac{1}{2} (1-\cos \phi)^2 \ln 
 \left( \frac{\kappa}{g^2 N_c} \right) \; , 
 \label{eq:IR_integral}
	\eea
where $x = \cos(\theta)$ is defined in Eq.~\eqref{eq:define_x}.  

{\color{black} 
In Eq.~\eqref{eq:IR_integral}, we introduce the parameter
$\kappa$. This represents the uncertainty in our computation, which is only done to leading
logarithmic order in weak coupling.
Perturbatively, it can be computed by including effects
at next to leading logarithmic order,
following Ref. \cite{Arnold:2003zc}.  
}

The angular integration becomes
\begin{align}
\int_{-1}^{1} dx \int \frac{d\phi}{2\pi}  
&\big[
F_{nm}^\mathrm{g}(p,x)+F_{nm}^\mathrm{g}(p',x)
\bigr]
(1-\cos \phi)^2\nn\\
&= \frac{20}{3}  \sum_{l=0}^{n{+}m}C^{\mathrm{g}}_{nm,l} \bigg[p^2\bigg(\frac{p}{T}\bigg)^{l}+p'^2\bigg(\frac{p'}{T}\bigg)^{l}\bigg] \;,
\end{align}
where 
\bea
C^{\mathrm{g}}_{nm,l}=\sum_{r=0}^{n}\sum_{s=0}^{m}b^{\mathrm{g}}_{nr} b^{\mathrm{g}}_{ms}\bigg(1+\frac{1}{10}(rs+2l)\bigg)\delta_{r+s,l} \;.
\label{Cnm}
\eea
Integrating over $p$ and $p'$, we obtain
\bea
\int  dp \,p^2 e^{-\beta p (n_1+n_3)} \int dp' p'^{4+l} \beta^l e^{-\beta p' (n_2+n_4)}=\frac{2T^8 \, \Gamma(5+l)}{(n_1+n_3)^3(n_2+n_4)^{5+l}} \;.
\eea
The matrix element becomes
\bea
\Lambda^{\mathrm{g}\mathrm{g}}_{nm}&=& 20g^4\frac{T^8}{\pi^5}\ln\bigg(\frac{\kappa}{g^2 N_c}\bigg) \prod_{i=1}^4 \sum_{a_i b_i} \sum_{n_1,n_2=1}^\infty \sum_{n_3, n_4=0}^\infty \mathcal{G}(Q,n_1,\ldots, n_4) \,t_{13}^{ab} \,t_{24}^{ba} \,t_{13}^{cd} \,t_{24}^{dc}\nn\\
&& \;\;\;\;\;\;\;\;\;\;\
\times \frac{1}{(n_1+n_3)^3} \sum_{l=0}^{m+n} \frac{C^{\mathrm{g}}_{nm,l}}{(n_2+n_4)^{5+l}}\frac{\Gamma(5+l)}{24} c^{\mathrm{g}}_nc^{\mathrm{g}}_m\;.
\eea
At large-$N_c$, this simplifies to
\bea
\Lambda^{\mathrm{g}\mathrm{g}}_{nm} &=&\frac{10 T^8 (g^2 N_c)^2 N_c^2}{\pi^5}\ln\bigg(\frac{\kappa}{g^2 N_c}\bigg)\mathcal{X}_{nm}^{\mathrm{g}\mathrm{g}}c^{\mathrm{g}}_nc^{\mathrm{g}}_m \; ,
\eea
with
\bea
\mathcal{X}_{nm}^{\mathrm{g}\mathrm{g}} &=& \sum_{m_1=1}^\infty\sum_{m_2=1}^\infty\sum_{m_3=1}^{m_1}\sum_{m_4=1}^{m_2}\sum_{l=0}^{m+n}\frac{\ell_{m_1}}{m_1^3} \frac{\ell_{m_2}}{m_2^{5+l}}\,C^{\mathrm{g}}_{nm,l}\frac{\Gamma(5+l)}{\Gamma(5)}\ell_{m_1,m_2,m_3,m_4}\;,
\eea
%
where we define
\bea
\ell_{m_1,m_2,m_3,m_4}=\ell_{m_4-m_3}\ell_{m_2+m_3-m_1-m_4}+\ell_{m_1-m_3+m_4}\ell_{m_2+m_3-m_4}\;.
\label{eq:lms}
\eea
\subsubsection{Gluon-teen scattering}
For gluon-teen scattering, Fig.~\ref{fig:ABCdiagram}b,
%
\begin{align}
\Lambda^{\mathrm{g} \begt}_{nm}&= {-\frac{1}{2}}2\frac{g^4}{(2\pi)^5} \prod_{i=1}^4 \sum_{a_i b_i} \sum_{n_1,n_2=1}^\infty \sum_{n_3, n_4=0}^\infty \mathcal{G}(Q,n_1,\ldots, n_4) \,t_{13}^{ab} \,t_{24}^{ba} \,t_{13}^{cd} \,t_{24}^{dc}  \int_0^{T_d} p_\perp dp_\perp  \int_0^\infty dp_\parallel \hspace*{3cm}\nn\\
&\times 
 \,e^{-\beta p (n_1+n_3)} \int_0^\infty dp' p'^2 e^{-\beta p' (n_2+n_4)}\int_{-1}^1 dx\, (F_{nm}^\begt(p,x)c^{\begt}_nc^{\begt}_m+F_{nm}^{\mathrm{g}}(p',x)c^{\mathrm{g}}_nc^{\mathrm{g}}_m) \int_{-\pi}^{\pi}\frac{d\phi}{2\pi}  \hspace{3cm} \nn\\
&\hspace{-0.1cm}\times \int_0^\infty\!\! dk\, k^3 \left[D_L^{ab}(K)+(1-x^2)\cos \phi D_T^{ab}(K)\right]\left[D_L^{cd}(K)+(1-x^2)\cos \phi D_T^{cd}(K)\right]^* \!. \hspace{-.2cm}
\end{align}
We put $-1/2$ factor in the above equation because the teen has the negative sign, $\mathcal{S}_\begt=-1$, and has no spin. The $k$ integration is the same as in Eq.~\eqref{eq:IR_integral}.
The angular integration becomes
\begin{align}
\int dx \int \frac{d\phi}{2\pi} &\bigl[F_{nm}^\begt(p,x)+F_{nm}^\mathrm{g}(p',x) \bigr] (1-\cos \phi)^2\nn\\
&= \frac{20}{3}  \sum_{l=0}^{n+m} \bigg[C^{\begt}_{nm,l}\bigg(\frac{p}
{T}\bigg)^{l}p^2+C^{\mathrm{g}}_{nm,l}\bigg(\frac{p'}{T}\bigg)^{l}p'^2\bigg] \;,
\end{align}
where $C^{\mathrm{g}}_{nm,l}$ is given in Eq.~\eqref{Cnm} and 
\bea
C^{\begt}_{nm,l}=\sum_{r=0}^{n}\sum_{s=0}^{m}b^{\begt}_{nr} b^{\begt}_{ms}\bigg(1+\frac{1}{10}(rs+2l)\bigg)\delta_{r+s,l} \;.
\eea
Integrating over $p$ and $p'$, we get
\bea
\int p_\perp dp_\perp\int  dp_\parallel e^{-\beta p_\parallel (n_1+n_3)} \int dp' p'^{4+l} \beta^l e^{-\beta p' (n_2+n_4)}=\frac{T^6 \,T_d^2\, \Gamma(5+l)}{2(n_1+n_3)(n_2+n_4)^{5+l}} \;.
\eea
In this expression we approximate $p \approx p_\parallel$.
We also require
\bea
\int p_\perp dp_\perp\int  dp_\parallel \,p_\parallel^{2+l} e^{-\beta p (n_1+n_3)} \int dp' p'^{2} \beta^l e^{-\beta p' (n_2+n_4)}=\frac{T^6 \,T_d^2\, \Gamma(3+l)}{(n_1+n_3)^{3+l}(n_2+n_4)^3} \;.
\eea
The matrix element becomes
\begin{align}
 \Lambda^{\mathrm{g} \begt}_{nm}&= {-\frac{1}{2}}20g^4\frac{T_d^2\, T^6}{\pi^5}\ln\bigg(\frac{\kappa}{g^2 N_c}\bigg) \prod_{i=1}^4 \sum_{a_i b_i} \sum_{n_1,n_2=1}^\infty \sum_{n_3, n_4=0}^\infty \mathcal{G}(Q,n_1,\ldots,n_4) \,t_{13}^{ab} \,t_{24}^{ba} \,t_{13}^{cd} \,t_{24}^{dc}\nn\\
 &\times\bigg[ \frac{1}{n_1+n_3} \sum_{l=0}^{m+n} \frac{C^{{\mathrm{g}}}_{nm,l}c^{\mathrm{g}}_nc^{\mathrm{g}}_m}{(n_2+n_4)^{5+l}}\frac{\Gamma(5+l)}{8\Gamma(5)}+\frac{1}{(n_2+n_4)^3} \sum_{l=0}^{m+n} \frac{C^{{\begt}}_{nm,l}c^{\begt}_nc^{\begt}_m}{(n_1+n_3)^{3+l}}\frac{\Gamma(3+l)}{48\Gamma(3)}\bigg] \;.
 \label{Lambdag3_mn_simp}
\end{align}
Equation~\eqref{Lambdag3_mn_simp} further simplifies to
\bea
\Lambda^{\mathrm{g} \begt}_{nm}&=& \frac{10 T_d^2 T^6 (g^2 N_c)^2 N_c^2}{\pi^5}\ln\bigg(\frac{\kappa}{g^2 N_c}\bigg)
\left( \mathcal{X}_{nm}'^{\mathrm{g}\mathrm{g}}c^{\mathrm{g}}_nc^{\mathrm{g}}_m
+\mathcal{X}_{nm}'^{\begt\begt}c^{\begt}_nc^{\begt}_m \right) \; ,  
\eea
with
\begin{subequations}
\bea
\mathcal{X}_{nm}'^{\mathrm{g}\mathrm{g}}&=&-{\frac{1}{2}}\sum_{m_1,m_2=1}^\infty
\sum_{m_3=1}^{m_1}\sum_{m_4=1}^{m_2}\sum_{l=0}^{m+n}
\frac{\ell_{m_1}}{m_1}\frac{\ell_{m_2}}{m_2^{5+l}}\,\frac{\Gamma(5+l)}{8\Gamma(5)}\, C^{\mathrm{g}}_{nm,l}\; \ell_{m_1,m_2,m_3,m_4}\;,
 \\
\mathcal{X}_{nm}'^{\begt\begt}&=&-{\frac{1}{2}}\sum_{m_1,m_2=1}^\infty\sum_{m_3=1}^{m_1}\sum_{m_4=1}^{m_2}\sum_{l=0}^{m+n}
\frac{\ell_{m_1}}{{m_1^{3+l}}}\frac{\ell_{m_2}}{{m_2^{3}}}\,\frac{\Gamma(3+l)}{48\Gamma(3)}
\, C^{{\begt}}_{nm,l}\;\ell_{m_1,m_2,m_3,m_4}\; .
\eea
\end{subequations}
\subsubsection{Teen-teen scattering}
We next turn to teen-teen scattering. The matrix elements in this case is
\bea
\Lambda^{{\begt\begt}}_{nm}&=& 2{\frac{1}{4}}\frac{g^4}{(2\pi)^5} \prod_{i=1}^4 \sum_{a_i b_i} \sum_{n_1,n_2=1}^\infty \sum_{n_3, n_4=0}^\infty \mathcal{G}(Q,n_1,\ldots, n_4) \,t_{13}^{ab} \,t_{24}^{ba} \,t_{13}^{cd} \,t_{24}^{dc}\,
\int_0^{T_d} p_\perp dp_\perp \int_0^\infty dp_\parallel \,e^{-\beta p (n_1+n_3)}\nn\\
&\times& \int_0^{T_d} p'_\perp dp'_\perp \int_0^\infty dp'_\parallel \,e^{-\beta p' (n_2+n_4)}\int_{-1}^1 dx\, (F_{nm}^\begt(p,x)+F_{nm}^\begt(p',x))c^{\begt}_nc^{\begt}_m \int_{-\pi}^{\pi}\frac{d\phi}{2\pi}  \int_0^\infty dk\, k^3\nn\\
&\times& \bigg\{D_L^{ab}(K)+(1-x^2)\cos \phi D_T^{ab}(K)\bigg\}\bigg\{D_L^{cd}(K)+(1-x^2)\cos \phi D_T^{cd}(K)\bigg\}^* \;.
\eea
We put the $1/4$ factor in front of the above equation because teen ghosts have no spin. Now, integrating over $p$ and $p'$,
\begin{align}
\int p_\perp dp_\perp\, \int p'_\perp dp'_\perp \int  dp_\parallel e^{-\beta p (n_1+n_3)} &\int dp' p'^{2+l} \beta^l e^{-\beta p' (n_2+n_4)}\nn\\
&=\frac{T^4 \,T_d^4\, \Gamma(3+l)}{4(n_1+n_3)(n_2+n_4)^{3+l}} \;.
\end{align}
The matrix element becomes
\begin{align}
\Lambda^{{\begt\begt}}_{nm}= 20{\frac{1}{4}}g^4\frac{T^4 \,T_d^4}{\pi^5}\ln\bigg(\frac{\kappa}{g^2 N_c}\bigg) \prod_{i=1}^4 \sum_{a_i b_i} &\sum_{n_1,n_2=1}^\infty \sum_{n_3, n_4=0}^\infty \mathcal{G}(Q,n_1,\ldots ,n_4) \,t_{13}^{ab} \,t_{24}^{ba} \,t_{13}^{cd} \,t_{24}^{dc}\nn\\
&\times \frac{1}{(n_1+n_3)} \sum_{l=0}^{m+n} \frac{C^{\begt}_{nm,l}c^{\begt}_nc^{\begt}_m}{(n_2+n_4)^{3+l}}\frac{\Gamma(3+l)}{96\,\Gamma(3)} \;.
\label{Lamba_nm^33}
\end{align}
%
Equation~\eqref{Lamba_nm^33} simplifies  as
%
\bea
\Lambda^{{\begt\begt}}_{mn}&=& 
\frac{10 T^4 \,T_d^4 (g^2 N_c)^2 N_c^2}{\pi^5}\ln\bigg(\frac{\kappa}{g^2 N_c}\bigg)\mathcal{X}_{nm}^{\begt\begt}c^{\begt}_nc^{\begt}_m \;,
\eea
with
\bea
\mathcal{X}_{nm}^{\begt\begt}&=&{\frac{1}{4}}\sum_{m_1,m_2=1}^\infty
\sum_{m_3=1}^{m_1}\sum_{m_4=1}^{m_2}\sum_{l=0}^{m+n}\frac{\ell_{m_1}}{m_1} \frac{\ell_{m_2}}{m_2^{3+l}}\,C^{\begt}_{nm,l}\frac{\Gamma(3+l)}{96\Gamma(3)}\ell_{m_1,m_2,m_3,m_4}\;.
\eea
In summary, Eq.~\eqref{eq:sum_cLc} reduces to
\begin{eqnarray}
\sum_{n,\begb}c_n^{\begb}(\mathcal{L}c)_n^{\begb}
&=&
\frac{10 (g^2 N_c)^2 N_c^2}{\pi^5}\ln\bigg(\frac{\kappa}{g^2 N_c}\bigg)\nonumber\\
&&\hspace{-1cm}\times\sum_{n,m}
\begin{bmatrix}
c^{\mathrm{g}}_n & c^{\begt}_n
\end{bmatrix}
\begin{bmatrix}
T^8\mathcal{X}_{nm}^{\mathrm{g}\mathrm{g}}+T^6T_d^2\mathcal{X}_{nm}^{'\mathrm{g}\mathrm{g}} & 0\\
0 & T^4T_d^4\mathcal{X}_{nm}^{\begt\begt}+T^6T_d^2\mathcal{X}_{nm}^{'\begt\begt}
\end{bmatrix}
\begin{bmatrix}
c^{\mathrm{g}}_m \\ c^{\begt}_m
\end{bmatrix}\; .
\end{eqnarray}
\subsection{Results for the shear viscosity}
	\begin{figure}[ht]
		\begin{center}
		\includegraphics[scale=1.4]{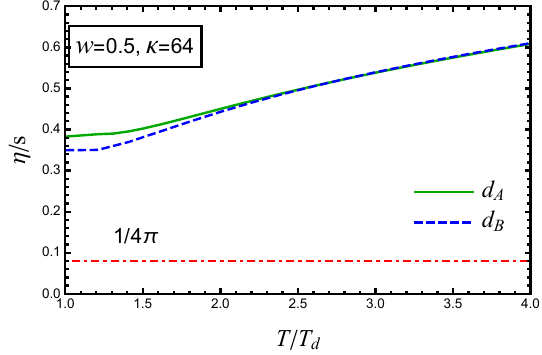}
			\caption{Ratio of the shear viscosity to the entropy with the ansatzes of Eq.~\eqref{improved_loop}.  The dash-dot line
            denotes the value from AdS/CFT, $1/4 \pi$.}
			\label{fig:shear_viscosity_1}
		\end{center}
	\end{figure}
	\begin{figure}[ht]
		\begin{center}
		\includegraphics[scale=1.4]{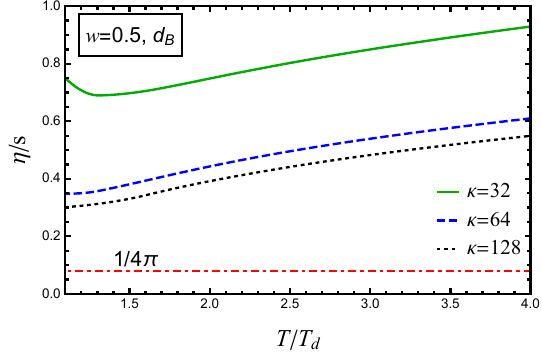}
			\caption{Ratio of the shear viscosity to the entropy for different parameter sets of $\kappa$.
            The values of the parameters $\kappa: 32, 64, 128$ are chosen so that $\eta/s$ is relatively small.
            }
			\label{fig:shear_viscosity_2}
		\end{center}
	\end{figure}
	\begin{figure}[ht]
		\begin{center}
  		\includegraphics[scale=1.4]{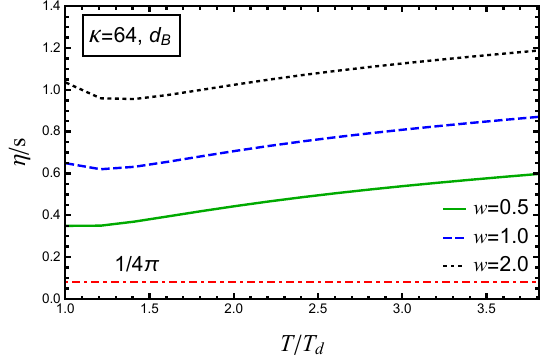}
			\caption{Ratio of the shear viscosity to the entropy for different parameter sets of $w$.}
			\label{fig:shear_viscosity_3}
		\end{center}
	\end{figure}
We make the extreme approximation of
considering only $n=0$. From Eq.~\eqref{eq:sum_cLc}
%
we write
\bea
\eta=\frac{2T}{5} \sum_{\begb}c_0^{\begb}(\mathcal{L}c)_0^{\begb} \; ,
\eea
where the sum over $\begb$ includes that over gluon and
teen fields.
%
%
%
%
%
%
The final expression for the shear viscosity becomes
\bea
\eta=\frac{4\pi^5 T^3}{225 (g(T)^2 N_c)^2 N_c^2\ln({\kappa}/{g^2 N_c})} \bigg[\frac{(d_0^\begt)^2}{\chi_{00}^{\begt\begt}+\frac{T^2}{T_d^2}\chi_{00}'^{\begt\begt} }+\frac{(d_0^\mathrm{g})^2}{\chi_{00}^{\mathrm{gg}}+\frac{T_d^2}{T^2}\chi_{00}'^{\mathrm{gg}} }\bigg]\; .
\label{Eq:eta_final}
\eea
%
Now we estimate the shear viscous coefficient $\eta$ for the small value of the Polyakov loop. Neglecting higher moments ($\ell_n$ for $n\ge 2$)  and power of $\ell_1$,
\bea\label{eta:small_l}
\eta&=&\frac{16\pi}{5} \frac{12800 T^4-3248 T^2 T_d^2+3 T_d^2}{64T^4-20 T^2 T_d^2+T^2_d}\,\frac{N_c^2 \,\,T^3}{10 (g^2 N_c)^2\ln (\kappa/g^2 N_c)}\;\ell_1^2 \;.
\eea
For pure gluonic cases without teen contribution, we have the following expression,
\bea\label{shear_gluon}
\eta_{\rm{glue}}&=&\frac{4\pi^5 T^3}{225 (g(T)^2 N_c)^2 N_c^2\ln({\kappa}/{g^2 N_c})} \frac{(d_0^{\mathrm{g}})^2}{\chi_{00}^{\mathrm{gg}} }\nn
\\
&=&\frac{\pi^6}{270\zeta^2(5)}\;\ell_1^2\;\eta_{\rm{pert}}\simeq 3.31\;\ell_1^2\;\eta_{\rm{pert}}\; ,
\eea
where $\eta_{\rm{pert}}$ is the shear viscosity at large $N_c$, in zero background field and given as
\bea
\eta_{\rm{pert}}&=& 540 \zeta^2(5)\left(\frac{2}{\pi}\right)^5\frac{T^3 N_c^2}{(g^2 N_c)^2 \ln (\kappa/g^2 N_c)}\; ,
\eea
which agrees with the previous works in the pure glue theory ~\cite{Baym:1990uj,Arnold:2000dr,Hidaka:2009ma}.
At small $\ell_1$, the shear viscosity is $\eta\sim\ell_1^2$ for the gluonic medium with and without teens. This can be understood as $\eta\sim d_0^2/\chi_{00}\sim \ell_1^2$, where $d_0\sim \ell_1^2$ and $\chi_{00}\sim \ell_1$.

In Fig~\ref{fig:shear_viscosity_1}, we plot Eq.~\eqref{Eq:eta_final} for two ansatzes mentioned in Eq.~\eqref{improved_loop}.
We adopt the functional form of the one-loop perturbative coupling for zero quark flavors $(N_f=0)$, as presented in Refs.~\cite{Fukushima:2013xsa,Madni:2022bea}, to define a non-perturbative running coupling characterized by a single parameter $w$.
For the running coupling, we take
	\bea
	g(T)^2 N_c=4\pi \frac{6\pi}{11 \log(w 2\pi T/T_d)}\; .
    \label{g2nc_exp}
	\eea
With the increasing temperature $T$, the coupling decreases, and the shear viscosity increases.
 With this form of the coupling constant, taking
the values $w = 0.5$ and $\kappa = 64$, at $T_d$
$\eta_{\rm{pert}}/s \approx 0.55$.
{\color{black} 
We comment that 
we {\it choose} rather large values for $\kappa = 32, 64, 128$, so that
that $\eta/s$ is relatively small.  
Needless to say, it would be useful to compute beyond leading logarithmic order
\cite{Arnold:2003zc} to see if these values 
are reasonable.
}

Here and henceforth, to be self-consistent we used 
the entropy as computed from the matrix model.  The difference from the entropy for $N_c = 3$ from the lattice is small.

In the matrix model, 
even with admittedly large values of $\kappa$, the ratio
$\eta/s$ is still well above that in Anti-de Sitter/Conformal Field Theory, AdS/CFT~\cite{Kovtun:2003wp} bound.
This is illustrated in the figures with a red dash-dotted line. 

The dependence of shear viscosity on the parameters $\kappa$ and $w$ are shown in Figs.~\ref{fig:shear_viscosity_2} and~\ref{fig:shear_viscosity_3}, respectively. The former is obtained by varying $\kappa$ with a fixed value of energy scale $\pi T$ i.e., $w=0.5$. With increasing $\kappa$, the value of the viscosity reaches towards the AdS/CFT bound. These results are shown for one ansatz $d_B$. One will obtain a similar result for another ansatz with a marginal difference.
As the QCD coupling is not well-known near the transition, we show the plots varying $w$.
Figure~\ref{fig:shear_viscosity_3} shows the results for different energy scales around $2\pi T$, i.e., for $w=0.5, 1, 2$. 

\begin{figure}[ht]
		\begin{center}
		\includegraphics[scale=1.4]{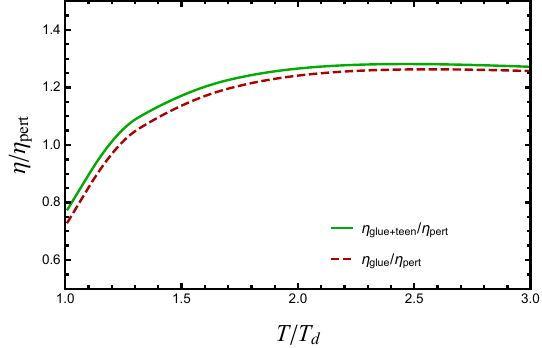}
			\caption{The green line denotes the ratio of the shear viscosity including teen-teen, 
            teen-gluon, and gluon-gluon scattering, to that with perturbative gluon-gluon scattering. The red-dashed line indicates the ratio of the shear viscosity including just gluon-gluon scattering, to that with perturbative. As a ghost  field, the contribution with teens increases the total shear viscosity.}
			\label{fig:shear_viscosity_4}
		\end{center}
	\end{figure}

In Fig.~\ref{fig:shear_viscosity_4}, we compare our results with zero background field case.
The green line represents the ratio of shear viscosity accounting for teen-teen, teen-gluon, and gluon-gluon scattering compared to gluon-gluon scattering in zero background field, whereas, the red line indicates the ratio of viscosity only including gluon-gluon scattering in presence and absence of the background fields. In the presence of background fields, these additional contributions, particularly from teens as ghost fields, lead to a rise in the overall viscosity. The inclusion of teen-teen and teen-gluon scattering contributes around $6\%$ to the viscosity near $T_d$. 
\section{Bulk Viscosity}
\label{sec:bulk_viscosity}
We next compute the bulk viscosity.  While of utmost importance
for phenomenology, in perturbation theory, it is more difficult to extract.  The bulk viscosity vanishes for any conformal theory,  so in gauge theory, it is strongly suppressed relative to the shear viscosity.  
As shown by Arnold, Dugan, and Moore~\cite{Arnold:2006fz}, the bulk viscosity
only arises because quarks and gluons develop a thermal ``mass'' $\sim g T$.
Even then, it is necessary to include the running of the
coupling constant with temperature.  Consequently, the
bulk viscosity is suppressed by $\sim g^8$ and is
$\zeta \sim g^4$, relative to the shear viscosity, $\eta \sim 1/g^4$.

In contrast, in our matrix model the theory is not conformally invariant if the Polyakov loops are less than unity, since then
there must be a non-trivial distribution for the 
eigenvalues of the thermal Wilson line.  
Thus while in perturbation theory it is involved
to measure the deviation from conformality through the bulk viscosity,
in our matrix model it is straightforward.  In particular,
we can compute in precisely the same way as for the shear
viscosity, to leading logarithmic order in the coupling constant.  As we show, the bulk viscosity is nonzero
because the speed of sound squared is not equal to $1/3$;
{\it i.e.}, the theory is not conformal.

For the bulk viscosity, 
from Eq.~\eqref{eq:BEfinal}, 
%

\begin{equation}\label{Eq:boltzmann_bulk}
 2P^{\mu, a_1 b_1} \partial_\mu^{(1)}f_{\begb, a_1 b_1}
 = \beta f^0_{a_1b_1}(1+f^0_{a_1b_1}) \mathrm{X}q(|\vb*{p}|) \; ,
\end{equation}
where
\begin{equation}
\mathrm{X}=
 \nabla\cdot \bm{u}\;,
 \end{equation}
and
\begin{equation}
q(p) = 2 p^2 \qty(\frac{1}{3} 
-  v_s^2) \equiv 2 p^2 \Delta v_s^2 \; ,
\end{equation}
with $v_s^2$ the speed of sound, squared, Eq.~\eqref{eq:speedofsound}. In a conformally invariant theory
$v_s^2 = 1/3$ and the bulk viscosity vanishes.  
In our matrix model, the dominant source
of conformal symmetry breaking is from $v_s^2 \neq 1/3$ as discussed below.

We expand the statistical distribution functions to linear order,
\begin{equation}
    f^{(n)}_{\begb, ab}=\ f^{0}_{ab} \;(1+f^{0}_{ab})\psi_{\begb,ab}^{(n)} \;.
    \end{equation}
Analogous to Eq.~(\ref{lhs=rhs}), then, the Boltzmann equation is written as
 \be
 \beta f^0_{a_1b_1} (1+f^0_{a_1b_1}) \mathrm{X} q(|\vb*{p}|)=-\signf_{\begb} (\bar{\mathcal{L}}\psi^{(1)})_{\begb,a_1b_1}\; ,
  \label{eq:Boltzmann_bulk} 
 \ee
 where $\bar{\mathcal{L}}$ is a linear operator,
 %

\bea\label{bulk_L}
&& (\bar{\mathcal{L}}
\psi^{(n)})_{\begb, a_1 b_1}=
\frac{1}{2}
\sum_{\begd, \begb', \begd'}
\signf_\begb \signf_\begd
\int_\begd d\Gamma_{a_2 b_2}
\,\int_{\begb'} d\Gamma_{a_3 b_3} \,\int_{\begd'}d\Gamma_{a_4 b_4}
\,(2\pi)^4 \delta^{(4)}(P^{a_1b_1}_1+P^{a_2b_2}_2-P^{a_3b_3}_3-P^{a_4b_4}_4)
 \nn\\
&&\;\;\;\;\;\;\;\;\;
\times |\mathcal{M}_{\begb \begd \rightarrow \begb' \begd'}|^2 \bigg[f_{a_1 b_1}^{0} f_{a_2 b_2}^{0} 
(1+f_{a_3 b_3}^{0})(1+f_{a_4 b_4}^{0})\bigg]
\bigg(\psi_{\begb, a_1b_1}^{(n)}+\psi_{\begd,a_2b_2}^{(n)}-
\psi_{\begb',a_3b_3}^{(n)}-\psi_{\begd',a_4b_4}^{(n)}\bigg) \;.
\eea
Here, the operator $\bar{\mathcal{L}}$ is the same as ${\mathcal{L}}$ in Eq.~\eqref{shear_L}, with the bar added to emphasize its role in describing the bulk viscosity.
The bulk viscosity is defined by the change in pressure caused by the deviation from equilibrium $\psi_\begb$.
The bulk viscosity, $\zeta$, is given by~\cite{Arnold:2006fz}
\bea
\zeta&=& -\sum_{\begb} \signf_{\begb}\; \int_{\begb} d\Gamma_{a_1b_1} \, q(|\vb*{p}|) f^0_{a_1b_1}(1+f^0_{a_1b_1}) \psi_{\begb} (p)\;.
\label{eq:zeta_1}
\eea
The source term, $ f^0_{a_1b_1} (1+f^0_{a_1b_1}) q(|\vb*{p}|)$, in Eq.~\eqref{eq:Boltzmann_bulk} can be written in terms of an expansion basis as
\bea \label{V_g0}
V_{\begb,n}=\signf_{\begb}\;  \int_{\begb} d\Gamma_{a_1b_1} \, 2p^2 \psi^\begb_{n} \beta f^0_{a_1b_1}(1+f^0_{a_1b_1})\, \Delta v_s^2 \;,
\eea
%
giving us $V_{\begb,n}=-(\bar{\mathcal{L}}\bar{c})_n^{\begb}$. 
Here, we expanded $\psi^{\begb}$ in an orthogonal basis equivalent to Eq.~\eqref{orthogonal_expansion} as $\psi^{\begd}=\sum_n\bar{c}_n^{\begd}\psi^{\begd}_n$.
Finally, the coefficient of bulk viscosity from Eq.~\eqref{eq:zeta_1} becomes
\bea
\label{eq:bulk_viscosity}
\zeta &=&T \sum_{\begb} \;\int_{\begb} d\Gamma_{a_1b_1} \psi_{\begb}(\bar{\mathcal{L}} \psi)_{\begb}\nonumber\\
&=&T \sum_{ \begb}\sum_{n=0}^\infty V_n^{\begb} \big(\bar{\mathcal{L}}^{-1}V\big)_n^{\begb}= T \sum_{\begb}\sum_{n=0}^{\infty}\bar{c}_n^{\begb}(\bar{\mathcal{L}}\bar{c})_n^{\begb}\; .
\eea

In Appendix \ref{appendix:teen_energy},
we show that if one computes the pressure for a
teen field from kinetic theory, then one obtains 
$\mathscr{P}_{\rm teen} = - T^2 T_d^2/72$, with the energy
$\mathscr{E} = - T^2 T_d^2/24$.  That is, in kinetic theory
teen quasiparticles still look three-dimensional, with
$\mathscr{E} = 3\mathscr{P}_{\rm teen}$.  
In Appendix~\ref{appendix:teen_energy}, we 
argue how a kinetic theory of two dimensional teen
particles can arise, with
$\mathscr{E} = \mathscr{P}_{\rm teen} = - T^2 T_d^2/24$
~\cite{Hidaka:2020vna}, as in thermodynamics.

In practice, we {\it define} our model such that in
Eq.~\eqref{eq:bulk_viscosity}, the deviation
from conformality, $\Delta v_s^2$, is computed
from {\it thermodynamics}.  After all, in our model 
the teen fields are not really two-dimensional,
but defined with the functions $d_A(T)$ and $d_B(T)$,
which are fit to obtain $(\mathscr{E}-3\mathscr{P})/T^4$ from the
lattice, Eq.~\eqref{improved_loop}.   Except
for the contribution to $\Delta v_s^2$, though,
we compute Eq.~\eqref{eq:bulk_viscosity} using
gluon and teen quasiparticles in the semi-QGP. 
We believe that this ansatz yields a reasonable estimate
of the bulk viscosity.
%

With this assumption, following Ref.~\cite{Arnold:2006fz}, we use a simple ansatz for the basis functions,
 $\psi^\mathrm{g}$,
 \be\label{basis_gl_bulk}
 \psi_m^{\mathrm{g}}=\frac{T}{(1+p/T)^{K-2}} \;
\(\frac{p}{T}\)^{m} , \qquad m=1,2,\cdots , K\;.
 \ee
    %
    Here, the $\psi^{\mathrm{g}}_m$ are not necessarily orthogonal.
We take a different basis for the $\psi^{\begt}_m$ as
%
 %
 \be\label{basis_teen_bulk}
 \psi_m^{\begt}=\frac{ T}{(1+p/T)^{K-4}} \; 
 \(\frac{p}{T}\)^{m} \; ,  \qquad m=1,2,\cdots , K\;.
 \ee
 By trial and error, we found that a different basis for
 the teen functions, Eq.~\eqref{basis_teen_bulk}, converges more quickly (at $K\sim 30$) than if
 we take the same basis as for gluons,
 Eq.~\eqref{basis_gl_bulk}.

 For a single order basis, i.e., $K=1$, we get
      \bea \label{eq:V_1}
    V_{\mathrm{g},1}&=& 2(\Delta v_s^2)\ \frac{24 T^4}{2\pi^2} \sum_{k=1}^{\infty} \frac{N_c^2 \ell_k^2-1}{k^5} (5+k) \;,\\
    V_{\begt,1}&=& - (\Delta v_s^2)\ \frac{T^2 T_d^2}{4\pi^2}\sum_{k=1}^{\infty} \frac{N_c^2 \ell_k^2-1}{k^5}(60+36k+9k^2+k^3)\;.
      \eea
%
Equation~\eqref{eq:V_1} shows the order of temperature $V_{\mathrm{g}}$ and $V_{\begt}$ contributes to the bulk viscosity computation. But 
beyond $K=1$, it is also necessary to treat the zero modes
in the operator for the bulk viscosity. This is discussed
in Appendix \ref{appendix:zero_mode}. 
In the next section, we have computed bulk viscosity results for $K\sim 20-30$, shown in Figs.~\ref{fig:bulk_viscosity},~\ref{fig:bulk_viscosity_withkw}, and \ref{fig:bulkbyshear_viscosity}.

\subsection{Matrix Elements at Leading logarithmic order}
Analogous to Eq.~\eqref{shear_matrix_1}, the matrix element form for the $\begb\text{-}\begd$ scattering becomes
\bea
&\bar{\Lambda}^{\begb,\begd}_{mn}&= 
\(2\frac{1}{4}\) \frac{1}{2} 
\signf_\begb \signf_\begd
\int_\begb d\Gamma_{{a_1b_1}}
\int_\begd d\Gamma_{a_2 b_2}
\,\int_{\begb} d\Gamma_{a_3 b_3} \,\int_{\begd}d\Gamma_{a_4 b_4}\notag\\
&&\times (2\pi)^4 \delta^{(4)}(P^{a_1b_1}_1+P^{a_2b_2}_2-P^{a_3b_3}_3-P^{a_4b_4}_4)|\mathcal{M}_{\begb \begd \rightarrow \begb \begd}|^2 f_{a_1 b_1}^{0}f_{a_2 b_2}^{0}\big(1+f_{a_3 b_3}^{0}\big)\big(1+f_{a_4 b_4}^{0}\big)\nn\\
&& \times \left[(\psi_1^{\begb,m}- \psi_3^{\begb,m})\bar{c}_n^{\begb} +(\psi_2^{\begd,m}- \psi_4^{\begd,m})\bar{c}_n^{\begd}\right]\left[(\psi_1^{\begb,n}- \psi_3^{\begb,n})\bar{c}_m^{\begb}+(\psi_2^{\begd,n} - \psi_4^{\begd,n})\bar{c}_m^{\begd}\right]\;.
\label{Lambdabd_nm}
\eea
If the momentum transfer in the scattering process is small, we get
\be
\psi^{\begb}_1 - \psi^{\begb}_3 \simeq \vb*{k}\cdot \partial_{\vb*{p}} \psi^{\begb} (p)\; , \quad  \psi^{\begb}_2 - \psi^{\begb}_4 \simeq -\vb*{k}\cdot \partial_{\vb*{p}'} \psi^{\begb} (p')\; ,
\ee
where we have used $\vb*{p}\equiv ({\vb*{p}_1+\vb*{p}_3})/{2}$ and $\vb*{p}' \equiv ({\vb*{p}_2+\vb*{p}_4})/{2}$, $x\equiv {k_0}/{k}\simeq \cos \theta_{\bm{pk}} \simeq \cos \theta_{\bm{p'k}}$ from Eqs.~(\ref{mom_variable})-(\ref{eq:define_x}). In the small momentum transfer limit, the matrix element from Eq.~\eqref{Lambdabd_nm} becomes
 \bea \label{Lambda_nm_2}
\bar{\Lambda}^{\begb,\begd}_{mn}&=&2 \frac{g^4}{(2\pi)^5}\frac{\sigma_\begb\sigma_\begd}{4}\signf_\begb \signf_\begd \prod_{i=1}^{4}\sum_{a_i,b_i} \sum_{n_1,n_2=1}^{\infty}\sum_{n_3,n_4=0}^{\infty} \mathcal{G}(Q,n_1,\cdots,n_4) t^{ab}_{13}\,t^{ba}_{24}\,t^{cd}_{13}\,t^{dc}_{24} \, \int_0^{\infty} dp\, p^2 e^{-\beta p (n_1+n_3)} \nn \\
&&\times \int_0^{\infty}dp' \, p'^2 \, e^{-\beta p' (n_2+n_4)} \int_0^{2\pi}\frac{d\phi}{2\pi} \,\int_{-1}^{1}dx\, \int_0^{\infty}dk\, k^3 \,|D_L^{ab}(K)+(1-x^2) \cos \phi \, D_T^{ab}(K)|^2 \nn \\
&&\times \[\bar{c}_m^{\begb}\,\hat{\bm{k}}\cdot \partial_{\vb*{p}} \psi^{\begb,m} (p) - \bar{c}_m^{\begd}\,\hat{\bm{k}}\cdot \partial_{\vb*{p}'} \psi^{\begd,m} (p')\] \[\bar{c}_n^{\begb}\, \hat{\bm{k}}\cdot \partial_{\vb*{p}} \psi^{\begb,n} (p) - \bar{c}_n^{\begd}\, \hat{\bm{k}}\cdot \partial_{\vb*{p}'} \psi^{\begd,n} (p')\]\;.
 \eea
 Here, $\sigma_\begb$ represents the spin degeneracy; $\sigma_\mathrm{g}=2$ and $\sigma_\begt=1$.
We assume rotational invariance,
$\psi^{\begb,n}(\vb*{p})=\psi^{\begb,n} (p)$, so that
 \bea\label{bulk_k_dot_p}
 \[\bar{c}_m^{\begb}\, \hat{\bm{k}}\cdot \partial_{\vb*{p}} \psi^{\begb,m} (p) - \bar{c}_m^{\begd}\, \hat{\bm{k}}\cdot \partial_{\vb*{p}'} \psi^{\begd,m} (p')\]= x \[\bar{c}_m^{\begb}\, (\psi_m^{\begb})^{'} (p)- \bar{c}_m^{\begd}\, (\psi_m^{\begd})^{'} (p')\]\;.
 \eea
Hence, using Eq.~(\ref{bulk_k_dot_p}) we can rewrite Eq.~(\ref{Lambda_nm_2}) as
\bea \label{matrix_elem_bulk}
\bar{\Lambda}^{\begb,\begd}_{mn}&=&\frac{g^4}{(2\pi)^5}\frac{\sigma_\begb\sigma_\begd}{4}\log\(\frac{\kappa}{g^2 N_c}\) \signf_\begb \signf_\begd \prod_{i=1}^{4}\sum_{a_i,b_i}\sum_{n_1,n_2=1}^{\infty}\sum_{n_3,n_4=0}^{\infty} \mathcal{G}(Q,n_1,\cdots,n_4) t^{ab}_{13}\,t^{ba}_{24}\, t^{cd}_{13}\,t^{dc}_{24}   \nn \\
&&\times\int_{0}^{\infty} dp\, p^2 \, e^{-\beta p (n_1+n_3)} \int_{0}^{\infty} dp'\, p'^2 \, e^{-\beta p' (n_2+n_4)} \int_0^{2\pi}\frac{d\phi}{2\pi} \(1-\cos \phi\)^2 \int_{-1}^{1} dx\, x^2 \,  \nn\\ 
&&\times \[\bar{c}_m^{\begb}\, (\psi_m^{\begb})^{'} (p)- \bar{c}_m^{\begd}\, (\psi_m^{\begd})^{'} (p')\] \[\bar{c}_n^{\begb}\, (\psi_n^{\begb})^{'} (p)- \bar{c}_n^{\begd}\, (\psi_n^{\begd})^{'} (p')\] \nn \\
&=&\frac{g^4}{(2\pi)^5}\frac{\sigma_\begb\sigma_\begd}{4}\log\(\frac{\kappa}{g^2 N_c}\) \signf_\begb \signf_\begd \prod_{i=1}^{4}\sum_{a_i,b_i}\sum_{n_1,n_2=1}^{\infty}\sum_{n_3,n_4=0}^{\infty} \mathcal{G}(Q,n_1,\cdots,n_4) t^{ab}_{13}\,t^{ba}_{24}\, t^{cd}_{13}\,t^{dc}_{24} \nn \\
&&\times  \int_{0}^{\infty} dp\, p^2 \, e^{-\beta p (n_1+n_3)} \int_{0}^{\infty} dp'\, p'^2 \, e^{-\beta p' (n_2+n_4)}\[\bar{c}_m^{\begb}\, (\psi_m^{\begb})^{'} (p)- \bar{c}_m^{\begd}\, (\psi_m^{\begd})^{'} (p')\] \nn \\
&&\times \[\bar{c}_n^{\begb}\, (\psi_n^{\begb})^{'} (p)- \bar{c}_n^{\begd}\, (\psi_n^{\begd})^{'} (p')\] \nn \\
&=& \frac{g^4}{(2\pi)^5}\frac{\sigma_\begb\sigma_\begd}{4}\log\(\frac{\kappa}{g^2 N_c}\) \signf_\begb \signf_\begd  \sum_{m_1,m_2=1}^{\infty}\sum_{m_3=1}^{m_1}\sum_{m_4=1}^{m_2}\frac{N_c^4}{2}\ell_{m_1}\ell_{m_2}(\ell_{m_1-m_2-m_3+m_4}\ell_{m_4-m_3} \nn \\
&& +\ell_{m_1-m_3+m_4}\ell_{m_2+m_3-m_4}) \int_{0}^{\infty} dp\, p^2 \, e^{-\beta p m_1} \int_{0}^{\infty} dp'\, p'^2 \, e^{-\beta p' m_2} \nn \\
&&\times \[\bar{c}_m^{\begb}\, (\psi_m^{\begb})^{'} (p)- \bar{c}_m^{\begd}\, (\psi_m^{\begd})^{'} (p')\]\[\bar{c}_n^{\begb}\, (\psi_n^{\begb})^{'} (p)- \bar{c}_n^{\begd}\, (\psi_n^{\begd})^{'} (p')\].
\label{Lambdabd_nm_simp}
\eea
Here we have used the color sum in the limit of
large $N_c$ and it becomes
   \bea
&&\prod_{i=1}^{4}\sum_{a_i,b_i}t^{ab}_{13}\,t^{ba}_{24}\,  t^{cd}_{13} \, t^{dc}_{24} \mathcal{G}(Q,n_1,\cdots,n_4)\nonumber\\
 &&  = \frac{N_c^4}{2} \ell_{n_1+n_3} \ell_{n_2+n_4} (\ell_{n_2-n_1} \ell_{n_4-n_3}+ \ell_{n_1+n_4}\ell_{n_2+n_3})\;,
   \eea
with $m_1\equiv n_1+n_3$, $m_2 \equiv n_2+n_4$, $m_3\equiv n_3$, $m_4 \equiv n_4$. 
\subsubsection{Gluon-gluon scattering}
For the gluon-gluon scattering, at nonzero holonomy, the matrix element from Eq.~\eqref{Lambdabd_nm_simp} becomes
   \bea
    \bar{\Lambda}^{\mathrm{gg}}_{mn} &=& \frac{g^4 N_c^4}{2(2\pi)^5}\log\(\frac{\kappa}{g^2 N_c} \)\sum_{m_1,m_2=1}^{\infty} \sum_{m_3=1}^{m_1}\sum_{m_4=1}^{m_2} \,\ell_{m_1}\ell_{m_2}\; \ell_{m_1,m_2,m_3,m_4}\; \int_{0}^{\infty} dp\, p^2  \nn \\
    && \times\int_0^{\infty}dp'\,p'^2 \text{e}^{-\beta (m_1 p +m_2 p')} \bar{c}_m^{\mathrm{g}}\bar{c}_n^{\mathrm{g}}[(\psi^{\mathrm{g}}_m(p))'-(\psi^{\mathrm{g}}_m(p'))'][(\psi^{\mathrm{g}}_n(p))'-(\psi^{\mathrm{g}}_n(p'))'] \nn \\
    &=& \frac{g^4 N_c^4}{2(2\pi)^5}\log\(\frac{\kappa}{g^2 N_c} \)\bar{c}_m^{\mathrm{g}}\bar{c}_n^{\mathrm{g}} T^6\,Y^{\mathrm{gg}}_{mn} \; ,
    \label{Lambdagg_mn}
   \eea
   where, $\ell_{m_1,m_2,m_3,m_4}$ is defined in Eq.~\eqref{eq:lms} and we define
   \bea
    Y^{\mathrm{gg}}_{mn} &=& \frac{1}{T^6}\sum_{m_1,m_2=1}^{\infty} \sum_{m_3=1}^{m_1}\sum_{m_4=1}^{m_2} \,\ell_{m_1}\ell_{m_2}\; \ell_{m_1,m_2,m_3,m_4}\; \int_{0}^{\infty} p^2 dp  \nn \\
    & \times&\int_0^{\infty}(p')^2dp'\text{e}^{-\beta (m_1 p +m_2 p')} [(\psi^{\mathrm{g}}_m(p))'-(\psi^{\mathrm{g}}_m(p'))'][(\psi^{\mathrm{g}}_n(p))'-(\psi^{\mathrm{g}}_n(p'))']\;.
   \eea
The factors related to temperature and the coupling constant remain consistent with those found in the perturbative Quark-Gluon Plasma (QGP) framework. The function $Y^{\mathrm{gg}}_{nm}$ encapsulates all the modifications or effects introduced by the background field in the system's behavior or properties. The diagram that plays a role in the contributions to the calculation is shown in Fig.~\ref{fig:ABCdiagram}a.
%
\subsubsection{Gluon-teen scattering }
Analogous to the gluon-gluon scattering, the matrix element for gluon-teen scattering at nonzero holonomy is
   \bea
   \bar{\Lambda}^{\mathrm{g}\begt}_{mn} &=& -\frac{1}{2}\frac{g^4 N_c^4}{2(2\pi)^5}\log\(\frac{\kappa}{g^2 N_c}\)\sum_{m_1,m_2=1}^{\infty} \sum_{m_3=1}^{m_1}\sum_{m_4=1}^{m_2} \ell_{m_1}\ell_{m_2} \;\ell_{m_1,m_2,m_3,m_4}\frac{T_d^2}{2}\int_{0}^{\infty} dp\,   \nn \\
    &&\hspace{-0.9cm}\times \int_0^{\infty}dp'\,p'^2 \text{e}^{-\beta (m_1 p +m_2 p')}[\bar{c}_m^{\begt}\,(\psi^{\begt}_m(p))'-\bar{c}_m^{\mathrm{g}}(\psi^{\mathrm{g}}_m(p'))'][\bar{c}_n^{\begt}(\psi^{\begt}_n(p))'-\bar{c}_n^{\mathrm{g}}(\psi^{\mathrm{g}}_n(p'))'] \nn \\
    &=&\frac{g^4 N_c^4}{2(2\pi)^5} \log \(\frac{\kappa}{g^2 N_c} \) T^4 T_d^2 (\bar{c}_m^{\mathrm{g}}\bar{c}_n^{\mathrm{g}}\overline{Y}^{\mathrm{gg}}_{mn}+\bar{c}_m^{\begt}\bar{c}_n^{\begt} \overline{Y}^{\begt \begt}_{mn}- \bar{c}_m^{\mathrm{g}}\bar{c}_n^{\begt} Y^{\mathrm{g}\begt}_{mn}-\bar{c}_m^{\begt} \bar{c}_n^{\mathrm{g}}Y^{\begt \mathrm{g}}_{mn}) \;,
    \label{eq:L_phi_ij}
  \eea
  with $\ell_{m_1,m_2,m_3,m_4}$ given in Eq.~\eqref{eq:lms} and
  \bea
    \overline{Y}^{\mathrm{gg}}_{mn} &=&-\frac{1}{2} \frac{1}{T^4 }\sum_{m_1,m_2=1}^{\infty} \sum_{m_3=1}^{m_1}\sum_{m_4=1}^{m_2} \,\ell_{m_1}\ell_{m_2}\; \ell_{m_1,m_2,m_3,m_4}\;  \nn \\
    & \times&\frac{1}{2}\int_{0}^{\infty}  dp \int_0^{\infty}(p')^2dp'\text{e}^{-\beta (m_1 p +m_2 p')} (\psi^{\mathrm{g}}_m(p'))'(\psi^{\mathrm{g}}_n(p))'\;,\nn\\ 
\overline{Y}^{\mathrm{g}\begt}_{mn} &=& -\frac{1}{2}\frac{1}{T^4 }\sum_{m_1,m_2=1}^{\infty} \sum_{m_3=1}^{m_1}\sum_{m_4=1}^{m_2} \,\ell_{m_1}\ell_{m_2}\; \ell_{m_1,m_2,m_3,m_4}\;   \nn \\
    & \times&\frac{1}{2}\int_{0}^{\infty}  dp \int_0^{\infty}(p')^2dp'\text{e}^{-\beta (m_1 p +m_2 p')} (\psi^{\mathrm{g}}_m(p'))'(\psi^\begt_n(p))'\;,\nn\\
        \overline{Y}^{\begt\mathrm{g}}_{mn} &=& -\frac{1}{2}\frac{1}{T^4}\sum_{m_1,m_2=1}^{\infty} \sum_{m_3=1}^{m_1}\sum_{m_4=1}^{m_2} \,\ell_{m_1}\ell_{m_2}\; \ell_{m_1,m_2,m_3,m_4}\;  \nn \\
    & \times&\frac{1}{2}\int_{0}^{\infty}  dp  \int_0^{\infty}(p')^2dp'\text{e}^{-\beta (m_1 p +m_2 p')} (\psi^\begt_m(p'))'(\psi^{\mathrm{g}}_n(p))'\;,\nn\\
        \overline{Y}^{\begt\begt}_{mn} &=&-\frac{1}{2} \frac{1}{T^4 }\sum_{m_1,m_2=1}^{\infty} \sum_{m_3=1}^{m_1}\sum_{m_4=1}^{m_2} \,\ell_{m_1}\ell_{m_2}\; \ell_{m_1,m_2,m_3,m_4}\;  \nn \\
    & \times&\frac{1}{2}\int_{0}^{\infty}  dp \int_0^{\infty}(p')^2dp'\text{e}^{-\beta (m_1 p +m_2 p')} (\psi^{\begt}_m(p'))'(\psi^\begt_n(p))'\;.
   \eea
The contributing diagram is presented in Fig.~\ref{fig:ABCdiagram}b.
\subsubsection{Teen-Teen scattering }
The matrix element for teen-teen scattering is
\bea
   \bar{\Lambda}^{\begt \begt}_{mn} &=& \frac{1}{4}\frac{g^4 N_c^4}{2(2\pi)^5}\log\(\frac{\kappa}{g^2N_c}\)\sum_{m_1,m_2=1}^{\infty} \sum_{m_3=1}^{m_1}\sum_{m_4=1}^{m_2} \ell_{m_1}\ell_{m_2} \;\ell_{m_1,m_2,m_3,m_4} \frac{T_d^2}{2}\int_{0}^{\infty} dp\,  \nn \\
    && \times \frac{T_d^2}{2}\int_0^{\infty}dp'\, \,e^{-\beta (m_1 p +m_2 p')}\bar{c}_m^{\begt}\bar{c}_n^{\begt}[(\psi^{\begt}_m(p))'-(\psi^{\begt}_m(p'))'][(\psi^{\begt}_n(p))'-(\psi^{\begt}_n(p'))'] \; \nonumber\\
    &=& \frac{g^4 N_c^4}{2(2\pi)^5} \log \(\frac{\kappa}{g^2 N_c} \)  \bar{c}_m^{\begt}\bar{c}_n^{\begt} T_d^4 T^2 Y^{\begt \begt}_{mn} \;,
   \eea
where 
\bea
 Y^{\begt\begt}_{mn} &=&\frac{1}{4} \frac{1}{T^2}\sum_{m_1,m_2=1}^{\infty} \sum_{m_3=1}^{m_1}\sum_{m_4=1}^{m_2} \,\ell_{m_1}\ell_{m_2}\; \ell_{m_1,m_2,m_3,m_4}\;   \nn \\
    & \times& \frac{1}{2}\int_{0}^{\infty}  dp \frac{1}{2}\int_0^{\infty}dp'\text{e}^{-\beta (m_1 p +m_2 p')} (\psi^{\begt}_m(p'))'(\psi^\begt_n(p))'\;.
\eea
Adding up the matrix elements from gluon-gluon, gluon-teen, and teen-teen scattering, one can rewrite the total matrix element as
\bea
\sum_{\begb,n}  \bar{c}_n^{\begb} (\bar{\mathcal{L}}\bar{c})_{n}^{\begb} &\simeq& 
\frac{g^4 N_c^4}{2(2\pi)^5}  \log \(\frac{\kappa}{g^2N_c}\) \nonumber\\
&&\hspace{-1cm}\times \sum_{m,n=1}^{K} \begin{bmatrix}
    \bar{c}_m^{\mathrm{g}} & \bar{c}_m^{\begt}
\end{bmatrix}
\begin{bmatrix}
   T^6 {Y}^{\mathrm{gg}}_{mn}+T^4\, T_d^2{\overline{Y}}^{\mathrm{gg}}_{mn} &  -T^4\,T_d^2 {Y}^{\mathrm{g}\begt}_{mn} \\  -T^4\, T_d^2 {Y}^{\begt \mathrm{g}}_{mn} &  T_d^4\, T^2 {Y}^{\begt \begt}_{mn}+ T^4\, T_d^2 {\overline{Y}}^{\begt \begt}_{mn}
   \end{bmatrix}
   \begin{bmatrix}
       \bar{c}_n^{\mathrm{g}} \\ \bar{c}_n^{\begt}
   \end{bmatrix}\;. 
   \label{cLc_mat}
   \eea
It is worth commenting about the terms from gluon-teen
scattering in this expression.  
Naively, one might expect that the only contribution to
the glue-glue term, $\bar{c}_m^{\mathrm{g}} \bar{c}_n^{\mathrm{g}}$, is from glue-glue scattering.
This is incorrect, since Eq.~\eqref{Lambdabd_nm_simp},
there are terms $\sim \bar{c}^{\begb} \bar{c}^\begd$,
where $\begb$ and $\begd$ can be {\it either} a gluon
or a teen field.  Hence gluon-teen scattering contributes
to $\bar{c}_m^{\mathrm{g}} \bar{c}_n^{\mathrm{g}}$ term, through the term  ${\overline{Y}}^{\mathrm{gg}}_{mn}$.  
Similarly, for the $\bar{c}_m^{\begt} \bar{c}_n^{\begt}$ term, there are terms
from gluon-teen scattering, $\sim  {\overline{Y}}^{\begt \begt}_{mn}$.

Because the scattering kernel in Eq. (\ref{cLc_mat})
has zero modes, it is necessary to add an additional term
to compute the bulk viscosity.  We discuss these details
in Appendix (\ref{appendix:zero_mode}).

\subsection{Results for the Bulk Viscosity}
We first compute $\zeta$ for small values of the loop.
Neglecting higher powers of the first loop, $\ell_1$,
and higher loops, $\ell_n$ for $n\geq 2$,
\bea\label{bulk_visc_moore}
\zeta\approx 
68.035\frac{( \Delta v_s^2)^2 \; T^3}{g^4 \ln({\kappa}/{g^2 N_c} )} 
\ell_1^2
 \;  .
\eea
Note, however, that in the pure glue theory the Polyakov loops are never small: near $T_d$,
$\langle \ell_1 \rangle \approx 0.5$ at $N_c = \infty$,
and $\langle \ell_1 \rangle \approx 0.4$ for $N_c = 3$.

Notice that the results with nonzero holonomy are very
different from perturbation theory.  In perturbation theory,
the theory is only nonconformal, with
$\Delta v_s^2 = 1/3 - v_s^2 \neq 0$, because of two
effects.  First, the gluons have a thermal mass, 
$\sim g T$.  Second, the coupling constant runs.
With $\eta \sim 1/(g^4 \log(1/g^2))$, this gives $\Delta v_s^2 \sim g^4$, and
$\zeta \sim (\Delta v_s^2)^2 \eta \sim g^4$,
up to logarithms of $\log(1/g^2)$.  
In contrast, our model is manifestly non-conformal,
with $\Delta v_s^2 = 1/3 - v_s^2 \neq 0$
from the nonperturbative equation of state.
Thus our bulk viscosity is 
$\zeta \sim (\Delta v_s^2)^2/(g^4 \log(1/g^2))$,
Eq.~\eqref{bulk_visc_moore}.

The entropy density $s(T)$ and $\Delta v_s^2$ are calculated using Eq.~\eqref{eq:pressure_N=3}, i.e., lattice pressure for three colors~\cite{Caselle:2018kap}. 
	\begin{figure}[ht]
		\begin{center}
        \includegraphics[scale=.7]{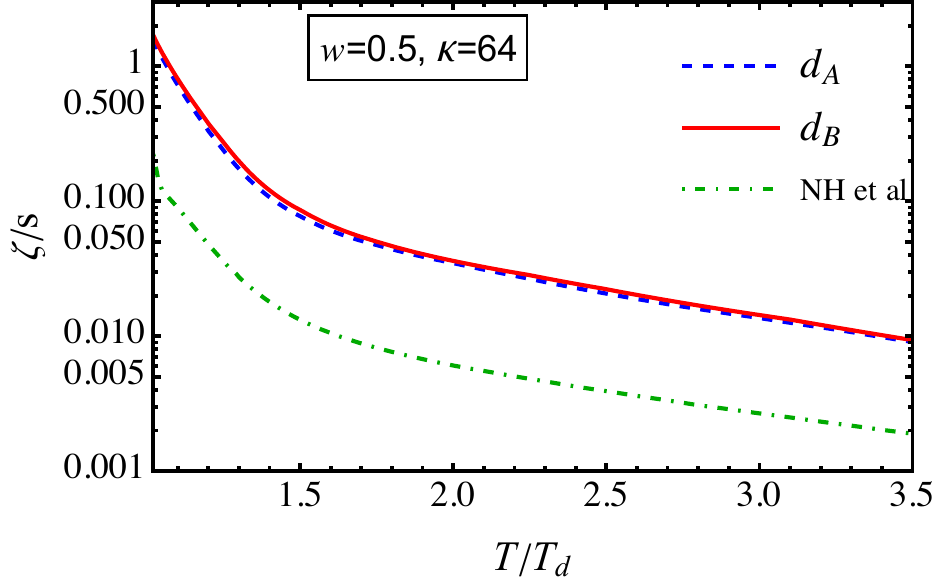}
			\caption{Ratio of the bulk viscosity to the entropy density with the ansatz of Eq.~\eqref{improved_loop}, with the parameters $w = 0.5$ and $\kappa = 64$. The dot-dashed line corresponds to the result in a Gribov-Zwanziger model~\cite{Jaiswal:2020qmj}.}
			\label{fig:bulk_viscosity}
		\end{center}
	\end{figure}
In Fig.~\ref{fig:bulk_viscosity}, we show the ratio of the bulk viscosity to the entropy,
 versus temperature, for the two ansatzes,
 $d_A$ and $d_B$.
 We choose the parameters to fix the coefficients in
 the logarithms as $w=1$ and $\kappa=64$. We compare our result with the Gribov-Zwanziger model of Ref.~\cite{Jaiswal:2020qmj},
 and find that our result for $\zeta/s$ is much larger,
 almost an order of magnitude. The value of $\zeta/s$ is $\sim 1.59$ and $1.71$ for $d_A$ and $d_B$. 

\begin{figure}[ht]
		\begin{center}
        \includegraphics[scale=.51]{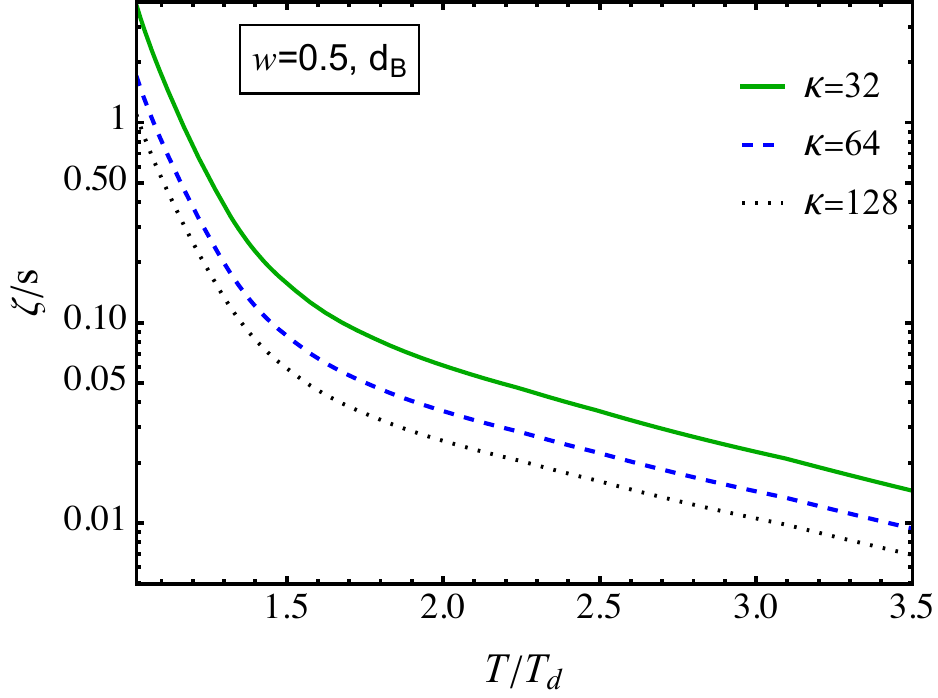}
             \includegraphics[scale=.51]{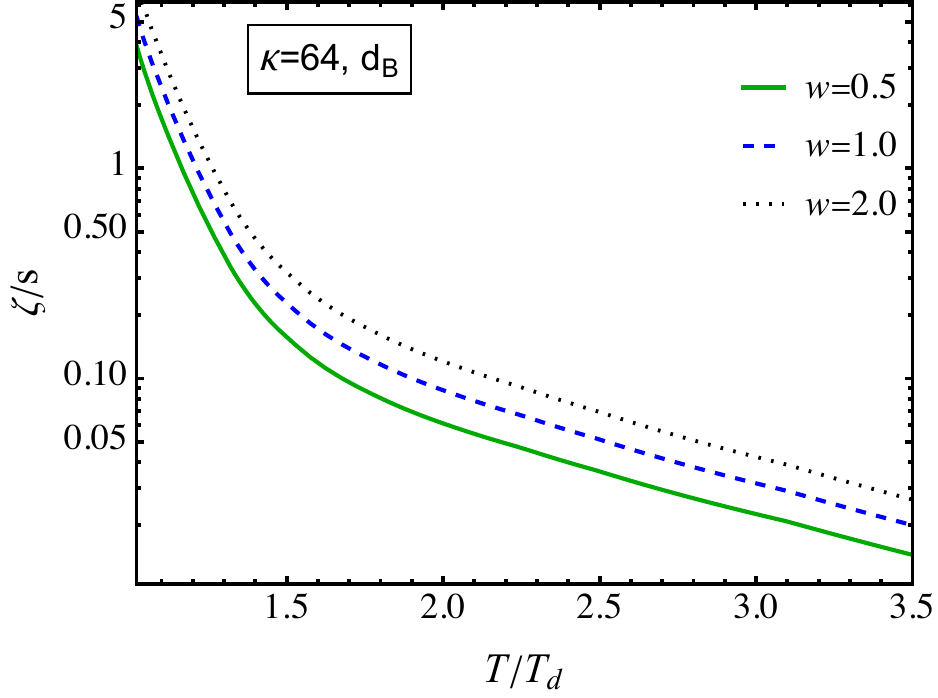}
			\caption{\textit{Left}: Ratio of the bulk viscosity to the entropy density for different parameter sets of $\kappa$. \textit{Right}: Ratio of the bulk viscosity to the entropy density for different parameter sets of $w$.}
        \label{fig:bulk_viscosity_withkw}
		\end{center}
	\end{figure}

 The dependencies of the parameters $\kappa$ and $w$ on $\zeta/s$ are shown in Fig.~\ref{fig:bulk_viscosity_withkw}. The left panel of Fig.~\ref{fig:bulk_viscosity_withkw} demonstrates a large sensitivity to the value of the parameter $\kappa$, which is greatest as $T \rightarrow T_d$. Similar to the parameter $\kappa$, there is a large uncertainty on $w$, especially as $T \rightarrow T_d$, although it is weaker than for $\kappa$.
%
The ratio of bulk viscosity $\zeta$ to shear viscosity $\eta$ has a thermodynamic significance as it reflects the type of dissipative response a medium exhibits in nonequilibrium conditions, particularly near phase transitions. 
	\begin{figure}[ht]
		\begin{center}
        \includegraphics[scale=.51]{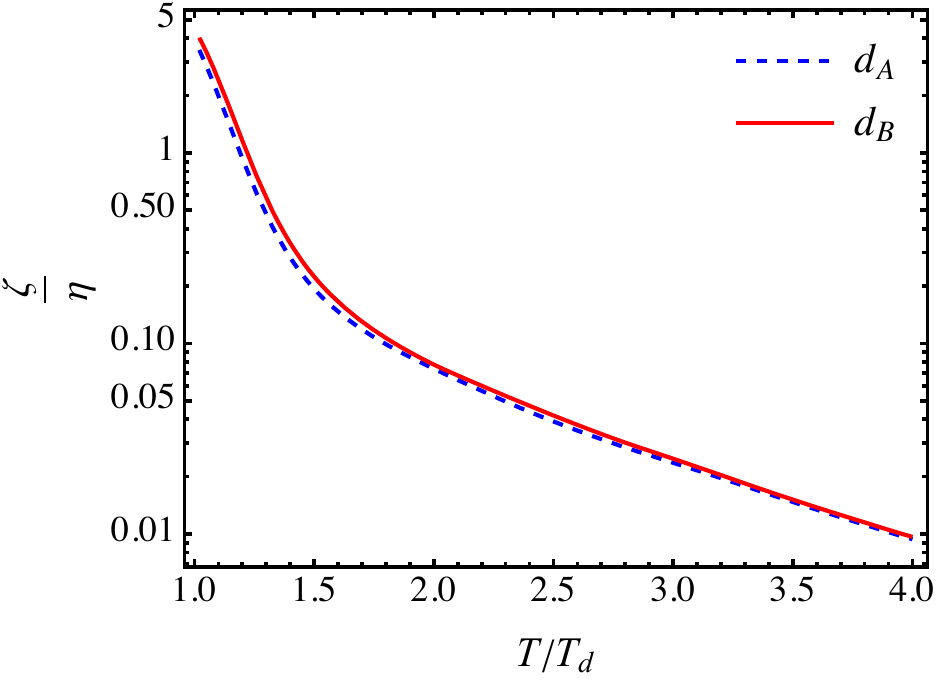}
        \includegraphics[scale=.515]{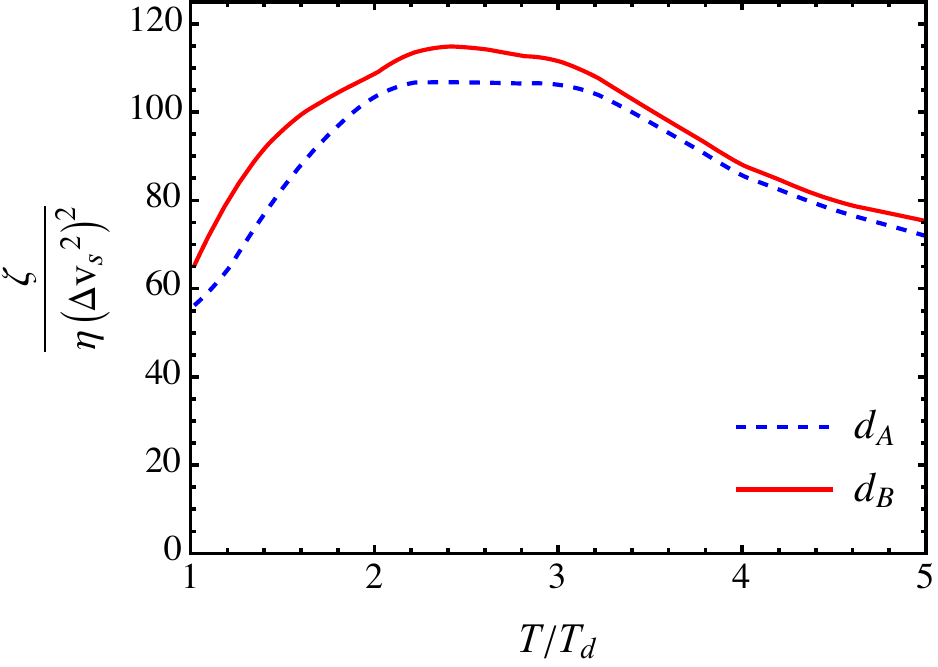}
			\caption{\textit{Left}: Ratio of the bulk viscosity to the shear viscosity with the ansatz of Eq.~\eqref{improved_loop}. \textit{Right}: Ratio of the bulk viscosity to the shear viscosity scaled with the square of the conformality measure ($\Delta v_s^2$).}
			\label{fig:bulkbyshear_viscosity}
		\end{center}
	\end{figure}

  Bulk viscosity, which relates to the resistance to volume changes, tends to increase near phase transitions due to critical slowing down, where the medium struggles to equilibrate under isotropic stress. This increase is tied to the medium's equation of state (EoS), which defines pressure and density relationships that influence the system's relaxation toward equilibrium.  Hence, the ratio $\zeta/\eta$ has an important physical significance. 
    
The left panel of Fig.~\ref{fig:bulkbyshear_viscosity} ~suggests a trend where the ratio decreases with increasing temperature. For $T/T_d$ values from around 1.5 to 3.5, $\zeta/\eta$ moves from values closer to 1 to significantly lower values.
Lastly, the right panel of Fig.~\ref{fig:bulkbyshear_viscosity} shows the temperature dependence of the ratio of the bulk viscosity to the shear viscosity scaled with the square of the conformality measure ($\Delta v_s^2$). 
We comment on the relationship to other models in
Eqs.~\eqref{buchel_bulk_viscosity} and \eqref{dusling_bulk_viscosity}.

\section{Summary}
\label{sec:summary}
The computation of the shear and bulk viscosities in our matrix model is admittedly involved, so we conclude
by summarizing our results, and comparing to those
obtained with different methods.

In heavy ion collisions at RHIC and LHC energies, experimentally
it is observed that a nearly ideal hydrodynamics works amazingly well~\cite{Luzum:2008cw,Casalderrey-Solana:2011dxg,Gale:2013da,Romatschke:2017ejr,Nagle:2018nvi}.
The standard paradigm for understanding this is the AdS/CFT correspondence: for a ${\cal N}=4$ 
$SU(N_c)$ gauge theory at $N_c = \infty$ and $g^2 N_c = \infty$ ~\cite{Policastro:2001yc,Kovtun:2004de,Son:2007vk},
\begin{equation}
\left. \frac{\eta}{s}\right|^{\rm AdS/CFT} = \frac{1}{4 \pi} \;\; , \;\;
\left. \frac{\zeta}{s}\right|^{\rm AdS/CFT} = 0 \; .
\label{AdS_visc}
\end{equation}
As a conformally invariant theory, the dimensionless
ratio $\eta/s$ is
independent of temperature, and the bulk viscosity
vanishes identically.

While our understanding of thermodynamics in equilibrium
rests upon computations in lattice QCD, measuring transport coefficients is extraordinarily difficult.  From the Kubo formula, they depend upon the relevant spectral densities as the frequency $\omega \rightarrow 0$.  However, this requires analytic continuation
from discrete, Euclidean $p_0$, to the continuous, Minkowski $\omega$.
The most recent results from the lattice are 
from Ref.~\cite{Altenkort:2022yhb}; at $1.5 \, T_d$, 
\begin{equation}
\left. \frac{\eta}{s}\right|_{ 1.5 T_d} \approx 0.15-0.48 \;\; ; \;\;
\left. \frac{\zeta}{s}\right|_{\rm 1.5 T_d} \approx 0.017-0.059 \; .
\label{lattice_visc}
\end{equation}
We can then compare results from AdS/CFT, Eq.~\eqref{AdS_visc},
and the lattice, Eq.~\eqref{lattice_visc}, to the results
at leading logarithmic order in matrix models. 
For the shear viscosity, $\eta/s$, see 
Figs.~\ref{fig:shear_viscosity_1},~\ref{fig:shear_viscosity_2},~\ref{fig:shear_viscosity_3}, and \ref{fig:shear_viscosity_4};
for the bulk viscosity, $\zeta/s$, see Figs.~\ref{fig:bulk_viscosity} and~\ref{fig:bulk_viscosity_withkw}.

In order to understand results from the lattice, it is essential to be able to model the spectral densities; see Refs. \cite{Moore:2008ws,Meyer:2008gt,Hong:2010at,Akamatsu:2017rdu}.

In a matrix model, the results
are rather insensitive to which ansatz we use
for the eigenvalue density, A or B in 
Eq.~\eqref{improved_loop}. They also depend
upon the parameter $w$, which determines the running
of the coupling, Eq.~\eqref{g2nc_exp}, and
the constant $\kappa$, which fixes the behavior beyond
leading logarithmic order in the coupling,
Eq.~\eqref{eq:IR_integral}. Using what we take
as the optimal values of $w = 0.5$ and $\kappa = 64$,
at $1.5 \, T_d$,from
Figs. \ref{fig:shear_viscosity_1} and
\ref{fig:bulk_viscosity} we find
\begin{equation}
\left. \frac{\eta}{s}\right|_{1.5 T_d} \approx  0.38-0.4 \; ; \;
\left. \frac{\zeta}{s}\right|_{1.5 T_d}
\approx 0.076 - 0.087 \; .
\label{matrix_model_1.5}
\end{equation}
The value for $\eta/s$ at $1.5 \, T_d$ is consistent with
Ref.~\cite{Altenkort:2022yhb}, although at the high end, while the bulk viscosity
is somewhat larger than that of Ref.~\cite{Altenkort:2022yhb}.
The matrix model, however, also provides results for the temperature dependence of each quantity.

If the coupling constant runs relatively slowly with temperature, as is true for the moderate values of the
coupling constant which we use, then as the temperature $T \rightarrow T_d$, the principle way by which  
$\eta/s$ decreases is
if the expectation value(s) of the Polyakov loops decrease
as well.  In the pure glue theory, for three colors the decrease in the expectation value of the first Polyakov loop is not very large, though:
from $\langle \ell_1 \rangle \rightarrow 1.0$ as
$T \rightarrow \infty$ to 
$\langle \ell_1 \rangle \approx 0.4$ at the
deconfining transition $T_d$; for $N_c = \infty$, 
$\langle \ell_1 \rangle \approx 0.5$ at $T_d$.
Thus at $T_d$, the decrease in the shear viscosity is
relatively moderate, as illustrated in Fig.~\ref{fig:shear_viscosity_1}. 
The precise temperature dependence is sensitive to varying $\kappa$, Fig.~\ref{fig:shear_viscosity_2},
or $w$, Fig.~\ref{fig:shear_viscosity_3}.
These changes tend to bring $\eta/s$ much closer to the
perturbative value.  

In contrast, for the bulk viscosity, $\zeta/s$ increases
sharply from $1.5 T_d$ to $T_d$. From Fig.~\ref{fig:bulk_viscosity}, with $w = 0.5$ and $\kappa = 64$,
at the transition temperature the bulk viscosity is large,
\begin{equation}
\left. \frac{\zeta}{s}\right|_{T_d} \approx 1.6-1.7 \; .
\end{equation}
The variation with respect to $\kappa$ and $w$ are shown
in Fig.~\ref{fig:bulk_viscosity_withkw}.   
Varying $\kappa$, at $T_d$ we find that $\zeta/s \approx
1.1 - 4$; varying $w$,
at $T_d$ the matrix model gives $\zeta/s \approx  4-7$.
The variation in these values and the sensitivity with
respect to $w$ and $\kappa$, indicate the limitations of our model and the attendant assumptions.  However, what is clear is that $\zeta/s$  increases {\it strongly} as $T \rightarrow T_d$.

It is also useful to compare our results with those obtained in other effective models.

Corrections to the perturbative results for the shear and
bulk viscosity were estimated in
Refs.~\cite{Jackson:2017hfz,Ghiglieri:2018dib,Moore:2020pfu,Danhoni:2022xmt,Danhoni:2024ewq,MacKay:2024jus,Xu:2007jv,Carrington:2009xf,Chen:2009sm,Chen:2011km}, and do tend to decrease $\eta/s$.

Another approach is to use a quasiparticle model, where the gluon
masses are $m_{\rm gl} \sim g(T) T$.  Taking $g(T)$ to
increase strongly as $T \rightarrow T_d$, a small pressure
is obtained by letting $g(T)$ become large
as $T \rightarrow T_d$, so 
the gluon quasiparticles are (very) heavy. 
This was done in Refs.~\cite{Bluhm:2009ef,Bluhm:2010qf},
with results that depend strongly on temperature.
As $T$ decreases, it reaches a minimum at 
$T_{\rm min} \approx 1.1 \, T_d$, and then increases sharply.
The shear viscosity is very small 
at $T_{\rm min}$, with $\eta/s(T_{\rm min}) \approx 0.1$,
close to the AdS/CFT bound.  For the bulk viscosity, it
is extremely small until $\approx 1.2 \, T_d$, where it
then increases sharply, and is large, $\zeta/s \approx 0.25$
at $T_{\rm min}$.  In contrast to our model, $\eta/s$ is
much smaller at $T_{\rm min}$, while our
values for $\zeta/s \approx 0.76- 0.81$ 
at $1.1 T_d$ (see Fig.~\ref{fig:bulk_viscosity}) is larger than the quasiparticle model.
Still, given the assumptions in both the quasiparticle model,
and our matrix model, the results are not that far apart.

There are also results for the bulk viscosity from
other models.  For the bulk viscosity, in ${\cal N}^* =2$ SUSY gauge theories Buchel~\cite{Buchel:2007mf}
established the bound
\begin{equation}
    \frac{\zeta}{\eta} \times \frac{1}{(1/3 - v_s^2)}
    \geq 2 \; .
    \label{buchel_bulk_viscosity}
    \end{equation}
However, this bound is not exact~\cite{Yarom:2009mw,Buchel:2011uj,Buchel:2011wx,Rebhan:2011vd}.

Bounds on the bulk viscosity were also obtained by Dusling and Schaefer in Ref.~\cite{Dusling:2011fd}.  In the relaxation time approximation to kinetic
theory,
\begin{equation}
    \frac{\zeta}{\eta} \times
    \frac{1}{ (1/3- v_s^2)^2 }\approx 15 \; .
    \label{dusling_bulk_viscosity}
\end{equation}
This is stronger than the bound from ${\cal N}^* = 2$ SUSY,
in that, it involves the square of the difference in the speed
of sound from the conformal limit, where $v_s^2 = 1/3$.  

In the right panel of Fig.~\ref{fig:bulkbyshear_viscosity}
we plot the same quantity as on the left-hand side
of Eq.~\eqref{dusling_bulk_viscosity}, $\zeta/\eta$
times $1/(1/3 - v_s^2)^2$.  Unlike the relaxation time
approximation, unsurprisingly, this is temperature dependent,
with a value $\approx  56-65$ at
$T_d$.  The ratio then increases to a maximum of
$\approx  105-114$ at $\approx 2.5 T_d$, and then
it falls off as $T$ increases further.
In this ratio, there is sensitivity to which ansatz is
used.

Still, the right panel of Fig.~\ref{fig:bulkbyshear_viscosity} shows
that the relaxation time approximation gives a good rule
of thumb, although the matrix model value is about five
time larger, with a non-trivial dependence upon temperature.

There are other models to which we can compare.
Using a Gribov-Zwanziger model, Refs.~\cite{Florkowski:2015rua,Jaiswal:2020qmj,Madni:2024xyj}
find that $\zeta/s \approx 0.2$ at $T_d$.  While large, ours
is approximately an order of magnitude larger, 
$\zeta/s \approx 1.7 $ at $T_d$.

There are also results from holographic models~\cite{DeWolfe:2018dkl}.
Anisotropic holographic models find that while the
transverse shear viscosity obeys the AdS/CFT bound,
that longitudinal to the direction of the anisotropy
can be smaller than the bound~\cite{Rebhan:2011vd}.
Reference~\cite{Mamo:2012sy} also finds that in holographic
models, $\eta/s$ can be below the AdS/CFT bound.
Reference~\cite{Cremonini:2012ny} finds $\eta/s$ has
a minimum at $T_d$, with a model dependent value
above $1/(4 \pi)$.  
Reference~\cite{Li:2014dsa} construct a model where $\eta/s$
is below the AdS/CFT bound, while $\zeta/s \approx 0.12$ at $T_d$.
Reference~\cite{Finazzo:2014cna} constructs a bottom-up
model where $\eta/s$ is constant, equal to the AdS/CFT
bound, with a small bulk viscosity, $\zeta/s \approx 0.045$ at $T_d$.
Reference~\cite{Yaresko:2014fia} found a large
value for the bulk viscosity at the transition,
$\zeta(T)/T^3 \approx 0.1$ at $T_d$.
Reference~\cite{Attems:2016ugt} finds a violation of the Buchel bound~\cite{Buchel:2007mf}, in a model-dependent fashion.
Reference~\cite{Mamo:2016oli} find $\zeta/s \approx 0.2$ at $T_d$.
At $T_d$, Ref.~\cite{Ballon-Bayona:2021tzw} find a small minimum
for the shear viscosity, $\eta/s \approx 0.2$,
and for the bulk viscosity, $\zeta/s \approx 0.12$.

In all, at $T_d$ holographic models give a value of $\eta/s$
near the AdS/CFT bound, while $\zeta/s$ is larger than that from the lattice, but much smaller than in a matrix model.

Of course, what is relevant to QCD, and real experiments, is
the theory with dynamical quarks.  In a matrix model,
the results for $\eta/s$ at the chiral
phase transition, $T_\chi$ will {\it certainly} be much
smaller than in the pure glue theory.  This is because
while the (first) Polyakov loop is relatively large
at $T_d$, at the chiral phase transition, at $T_\chi$, the expectation value of the (first) Polyakov loop is rather small, 
$\langle \ell_1 \rangle(T_\chi) \approx 0.04$, Fig.~1
of Ref.~\cite{Petreczky:2015yta}; see, also,
Ref.~\cite{Bazavov:2016uvm}.  This suggests that QCD
enters a regime where, at $T_\chi$, chiral symmetry restoration
occurs in a regime with nearly perfect confinement.  In a matrix
model, inexorably this will generate a {\it much} smaller value of
$\eta/s$~\cite{Pisarski:2016ixt}. The implications for the bulk viscosity are less clear, and can only be determined by computation, which is actively underway.

\acknowledgments{We thank Daisuke Satow for his collaboration in the initial stages of this project, and for his notes on the bulk viscosity. Our thanks go to Palash Baran Pal, who developed the Bangtex package, and Aritra Ghosh, who modified it for pdfLaTeX~\cite{Ghosh_pdfLaTeX_Bengali_2022}, enabling us to insert Bengali characters.  M.D. and N.H. are supported by the DAE, Govt. of India. The work of R.G. was partially supported by the US National Science Foundation under Grant No.~PHY-2209470. N.H. acknowledges support from the Alexandar von Humboldt Foundation, Germany, and Justus
Liebig University Giessen, Germany where part of the work is done.  R.D.P. was supported by the U.S. Department of Energy under contract DE-SC0012704, and by the Alexander von Humboldt Foundation.}
 
\appendix
\section{Sign in the Boltzmann equation with teen ghosts}
\label{sec:ghost_sign}
The sign of the Boltzmann equation with teen ghosts is a bit non-trivial.
Here we explain how to obtain the sign in Eqs.~\eqref{eq:coll_term} and \eqref{eq:collision_term}.
The point is that the teen ghost is equivalent to a fermion with periodic boundary conditions. 
In other words, it is equivalent to a fermion with an imaginary chemical potential $i\pi T$,
which corresponds to the replacement of the fermion distribution function $f_f$ with $-f_\begt$.
This correspondence is more obvious in the case of equilibrium systems:
\begin{equation}
     \frac{1}{\mathrm{e}^{(E_p - i (Q^a-Q^b + \pi T))/T} + 1}
    = -\frac{1}{\mathrm{e}^{(E_p - i (Q^a-Q^b))/T} - 1}=- f^0_{ab} \; .
    \label{eq:sign_teen_function2}
\end{equation}
From this correspondence, the sign of the left-hand side of the Boltzmann equation in Eq.~\eqref{eq:coll_term} should be obvious.
Let us see what happens to the sign of the collision term. 
In gluon-fermion scattering, we have a combination of distribution functions,
$f_{\mathrm{g}}f_{f}(1+f_{\mathrm{g}})(1-f_{f})-
(1+f_{\mathrm{g}})(1-f_{f})f_{\mathrm{g}}f_{f}$.
This turns to
\bea
&&f_{\mathrm{g}}f_{f}(1+f_{\mathrm{g}})(1-f_{f})-
(1+f_{\mathrm{g}})(1-f_{f})f_{\mathrm{g}}f_{f}\notag\\
&&\to 
(-)\Bigl[f_{\mathrm{g}}f_{\begt}(1+f_{\mathrm{g}})(1+f_{\begt})-
(1+f_{\mathrm{g}})(1+f_{\begt})f_{\mathrm{g}}f_{\begt}\Bigr]
\eea
in the gluon-teen scattering. We obtain the negative overall sign $(-)$.
Similarly, for teen-teen scattering, we have
\bea
&&f_{f}f_{f}(1-f_{f})(1-f_{f})-
(1-f_{f})(1-f_{f})f_{f}f_{f}\notag\\
&&\to 
(+)\Bigl[f_{\begt}f_{\begt}(1+f_{\begt})(1+f_{\begt})-
(1+f_{\begt})(1+f_{\begt})f_{\begt}f_{\begt}\Bigr] \,.
\eea
The sign is positive. 
These considerations show that the sign is determined by the evenness of the incoming ghost number,
which corresponds to $\signf_\begb \, \signf_\begd$ in Eq.~\eqref{eq:collision_term}.

\section{Energy and pressure for teen particles}
\label{appendix:teen_energy}
In kinetic theory, the energy-momentum tensor for teen ghost is given by
\begin{eqnarray}
 T^{\mu\nu}=-\int \frac{d\Omega}{4\pi}\int \frac{d^3 p}{(2\pi)^3}\frac{p^\mu p^\nu}{E_p}\frac{1}{e^{\beta E_p}-1}\,.
\end{eqnarray}
Now,
\begin{eqnarray}
 T^{00}&=&-\int \frac{d\Omega}{4\pi}\int \frac{d^3 p}{(2\pi)^3}\frac{p^0 p^0}{E_p}\frac{1}{e^{\beta E_p}-1}\,,\nonumber\\
 \mathscr{E}^{\mathrm{kt}}&\simeq& -2\int_0^\infty \frac{dp_\parallel}{2\pi}\int\frac{p_\perp dp_\perp}{2\pi}|p_\parallel| \frac{1}{e^{\beta |p_\parallel|}-1}\,,\nonumber\\
 \mathscr{E}^{\mathrm{kt}}&=&-\frac{T^2 T_d^2}{24}\,.
\end{eqnarray}
Take $\hat n\equiv \hat z$ and 
$p_\perp\sim T_d \ll p_\parallel$.
The longitudinal pressure along the z-direction is
\begin{eqnarray}
 T^{33}&=&-\int \frac{d\Omega}{4\pi}\int \frac{d^3 p}{(2\pi)^3}\frac{p^3 p^3}{E_p}\frac{1}{e^{\beta E_p}-1}\nn\\
 &\simeq& -2\int_0^\infty \frac{dp_\parallel}{2\pi}\int\frac{p_\perp dp_\perp}{2\pi}|p_\parallel| \frac{1}{e^{\beta |p_\parallel|}-1}\,,\nn\\
 \mathscr{P}^{\mathrm{kt}}_z&=&-\frac{T^2 T_d^2}{24}\,.
 \label{pz_tmunu}
\end{eqnarray}
The perpendicular pressure is
\begin{eqnarray}
 T^{11}=T^{22}&=&-\int \frac{d\Omega}{4\pi}\int \frac{d^3 p}{(2\pi)^3}\frac{p_\perp^2}{E_p}\frac{1}{e^{\beta E_p}-1}\,,\nn\\
 \mathscr{P}^{\mathrm{kt}}_\perp&\simeq& 0\,.\label{pxy_tmunu}
\end{eqnarray}
The difficulty is that in Eq.~\eqref{pz_tmunu}, one should
average over the direction of the teen field, which gives
$\mathscr{P}^{\mathrm{kt}} = - T^2 T_d^2/72$, so that in kinetic
theory, the teen field looks three dimensional, with
$\mathscr{E}^{\mathrm{kt}} = 3 \mathscr{P}^{\mathrm{kt}}$.  If we compute
directly from thermodynamics, though, we obtain
$\mathscr{P}^{\mathrm{th}} = \mathscr{E}^{\mathrm{th}} 
= - T^2 T_d^2/24$, as expected
for a two dimensional field.

Thus, as noted in Sec. (\ref{sec:bulk_viscosity}),
kinetic theory gives the wrong pressure for a teen field.  
In computing the bulk viscosity, we avoided this
by taking $\Delta v_s^2$
from thermodynamics.  In turn this is fit
to the lattice results, through the functions $d_A(T)$ and $d_B(T)$, Eq.~\ref{improved_loop}.

We next discuss how to obtain a kinetic theory for teen particles which is consistent with a direct computation in thermodynamics.
This considers the analogy to a magnetic field,
where there is a difference between computing at constant magnetic field or constant flux.

%
We start with the energy density and pressure, defined
from the partition function $\mathcal{Z}$, as
\bea
\mathscr{E}=-\frac{1}{V} \frac{\mathrm{~d} \log \mathcal{Z}}{\mathrm{~d}(1 / T)}\,, \quad \quad \mathscr{P}_i=\frac{T}{V} L_i \frac{\mathrm{~d} \log \mathcal{Z}}{\mathrm{~d} L_i}\,,\label{deri_part}
\eea
where $L_i$ is the length of the system along i-th direction. The energy density in Eq.~\eqref{deri_part} becomes
\bea
\mathscr{E}&=&(2-1)T \int \frac{d \Omega_n}{4 \pi} \int \frac{d p_{\|}}{2 \pi} \int \frac{d^2 p_{\perp}}{(2 \pi)^2} \ln \left(1-e^{-\beta\left|p_{\|}\right|}\right)\nn\\
&=&T \frac{1}{4 \pi^2} \int_0^{T_d} d p_{\perp} p_{\perp} \int_{-\infty}^{\infty} d p_{\|} \ln \left(1-e^{-\beta\left|p_{\|}\right|} \right)\nn\\
&=&-\frac{1}{24} T_d^2 T^2\,.
\eea
Note that the energy density is the same, whether computed
in kinetic theory, or from the partition function.

For the pressure $\mathscr{P}_i$, one can calculate the derivatives of the partition function in in two ways:
\begin{enumerate}
\item Keeping the energy density $\mathscr{E}$ constant,
analogous to constant magnetic field $B$.
\item Keeping the longitudinal energy density $\mathscr{E}L_xL_y$ constant,
which is analogous to constant magnetic flux $\Phi$.
\end{enumerate}

In the first case, at constant $\mathscr{E}$, the pressure is
\bea
\mathscr{P}^{\mathrm{th}}_i=\frac{T}{V} \frac{\mathrm{~\partial} \log \mathcal{Z}}{\mathrm{\partial}\log L_i}=-\frac{1}{24} T_d^2 T^2\,.
\eea

The second case, with constant longitudinal energy density,
corresponds to the conventional energy momentum tensor, in which the pressure is given by Eqs.~\eqref{pz_tmunu} and~\eqref{pxy_tmunu}. 
To see this, note that
\begin{eqnarray}
\mathscr{P}_i^{\mathrm{kt}}&=&\frac{T}{V} \frac{\partial \log \mathcal{Z}}{\partial \log L_i}+\left.\frac{T}{V} \frac{\partial \log \mathcal{Z}}{\partial \mathscr{E}} \cdot \frac{\partial\mathscr{E}}{\partial \log L_i}\right|_{\mathrm{kt}}=\mathscr{P}_i^{\mathrm{th}}+\left.\frac{T}{V} \frac{\partial \log \mathcal{Z}}{\partial \mathscr{E}} \cdot \frac{\partial\mathscr{E}}{\partial \log L_i}\right|_{\mathrm{kt}}\,.
\end{eqnarray}
Since $\mathscr{E}$ decreases with $L_x$ and $L_y$,
\begin{eqnarray}
\mathscr{P}_z^{\mathrm{kt}}&=&\mathscr{P}_z^{\mathrm{th}}=-\frac{1}{24}T^2T_d^2\qquad \mbox{and}\nonumber\\
\mathscr{P}_\perp^{\mathrm{kt}}&=& \mathscr{P}_\perp^{\mathrm{th}}+\frac{\partial \mathscr{P}_\perp^{\mathrm{th}}}{\partial\mathscr{E}}(-\mathscr{E})
=\mathscr{P}_\perp^{\mathrm{th}}+(1)(-\mathscr{P}_\perp^{\mathrm{th}})=\mathscr{P}_\perp^{\mathrm{th}}-\mathscr{P}_\perp^{\mathrm{th}}=0\,.
\nonumber
\end{eqnarray}
Thus, the results from $T^{\mu\nu}$ and thermodynamics are consistent with each other, whether one computes at
constant longitudinal energy density, or at constant energy
density.
%

\section{Zero mode}
\label{appendix:zero_mode}
The analysis of the collision kernel for the bulk viscosity,
Eq.~\eqref{cLc_mat}, involves zero modes
that do not arrive for the shear viscosity.
Since only $2\leftrightarrow 2$ scattering processes are
considered, there are zero modes for the conservation
of particle number.  There are also zero modes
for energy conservation.
Because of this, finding the inverse of the $\bar{\mathcal{L}}$ matrix is ill-defined.

If only gluons scatter, in Eq. 
(\ref{cLc_mat}) only the term $\sim Y^\mathrm{gg}_{mn}$ 
contributes, and has
zero modes. Reference~\cite{Arnold:2006fz}
adds a term to lift this degeneracy
and computes the inverse of $\bar{\mathcal{L}}$, proportional to a parameter
$\lambda$.  They then check, numerically, that the result is independent of $\lambda$. 

Adding teen fields complicates the analysis, as teens
contribute to a variety of processes.  
The scattering of gluons into themselves, $\mathrm{gg }\rightarrow \mathrm{gg}$,
gives a term $\sim T^6 Y^\mathrm{gg}_{mn}$;
that from teens with themselves, 
$\begt \begt \rightarrow \begt \begt$, gives a term
$T_d^4 T^2 Y^{\begt \begt}_{mn}$.
These contributions have zero modes.

However, there are additional scattering from
glue teen scattering, $\begt \mathrm{g} \rightarrow \begt \mathrm{g}$.
For the
glue-glue term in the upper left hand corner of Eq.
(\ref{cLc_mat}), the total is 
$T^6 \,{Y}^{\mathrm{gg}}_{mn}+T^4\,T_d^2{\overline{Y}}^{\mathrm{gg}}_{mn}$.
Similarly, for the teen-teen term in the 
lower right hand corner, the total is 
$T_d^4\, T^2 {Y}^{\begt \begt}_{mn}+ T^4\, T_d^2 {\overline{Y}}^{\begt \begt}_{mn}$.
Because of the additional contributions, these terms no longer have zero modes.

However, there are also off-diagonal terms in
Eq.~\eqref{cLc_mat}. These are
$T^4 T_d^2 Y^{\mathrm{g}\begt}_{mn}$ and $T^4 T_d^2 Y^{\begt \mathrm{g}}_{mn}$,
and each has zero modes.  
By trial and error, we find that a term which lifts the
degeneracy of the scattering kernel in Eq. (\ref{cLc_mat}) is
\bea\label{zero_mode_removal}
\bar{\Lambda}^{\begb, \begd}_{mn}\rightarrow \bar{\Lambda}^{\begb, \begd}_{mn}+(1-\delta^{\begb, \begd})\; \frac{\lambda g^4 N_c^4}{2(2\pi)^5}  \; \log \(\frac{\kappa}{g^2N_c}\) \bigg(\int \frac{d^3p}{(2\pi)^3} p^2 \frac{\partial \psi_{\begb,m}}{\partial p} \bigg) \bigg(\int \frac{d^3k}{(2\pi)^3} k^2 \frac{\partial \psi_{\begd,m}}{\partial k} \bigg)\,.\ \ \hspace{.5cm}
\eea
We checked numerically that for any positive $\lambda$,
$T \sum_{ \begb}\sum_{n=0}^\infty V_n^{\begb} \big(\bar{\mathcal{L}}^{-1}V\big)_n^{\begb}$ is independent of $\lambda$. 

\begin{figure}[ht]
    \centering
    \includegraphics[width=0.7\linewidth]{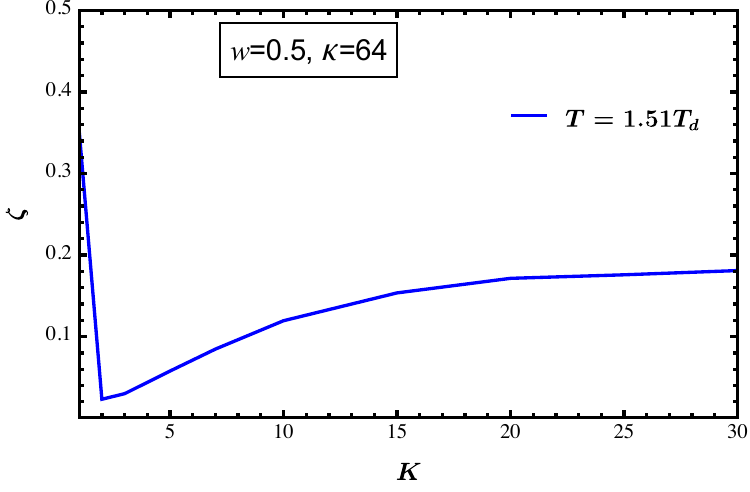}
    \caption{Variation of bulk viscosity with increasing basis number at $T/T_d=1.51$, with $w=0.5$ and $\kappa=64$.}
    \label{fig:bulk_basis_saturation}
\end{figure}

In Fig.~\ref{fig:bulk_basis_saturation} 
we illustrate the convergence of bulk viscosity versus $K$,
the dimension of the basis functions, for one typical value
of the temperature and other parameters.
The convergence for other values was
consistently similar.  While $K$ is in principle infinite,
in Ref. \cite{Arnold:2006fz} they found good convergence
for small values of $K \leq 5$.  
In contrast, we
uniformly found convergence by values of $K \sim 20 - 30$,
and we computed using $K=30$.

%

\end{document}